\documentclass[article]{IEEEtran}
\usepackage{graphicx}
\usepackage{setspace}
\usepackage{longtable}
\usepackage{url}
\usepackage{cite}
\usepackage{multirow}
\usepackage[symbol]{footmisc}

\usepackage[usenames, dvipsnames]{color}

\newcommand{\astfootnote}[1]{
	\let\oldthefootnote=\thefootnote
	\setcounter{footnote}{0}
	\renewcommand{\thefootnote}{\fnsymbol{footnote}}
	\footnote{#1}
	\let\thefootnote=\oldthefootnote
}

\title{Ultra-Low Latency (ULL) Networks: The IEEE TSN
  and IETF DetNet Standards and Related 5G ULL Research
  \thanks{Please direct correspondence to M.~Reisslein.}}

\author{Ahmed Nasrallah, Akhilesh S. Thyagaturu, Ziyad Alharbi,
  Cuixiang Wang, Xing Shao, Martin Reisslein, and Hesham ElBakoury
\thanks{A.~Nasrallah, A.S.~Thyagaturu, Z. Alharbi, and M.~Reisslein are with the
 School of Electrical, Computer, and Energy Eng., Arizona State University,
  Tempe, AZ 85287-5706, USA,
Phone: 480-965-8593, Fax: 480-965-8325,
(e-mail: \{ahnasral, athyagat, zalharbi, reisslein\}@asu.edu).}
\thanks{C.~Wang and X.~Shao are with the Dept. of Internet of Things Eng.,
  Yancheng Inst. of Techn., Yancheng, Jiangsu Province, P.R. China
  (e-mail: cwang288@asu.edu, xshao13@asu.edu).
  C.~Wang and X.~Shao visited Arizona State
  Univ., Tempe, while contributing to this study.
  C.~Wang was supported by the Overseas Training Program of the Yancheng
  Institute of Technology.
X.~Shao was supported by a Jiangsu Provincial Government Scholarship for Overseas Learning.}
\thanks{H.~ElBakoury is with Futurewei Technologies Inc.,
2330 Central Expressway, Santa Clara, CA, 95050, USA
    (e-mail: Hesham.ElBakoury@huawei.com).}  }

\begin{document}

\maketitle

\begin{abstract}
Many network applications, e.g., industrial control, demand Ultra-Low
Latency (ULL).  However, traditional packet networks can only reduce
the end-to-end latencies to the order of tens of milliseconds.
The IEEE 802.1 Time Sensitive Networking (TSN) standard and
  related research studies have sought to provide link layer support
  for ULL networking, while the emerging IETF Deterministic Networking
  (DetNet) standards seek to provide the complementary network layer
  ULL support.  This article provides an up-to-date comprehensive
  survey of the IEEE TSN and IETF DetNet standards and the related
  research studies.  The survey of these standards and research
  studies is organized according to the main categories of flow
  concept, flow synchronization, flow management, flow control, and
  flow integrity.  ULL networking mechanisms play a critical role in
  the emerging fifth generation (5G) network access chain from
  wireless devices via access, backhaul, and core networks.  We survey
  the studies that specifically target the support of ULL in 5G
  networks, with the main categories of fronthaul, backhaul, and
  network management. Throughout, we identify the pitfalls and
limitations of the existing standards and research studies.  This
survey can thus serve as a basis for the development of standards
enhancements and future ULL research studies that address the
identified pitfalls and limitations.
\end{abstract}


\section{Introduction}

\subsection{Motivation}
Traditional networks which provide end-to-end connectivity to the
users have only been successful in reducing the operating end-to-end
latencies to the order of tens of milliseconds. However, present and
future applications demand Ultra-Low Latency (ULL).  For instance, the
end-to-end latencies should be on the order of a few
microseconds to a few milliseconds for
industrial applications~\cite{Wollschlaeger2017}, around 1~millisecond
for the tactile Internet~\cite{fet2014tac,Maier2016}, and on the order
of 100~microseconds for the one-way fronthaul in wireless cellular
networks.  For example, critical healthcare applications, e.g., for
tele-surgery, and transportation applications~\cite{fin2017inc}
require near real-time connectivity. Throughput requirements largely
dependent on the application needs, which may vary widely from small
amounts of IoT data to large exchanges of media data transfers to and
from the cloud (or the fog to reduce latency)~\cite{sch2017lat}.  Additionally,
autonomous automotive vehicles~\cite{sam2018lev}, augmented and virtual reality (AR/VR),
as well as robotic applications, which are essential for Industrial IoT
(IIoT), may require both high data rates as well as ULL~\cite{del2006con,del2006con2,gut2018tim,pri2018dyn}. The high data
rates may be required for transporting video feeds from cameras that
are used to control vehicles and robots~\cite{bac2017min}.  Therefore,
in such heterogeneous environments and applications, a dedicated
mechanism to universally accommodate a diverse range of ULL requirements
would be very helpful~\cite{gar2018avn}.

\subsection{Contributions and Organization of this Survey}
This article provides a comprehensive up-to-date survey of standards
and research studies addressing networking mechanisms for ULL
applications.  Section~\ref{tsn:std:sec} covers the IEEE TSN standards
that have grown out of the AVB standards and focus primarily on the
link layer, while Section~\ref{tsn:res:sec} covers the ULL research
studies related to TSN.  Section~\ref{detnet:std:sec} covers the
Internet Engineering Task Force (IETF) Deterministic Networking
(DetNet) standards developments, while Section~\ref{detnet:res:sec}
covers the ULL research studies related to DetNet.  This sequence of
the section on standards followed by the section on related research
studies is inspired by the temporal sequence of the development of the
ULL field, where standard development has typically preceded research
studies.

A large portion of the ULL applications will likely involve
  wireless communications, whereby the fifth generation (5G) wireless
  systems will play a prominent role.  In particular, the emerging
  tactile Internet paradigm with end-to-end target latencies below
  1~ms is tightly coupled to the ongoing 5G
  developments~\cite{aij2017sha,aij2017rea,ant2018tow,del2018net,sch2017lat}.
  The support of 5G wireless ULL communications services will likely
  heavily rely on the TSN and DetNet standards and research results.
  On the other hand, due to prevalence and importance of wireless
  communications in today's society, the particular 5G wireless
  context and requirements will likely influence the future
  development of ULL standards development and research.  We believe
  that for a thorough understanding of the complete ULL research area
  it is vital to comprehensively consider the ULL standards,
  namely TSN and DetNet, as well as a main ``application domain'' of
  ULL standards and research results.
  We anticipate that 5G wireless communications will emerge as a highly
  important application domain of ULL standards and research results and
  we therefore survey ULL related standards and
  research studies for 5G wireless systems in Section~\ref{sec:5g}.

Section~\ref{fut:res} identifies the main gaps and limitation of the
existing TSN and DetNet standards as well as
ULL related 5G standards and research studies and outlines future research
directions to address these gaps and limitations.

\subsection{Related Literature}

While to the best of our knowledge there is no prior survey on time
sensitive networking (TSN), there are prior surveys on topics that
relate to TSN. We proceed to review these related surveys and
differentiate them from our survey.

A survey on general techniques for reducing latencies in Internet
Protocol (IP) packet networks has been presented
in~\cite{briscoe2016reducing}. The survey~\cite{briscoe2016reducing}
gave a broad overview of the sources of latencies in IP networks and
techniques for latency reduction that have appeared in publications up
to August 2014. The range of considered latency sources included the
network structure, e.g., aspects of content placement and service
architectures, the end point interactions, e.g., aspects of transport
initialization and secure session initialization, as well as the
delays inside end hosts, e.g., operating system delays.  In contrast,
we provide an up-to-date survey of the IEEE Time Sensitive Networking
(TSN) standard for the link layer and the Deterministic Networking
(DetNet) standard for the network layer, and related research studies.
Thus, in brief, whereas the survey~\cite{briscoe2016reducing} broadly
covered all latency sources up to 2014, we comprehensively cover the
link and network layer latency reduction standards and studies up to
July 2018.

A few surveys have examined specific protocol aspects that relate to
latency, e.g., time synchronization protocols have been surveyed
in~\cite{levesque2016survey,Luo2012standardization},
routing protocols have been surveyed
in~\cite{guc2017uni,lau2016rou,sil2017sur}, while congestion control protocols
have been covered in~\cite{luo2017standardization,xu2016con}.  Several surveys have covered
latency reduction through mobile edge and fog computing, e.g.,
see~\cite{mac2017mob,mou2018com,ni2018sec,tal2017mul}.  Also, the impact of
wireless protocols on latency has been covered in a few
surveys~\cite{al2016sur,ben2018ult,dou2013sur,ford2017achieving,par2018wir,parvez2018survey,sut2018ena}, while smart grid
communication has been covered in~\cite{ero2015ene}.  Low-latency
packet processing has been surveyed in~\cite{cer2018fas}, while
coding schemes have been surveyed in~\cite{bad2017per,dou2017ins}.  A
comprehensive guide to stochastic network calculus, which can be
employed to analyze network delays has appeared in~\cite{fid2015gui}.

Several surveys have covered the Tactile Internet
paradigm~\cite{aij2017sha,fet2014tac,Maier2016,Simsek2016},
which strives for latencies on the order of one millisecond.  The AVB
standard, which is a predecessor to the IEEE TSN standards development was
surveyed in~\cite{Eveleens2014,teener2013heterogeneous}.
In contrast to these existing surveys we provide a comprehensive
up-to-date survey of the IEEE TSN standards development
and the related research studies.

\section{Background}

\subsection{Latency Terminology}
Generally, latency refers to the total end-to-end packet delay from
the instant of the beginning of transmission by the sender (talker) to
the complete reception by the receiver (listener).  The term ultra-low
latency (ULL) commonly refers to latencies that are very short, e.g.,
on the order of a few milliseconds or less than one millisecond.  ULL
applications often require deterministic latency, i.e., all frames of
a given application traffic flow (connection) must not exceed a
prescribed bound~\cite{qian2017xpresseth}, e.g., to ensure the proper
functioning of industrial automation systems.  It is also possible
that applications may require probabilistic latency, i.e., a
prescribed delay bound should be met with high probability, e.g., for
multimedia streaming systems~\cite{amj2018wir,tre2018sea}, where rare
delay bound violations have negligible impact of the perceived quality
of the multimedia.

Latency jitter, or jitter for short, refers to the packet latency
variations. Often ULL systems require very low jitter.  Latency and
jitter are the two main quality of service (QoS) metrics for ULL
networking.  We note that there are a wide range of ULL applications
with vastly different QoS requirements, see
Table~\ref{lat_jit:tab}. For instance, some industrial control
applications have very tight delay bounds, e.g., only a few
microseconds, while other industrial control applications have more
relaxed delay bounds up to a millisecond.

\begin{table*}[t]
	\caption{End-to-end latency and jitter requirements for typical ULL applications} \label{lat_jit:tab} \footnotesize \centering
	\begin{tabular}{llrr}
	\multirow{2}{*}{Area} & \multirow{2}{*}{Application} & {QoS Requirements} \\
	& & Latencies & Jitter \\ \hline
	\multirow{1}{*}{\textbf{Medical}~\cite{ciz2017mul,Xu2014,zha2018tow}} & Tele-Surgery, Haptic Feedback & 3--10~ms & $<2$~ms \\  \hline
	\multirow{3}{*}{\textbf{Industry}~\cite{UCIEC2018}} & \multirow{2}{*}{Indust. Automation, Control Syst.} & 0.2~$\mu$s--0.5~ms for netw. with 1 Gbit/s link speeds  & meet lat. req. \\
	& & 25~$\mu$s--2~ms for netw. with 100 Mbit/s link speeds & meet lat. req. \\
	& Power Grid Sys. & approx. 8ms & few~$\mu$s \\  \hline
	\multirow{1}{*}{\textbf{Banking}~\cite{moa2013or}}  & High-Freq. Trading & $<1$~ms& few~$\mu$s \\ \hline
	\multirow{1}{*}{\textbf{Avionics}~\cite{schneele2012comparison}} & AFDX Variants & 1--128ms& few~$\mu$s \\ \hline
	\multirow{3}{*}{\textbf{Automotive}~\cite{aenreq,avbsys,osseiran2014scenarios,huang2018ivn}} & Adv. Driver. Assist. Sys. (ADAS) & 100--250~$\mu$s& few~$\mu$s \\
	& Power Train, Chassis Control & $<10\mu$s & few~$\mu$s \\
	& Traffic Efficiency \& Safety & $<5$~ms & few~$\mu$s \\ \hline
	\multirow{2}{*}{\textbf{Infotainment}~\cite{chatzopoulos2017mobile}} & Augmented Reality & 7--20~ms & few~$\mu$s \\
	 & Prof. Audio/Video & 2--50~ms & $<100~\mu$s
	\end{tabular}
\end{table*}

\subsection{IEEE 802.1 Overview}
Before we delve into the standardization efforts of the IEEE
Time-Sensitive Network (TSN) Task Group (TG), we briefly explain the
organizational structure of the IEEE 802.1 Working Group (WG). The
802.1 WG is chartered to develop and maintain standards and
recommended practices in the following areas: 1) 802 LAN/MAN
architecture, 2) internetworking among 802 LANs, MANs, and other wide
area networks, 3) security, 4) 802 overall network management, and 5)
protocol layers above the MAC and LLC layers. Currently, there are
four active task groups in this WG: 1) Time Sensitive Networking, 2)
Security, 3) Data Center Bridging, and 4) OmniRAN.

The main IEEE 802.1 standard that has been continuously revised and
updated over the years is IEEE~802.1Q-2014~\cite{IEEE8021Q}, formally
known as the IEEE~802.1D standard.  That is, IEEE 802.1Q-2014, which
we abbreviate to IEEE~802.1Q, is the main Bridges and Bridged Networks
standard that has incorporated all 802.1Qxx amendments, where ``xx''
indicates the amendment to the previous version of 802.1Q.

\subsubsection{IEEE 802.1 Bridge}
IEEE~802.1Q extensively utilizes the terminology ``IEEE~802.1
bridge'', which we abbreviate to ``bridge''.  A bridge is defined as
any network entity within an 802.1 enabled network that conforms to
the mandatory or optional/recommended specifications outlined in the
standard, i.e., any network node that supports the IEEE~802.1Q
functionalities.  IEEE~802.1Q details specifications for VLAN-aware
bridges and bridged LAN networks.  More specifically, IEEE~802.1Q
specifies the architectures and protocols for the communications
between interconnected bridges (L2 nodes), and the inter-process
communication between the layers and sublayers adjacent to the main
802.1 layer (L2).

\subsubsection{802.1Q Traffic Classes}
The  IEEE 802.1Q standard specifies
traffic classes with corresponding priority values that characterize the
traffic class-based forwarding behavior, i.e, the Class of Service
(CoS)~\cite[Annex~I]{IEEE8021Q}.
Eight traffic classes are specified in the 802.1Q standard,
whereby the priority level ranges from
lowest priority (0) to highest priority (7), as summarized in
Table~\ref{tracla:tab}.
\begin{table}[t]
  \caption{IEEE 802.1 Traffic Classes}  \label{tracla:tab}
  \footnotesize
  \centering
  \begin{tabular}{ll}
    \multicolumn{1}{c}{\textbf{Priority}} & \multicolumn{1}{c}{\textbf{Traffic Class}} \\
    \multicolumn{1}{c}{0} & Background \\
    \multicolumn{1}{c}{1} & Best effort \\
	\multicolumn{1}{c}{2} & Excellent effort \\
	\multicolumn{1}{c}{3} & Critical application\\
	\multicolumn{1}{c}{4} & ``Video'' $<$ 100 ms latency and jitter \\
	\multicolumn{1}{c}{5} & ``Voice'' $<$ 10 ms latency and jitter \\
	\multicolumn{1}{c}{6} & Internetwork control, e.g., OSPF, RIP \\
	\multicolumn{1}{c}{7} & Control Data Traffic (CDT), e.g., from IACSs \\
  \end{tabular}
  \end{table}

\subsection{General Development Steps from Ethernet Towards TSN}
Ethernet has been widely adopted as a common mode of networking
connectivity due to very simple connection mechanisms and protocol
operations.  Since its inception in the
1970s~\cite{met1976eth,met1993com} and first standardization in the
IEEE 802.3 standard in 1983~\cite{wri2018his}, Ethernet has kept up with the ``speed
race'' and today's Ethernet definitions support connections up to
400~Gbps. Due to the ever increasing demands, there is an ongoing
effort to advance Ethernet connectivity technologies to reach speeds
up to 1~Tbps.  The best-effort Ethernet service reduces the network
complexity and keeps protocol operations simple, while driving down
the product costs of Ethernet units.  Despite the enormous successes
and wide-spread adoption of Ethernet, the Ethernet definitions
fundamentally lack deterministic quality of service (QoS) properties
of end-to-end flows.  Prior to the development of the TSN standards,
ULL applications, e.g., industrial communications, deployed
point-to-point communication and circuit switching or
specialized/semi-proprietary specifications, such as, fieldbus
communication, e.g., IEEE 1394 (FireWire), Process Field Network
(Profinet), or Ethernet for Controlled Automation Technology
(EtherCAT).  In general, the Ethernet definitions lack the following
aspects for supporting ULL applications:
\begin{itemize}
\item[i)] Lack of QoS mechanisms to deliver packets in real time for
  demanding applications, such as real time audio and video delivery.
\item[ii)] Lack of global timing information and synchronization in network
	elements.
\item[iii)] Lack of network management mechanisms, such as bandwidth
  reservation mechanisms.
\item[iv)] Lack of policy enforcement mechanisms, such as packet
  filtering to ensure a guaranteed QoS level for an end-user.
\end{itemize}

Motivated by these Ethernet shortcomings, the Institute of Electronics
and Electrical Engineers (IEEE) and the Internet Engineering Task
Force (IETF) have proposed new definitions to introduce deterministic
network packet flow concepts. The IEEE has pursued the Time Sensitive
Networking (TSN) standardization~\cite{far2018tim} focusing mainly on physical layer
(layer one, L1) and link layer (layer two, L2) techniques within the
TSN task group in the IEEE 802.1 working group (WG).  The IETF has
formed the DETerministic NETwork (DETNET) working group focusing on
the network layer (L3) and higher layer techniques.

\section{IEEE TSN Standardization}   \label{tsn:std:sec}
This section surveys the standardization efforts of the IEEE 802.1 TSN
TG.  IEEE 802.1 TSN TG standards and protocols extend the traditional
Ethernet data-link layer standards to guarantee data transmission with
bounded ultra-low latency, low delay variation (jitter), and extremely
low loss, which is ideal for industrial control and automotive
applications~\cite{fin2018int,mes2018tim}. TSN can be deployed over
Ethernet to achieve the infrastructure and protocol capabilities for
supporting real-time Industrial Automation and Control System (IACS)
applications.

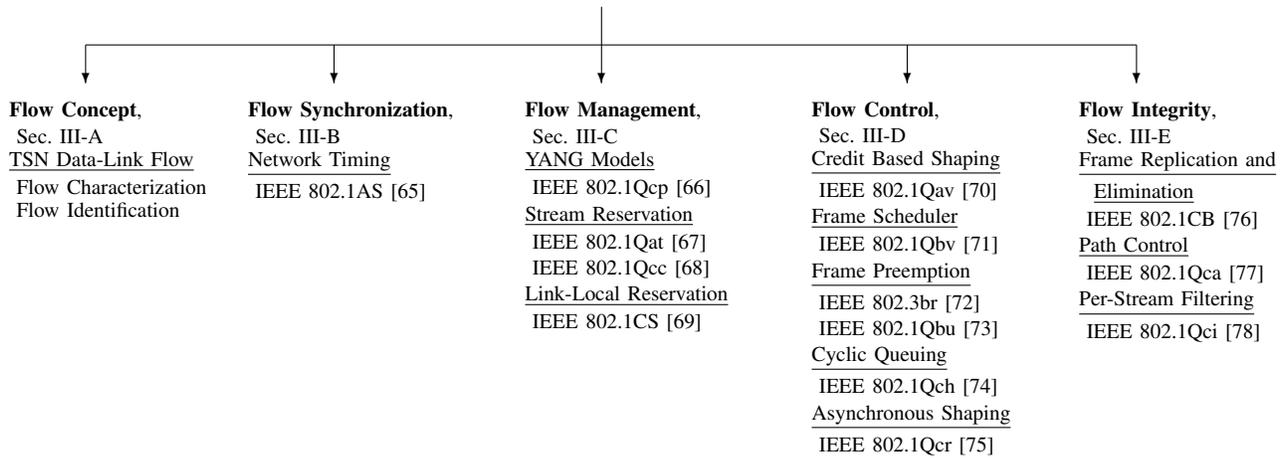
\begin{figure*}[t!]
	\footnotesize
	\setlength{\unitlength}{0.10in}
	\centering
	\begin{picture}(33,33)
	\put(8,33){\textbf{TSN Standardization, Sec.~\ref{tsn:std:sec}}}
	\put(-13,30){\line(1,0){55}}
    \put(-13,30){\vector(0,-1){2}}
	\put(-17,27){\makebox(0,0)[lt]{\shortstack[l]{			
				\textbf{Flow Concept}, \\
				\ Sec.~\ref{tsn:flow_def:sec} \\
				\underline{TSN Data-Link Flow} \\
				\ Flow Characterization \\
				\ Flow Identification
	}}}
	\put(0,30){\vector(0,-1){2}}
	\put(-4.5,27){\makebox(0,0)[lt]{\shortstack[l]{			
				\textbf{Flow Synchronization},\\
				\ Sec.~\ref{tsn:flow_sync:sec} \\
				\underline{Network Timing} \\
				\ IEEE 802.1AS~\cite{IEEE8021AS2011}
	}}}
	\put(14,30){\line(0,1){2}}
	\put(14,30){\vector(0,-1){2}}		
    \put(10,27){\makebox(0,0)[lt]{\shortstack[l]{
			\textbf{Flow Management},\\
			\ Sec.~\ref{tsn:flow_mgt:sec} \\
			\underline{YANG Models} \\
			\ IEEE 802.1Qcp~\cite{IEEE8021Qcp} \\
			\underline{Stream Reservation} \\
			\ IEEE 802.1Qat~\cite{IEEE8021Qat} \\
			\ IEEE 802.1Qcc~\cite{IEEE8021Qcc} \\
			\underline{Link-Local Reservation} \\
			\ IEEE 802.1CS~\cite{IEEE8021CS}
    }}}
	\put(30,30){\vector(0,-1){2}}
	\put(25,27){\makebox(0,0)[lt]{\shortstack[l]{
				\textbf{Flow Control},\\
				\ Sec.~\ref{tsn:flow_ctrl:sec} \\
				\underline{Credit Based Shaping} \\
				\ IEEE 802.1Qav~\cite{IEEE8021Qav} \\
				\underline{Frame Scheduler} \\
				\ IEEE 802.1Qbv~\cite{IEEE8021Qbv} \\
				\underline{Frame Preemption} \\
				\ IEEE 802.3br~\cite{IEEE8023br} \\
				\ IEEE 802.1Qbu~\cite{IEEE8021Qbu} \\
				\underline{Cyclic Queuing} \\
				\ IEEE 802.1Qch~\cite{IEEE8021Qch} \\
				\underline{Asynchronous Shaping} \\
				\ IEEE 802.1Qcr~\cite{IEEE8021Qcr}
	}}}		
	\put(42,30){\vector(0,-1){2}}
	\put(39,27){\makebox(0,0)[lt]{\shortstack[l]{
				\textbf{Flow Integrity},\\
				\ Sec.~\ref{tsn:flow_int:sec} \\
				\underline{Frame Replication and} \\
				\ \ \underline{Elimination} \\
				\ IEEE 802.1CB~\cite{IEEE8021CB} \\
				\underline{Path Control} \\
				\ IEEE 802.1Qca~\cite{IEEE8021Qca} \\
				\underline{Per-Stream Filtering} \\
				\ IEEE 802.1Qci~\cite{IEEE8021Qci}
	}}}
	\end{picture}
	\vspace{-2.5cm}	
        \caption{Classification taxonomy of Time Sensitive Networking
          (TSN) standardization.}
	\label{ctl_class:fig}
\end{figure*}

\begin{figure*}[t!]  \centering
	\includegraphics[width=\textwidth]{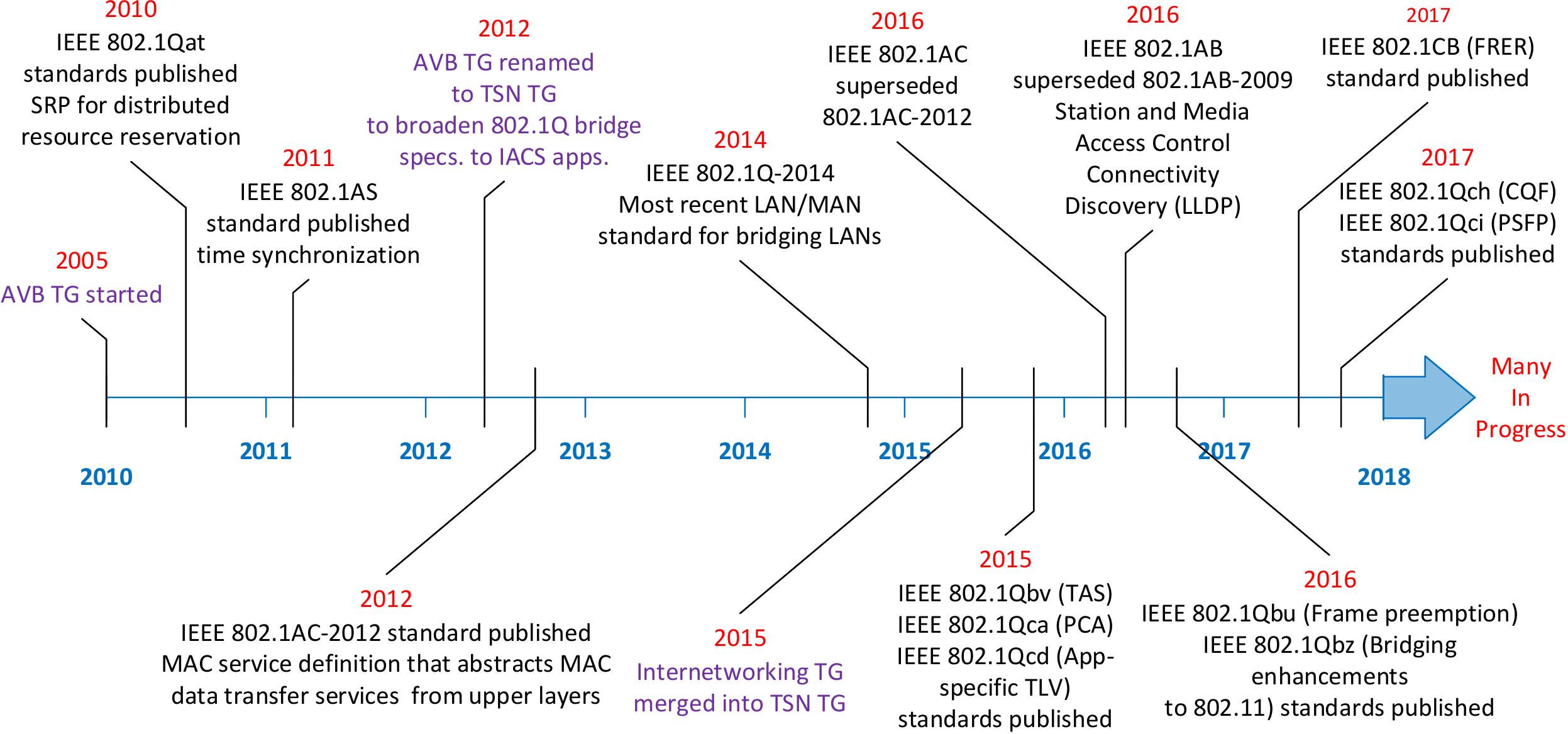}
\caption{Timeline of IEEE TSN task group (TG), highlighting
  significant milestones and depicting the shift from Audio Video
  Bridging (AVB) to TSN.}
	\label{fig_tsn_TSNtimeline}
\end{figure*}

In order to give a comprehensive survey of the current state of the
art of TSN standardization, we categorize the standardization efforts
for the network infrastructure supporting ULL applications.  We have
adopted a classification centered around the notion of the TSN flow,
which is defined as follows.  An end-to-end unicast or multicast
network connection from one end station (talker, sender) to another
end station(s) (listener(s), receiver(s)) through the time-sensitive
capable network is defined as a \textit{TSN flow}, which we often
abbreviate to ``flow'' and some publications refer to as ``stream''.
We have organized our survey of the standardized TSN mechanisms and
principles in terms of the TSN flow properties, as illustrated in
Fig.\ref{ctl_class:fig}.  Complementarily to the taxonomy in
Fig.\ref{ctl_class:fig}, Fig.~\ref{fig_tsn_TSNtimeline} provides a
historical perspective of the TSN standards and the ongoing
derivatives and revisions.

\subsection{Flow Concept: PCP and VLAN ID Flow Identification}
\label{tsn:flow_def:sec}
\begin{figure*}[t!]  \centering
\includegraphics[width=\textwidth]{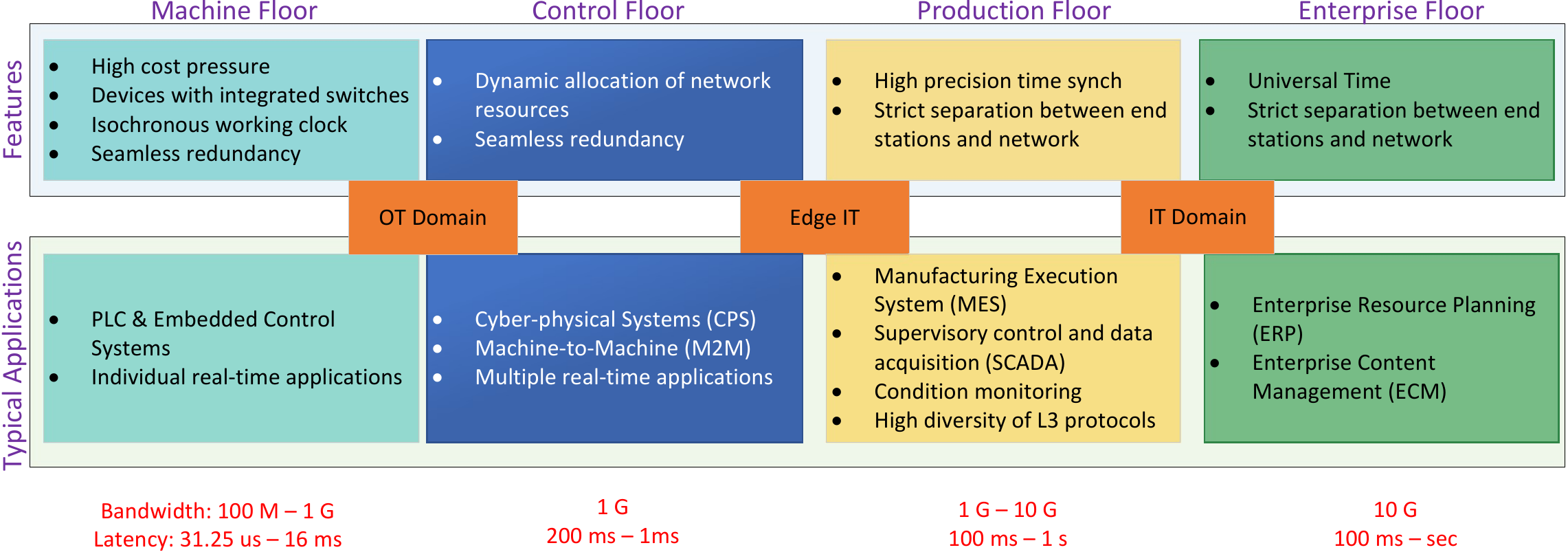}
	\vspace{-0.5cm}
\caption{Illustration of the broad range of QoS requirements according
  to the network setting (floor), whereby the machine floor requires
  the highest level of determinism and the lowest latency. Traditional
  networking is deployed on the enterprise floor. The top row
  summarizes the features required at each floor, while the bottom row
  illustrates typical example applications.  TSN can, in principle, be
  deployed everywhere, but typically, TSN is most attractive for the
  real-time systems in the OT Domain, i.e., the machine and control
  floors.}
	\label{fig_tsn_flowConcept}
\end{figure*}

A TSN flow (data link flow) is characterized by the QoS
properties (e.g., bandwidth and latency) defined for the traffic class
to which the flow belongs. In particular, a TSN flow is defined by the
priority code point (PCP) field and VLAN ID (VID) within the 802.1Q
VLAN tag in the Ethernet header. The PCP field and VID are
assigned based on the application associated with the flow.
Fig.~\ref{fig_tsn_flowConcept} outlines the general QoS
characteristics of the traffic classes related to the Informational
Technology (IT) and Operational Technology (OT) domains. Furthermore,
Fig.~\ref{fig_tsn_flowConcept} provides the main features for each
block, including typical applications used. As IT and OT establish a
converged interconnected heterogeneous network, the delay bottleneck must be
diminished to tolerable levels for IACS applications, i.e., the
machine and control floor networks.

\subsection{Flow Synchronization}  \label{tsn:flow_sync:sec}
\subsubsection{IEEE 802.1AS Time Synchronization for Time-Sensitive
  Applications} 
\begin{figure}[t!]  \centering
\includegraphics[width=3in]{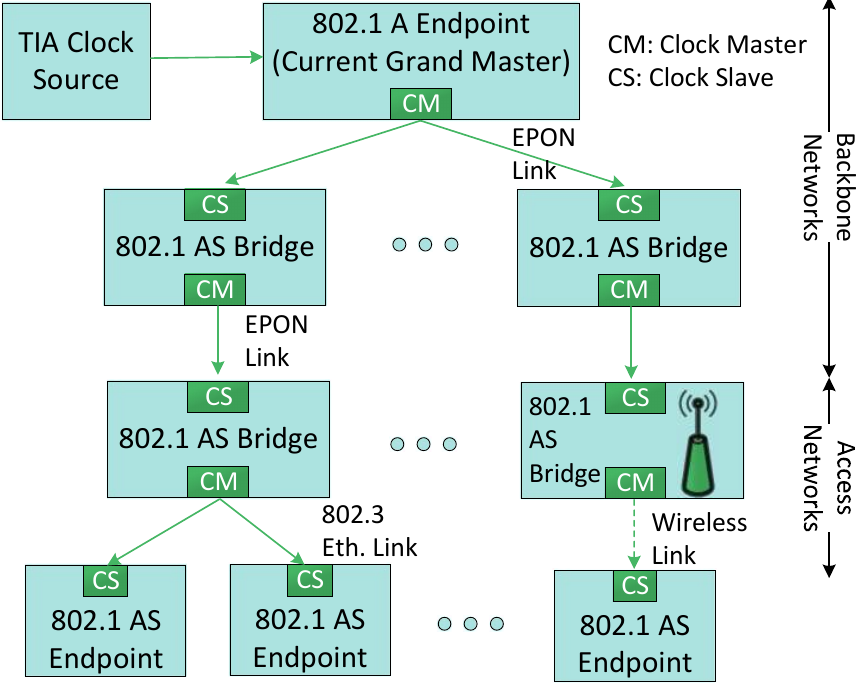}   \vspace{-0.1cm}
\caption{Illustration of a typical gPTP domain operation and time
  sharing where the selected GM source distributes timing information
  to all downstream 802.1AS bridges. Each bridge corrects the delay
  and propagates the timing information on all downstream ports,
  eventually reaching the 802.1AS end points (end stations).  The
  International Atomic Time (TIA) is the GM's source for timing
  information.}
	\label{fig_tsn_sync}
\end{figure}

Many TSN standards are based on a network-wide precise
time synchronization, i.e., an established common time reference that
is shared by all TSN network entities.  The time synchronization is,
for instance, employed to determine opportune data and control
signaling scheduling.  Time synchronization is accomplished through
the IEEE 802.1AS stand-alone standard~\cite{IEEE8021AS2011,sta2018dis},
which uses a specialized
profile (selection of features/configuration) of IEEE 1588-2008
(1588v2)~\cite{IEEE1588}, the generic Precision Time Protocol (gPTP).
The gPTP synchronizes clocks between network devices by passing
relevant time event messages~\cite{levesque2016survey}.
The message passing between the Clock Master (CM) and the Clock Slaves
(CSs) forms a time-aware network, also referred to as gPTP domain, as
illustrated in Fig.~\ref{fig_tsn_sync}.
The time-aware network utilizes the peer-path
delay mechanism to compute both the residence time, i.e., the
ingress-to-egress processing, queuing, and transmission time within a
bridge, and the link latency, i.e., the single hop propagation delay
between adjacent bridges within the time-aware network hierarchy with
reference to the GrandMaster (GM) clock at the root of the
hierarchy~\cite[Section 11]{IEEE8021AS2011}.  The GM clock is defined
as the bridge with the most accurate clock source selected by the
Best Master Clock Algorithm (BMCA)~\cite{IEEE1588}.

For example, in Fig.~\ref{fig_tsn_sync}, the bottom left-most 802.1AS
end point receives time information from the upstream CM which
includes the cumulative time from the GM to the upstream CM. For
full-duplex Ethernet LANs, the path delay measurement between the
local CS and the direct CM peer is calculated and used to correct the
received time. Upon adjusting (correcting) the received time, the
local clock should be synchronized to the gPTP domain's GM clock.

In general, gPTP systems consist of distributed and interconnected
gPTP and non-gPTP devices.  Time-aware bridges and end points are gPTP
devices, while non-gPTP devices include passive and active devices
that do not contribute to time synchronization in the distributed
network. gPTP is a distributed protocol that uses a master-slave
architecture to synchronize real-time clocks in all devices in the
gPTP domain with the root reference (GM) clock.  Synchronization is
accomplished through a two-phase process: The gPTP devices 1)
establish a master-slave hierarchy, and 2) apply clock synchronization
operations. In particular, gPTP establishes a master-slave hierarchy
using the BMCA~\cite{IEEE1588}, which consists of two separate
algorithms, namely data set comparison and state decision. Each gPTP
device operates a gPTP engine, i.e., a gPTP state machine, and employs
several gPTP UDP IPv4 or IPv6 multicast and unicast messages to
establish the appropriate hierarchy and to correctly synchronize
time~\cite{IEEE8021AS2011}. Any non-time aware bridge that cannot
relay or synchronize timing messages does not participate in
the BMCA clock spanning tree protocol.

The time synchronization accuracy depends mainly on the accuracy of
the residence time and link delay measurements. In order to achieve
high accuracy, 802.1AS time-aware systems correct the received
upstream neighbor master clock's timing information through the GM's
frequency ratio, this process is called logical syntonization in the
standard.  In the synchronization context, frequency refers to the
clock oscillator frequency.  The frequency ratio is the ratio of the
local clock frequency to the frequency of the time-aware system at the
other end of an attached link.  802.1AS achieves proper
synchronization between time-aware bridges and end systems using both
the frequency ratio of the GM relative to the local clock to compute
the synchronized time, and the frequency ratio of the neighbor CM
relative to the local CS to correct any propagation time measurements.

IEEE802.1AS-REV introduces new features needed for time-sensitive
applications. These features include the ability to
support multiple time domains to allow rapid switchover should a GM
clock fail, and improved time measurement accuracy.

\subsubsection{Summary and Lessons Learned}
IEEE 802.1AS provides reliable accurate network wide time
synchronization.  All gPTP systems compute both the residence time and
the link latency (propagation delay) and exchange messages along a
hierarchical structure centered around the selected GM clock to
accurately synchronize time.  Flow control and management components,
e.g., IEEE 802.1Qbv and 802.1Qcc (see Sections~\ref{tsn:flow_ctrl:sec}
and~\ref{tsn:flow_mgt:sec}), can utilize the 802.1AS timing
synchronization to provide accurate bounded latency and extremely low
loss and delay variation for TSN applications.

An open aspect of time synchronization is that the frequent periodic
exchange of timing information between the individual network entities
can stress and induce backpressure on the control plane.  The control
plane load due to the time synchronization can ultimately impact ULL
applications. A centralized time synchronization system, e.g., based
on a design similar to software defined networking (SDN)~\cite{alv2017com,nun2014sur},
with message exchanges only between a
central synchronization controller and individual network entities
could help mitigate the control plane overhead.  However, such a
centralized synchronization approach may create a single-point of
failure in the time synchronization process.  The detailed
quantitative study of these tradeoffs is an interesting direction for
future research.

\subsection{Flow Management}  \label{tsn:flow_mgt:sec}
Flow management enables users or operators to dynamically discover,
configure, monitor, and report bridge and end station capabilities.

\subsubsection{IEEE 802.1Qcp YANG Data Model} \label{IEEE802.1Qcp:sec}
The TSN TG has proposed the IEEE 802.1Qcp TSN Configuration YANG model
standard to achieve a truly universal Plug-and-Play (uPnP) model.  The
IEEE 802.1Qcp standard utilizes the Unified Modeling Language (UML),
specifically the YANG~\cite{bjorklund2010rfc, rfc6087} data model.  The YANG
data model provides a framework for periodic status reporting as well
as for configuring 802.1 bridges and bridge components, including
Media Access Control (MAC) Bridges, Two-Port MAC Relays (TPMRs),
Customer Virtual Local Area Network (VLAN) Bridges, and Provider
Bridges~\cite{IEEE8021Qcp}.  Additionally, IEEE 802.1Qcp is used to support
other TSN standard specifications, such as the Security and Datacenter
Bridging TG standards 802.1AX and 802.1X.

YANG~\cite{bjorklund2010rfc, rfc6087} is a data modeling language for
configuration data, state data, remote procedure calls, and
notifications for network management protocols, e.g., NETCONF and
RESTCONF. NETCONF is the Network Configuration
Protocol~\cite{enns2006netconf} that provides mechanisms to install,
manage, and delete the configurations of network devices.  The
industry wide adoption of the YANG formalized data modeling language,
e.g., by the IETF and the Metro Ethernet Forum (MEF), is an important
motivation for integrating, automating, and providing support for YANG
data modeling in 802.1 bridges and related services for upper layer
components.

\subsubsection{IEEE 802.1Qat Stream Reservation Protocol (SRP) and
  IEEE 802.1Qcc Enhancements to SRP and Centralization Management}
\label{802.1Qat:sec}
The IEEE 802.1Qat Stream Reservation Protocol
(SRP)~\cite{IEEE8021Qat}, which has been merged into 802.1Q, provides
a fundamental part of TSN.  In particular, IEEE 802.1Qat specifies the
admission control framework for admitting or rejecting flows
based on flow resource
requirements and the available network resources.
Moreover, IEEE 802.1Qat specifies the framework
for reserving network resources and
advertising streams in packet switched networks over full-duplex
Ethernet links. Most of the standards that use priorities, frame
scheduling, and traffic shaping protocols depend on
SRP~\cite{IEEE8021Qat}, since these protocols work correctly only if
the network resources are available along the entire path from the
sender (talker) to the receivers (listeners).
IEEE~802.1Qat is a
distributed protocol that was introduced by the AVB TG to ensure that
the AVB talker is guaranteed adequate network resources along its
transmission path to the listener(s). This is accomplished using the
Multiple Registration Protocol (MRP)~\cite[Section~10]{IEEE8021Q},
where the traffic streams are identified and registered using a 48-bit
Extended Unique Identifier (EUI-48). The EUI-48 is usually the MAC
source address concatenated with a 16-bit handle to differentiate
different streams from the same source and is also referred to as
\textit{StreamID}. The SRP reserves resources for a stream based on
the bandwidth requirement and the latency traffic class using three
signaling protocols, namely 1) the Multiple MAC Registration Protocol
(MMRP), 2) the Multiple VLAN Registration Protocol (MVRP), and 3) the
Multiple Stream Registration Protocol
(MSRP)~\cite[Section~35]{IEEE8021Qat,IEEE8021Q}.

\begin{figure}[t!]  \centering
\includegraphics[width=3.5in]{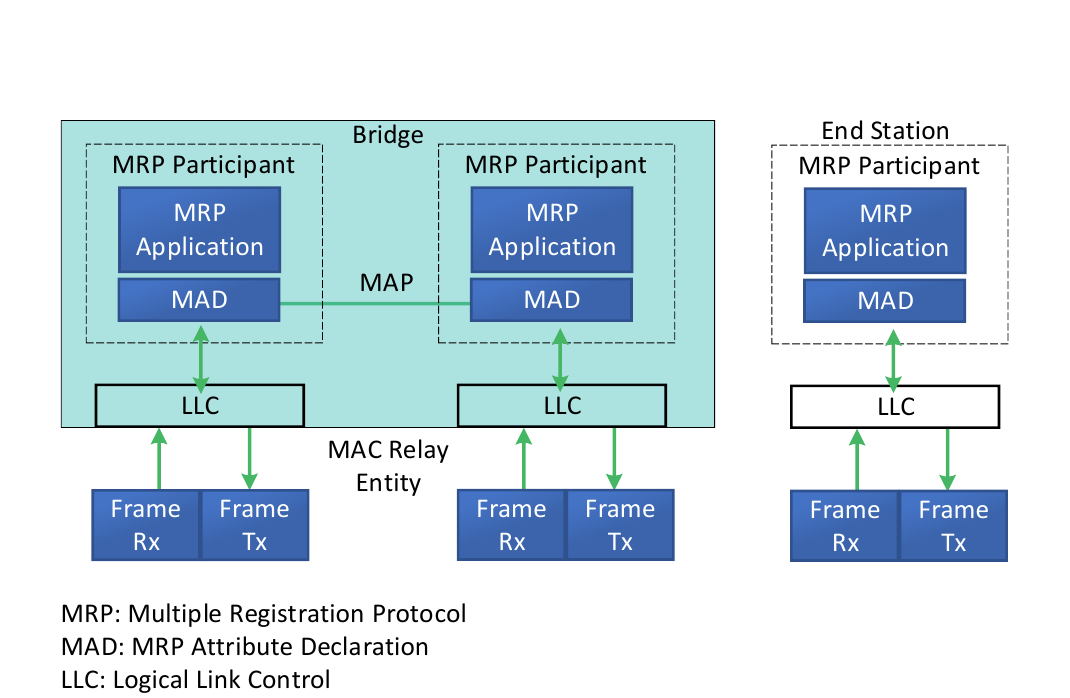} \vspace{-0.5cm}
\caption{Illustration of Multiple Registration Protocol (MRP)
  architecture: Each end station (illustrated on the right) declares
  the propagation attributes using the MRP Attribute Declaration (MAD)
  and the MRP Applications encapsulated as an MRP participant which
  gives end stations the ability to register resources. The MRP
  participant entry is stored in bridges and mapped between all
  required ports using MRP Attribute Propagation (MAP). A bridge
  mapping between two different interfaces in the LAN is illustrated
  on the left.}
	\label{fig_tsn_MRP}
\end{figure}

MMRP and MVRP control the group registration propagation and the VLAN
membership (MAC address information~\cite[Sections~10
  and~11]{IEEE8021Q}), while MSRP conducts the distributed network
resource reservation across bridges and end stations. MSRP registers
and advertises data stream characteristics and reserves bridge
resources to provide the appropriate QoS guarantees according to the
talker's declared propagation attributes, which include the SRP
parameters that are sent by the end station in MSRP PDUs (MSRPDUs).  A
station (talker) sends a reservation request with the MRP, i.e., the
general MRP application which registers the stream resource
reservation.  The 802.1 TSN TG has developed the MRP Attribute
Declaration (MAD) for describing the request based on the stream
characteristics.  All participants in the stream have an MSRP
application and MAD specification and each bridge within the same SRP
domain can map, allocate, and forward the stream with the necessary
resources using the MRP attribute propagation
(MAP)~\cite{IEEE8021Qat}. Fig.~\ref{fig_tsn_MRP} illustrates the MRP
architecture.

In essence, the SRP protocol ensures QoS constraints for each stream
through the following steps:
\begin{enumerate}
	\item Advertise stream
	\item Register paths of stream
	\item Calculate worst-case latency
	\item Establish an AVB domain
	\item Reserve the bandwidth for the stream.
\end{enumerate}
Since the existing IEEE 802.1Qat (802.1Q~Section 35) SRP features a
decentralized registration and reservation procedure, any changes or
new requests for registrations or de-registrations can overwhelm the
network and result in intolerable delays for critical traffic
classes. Therefore, the TSN TG has introduced the IEEE 802.1Qcc
standard to improve the existing SRP by reducing the size and
frequency of reservation messages, i.e., relaxing timers so that
updates are only triggered by link state or reservation changes.

\begin{figure}[t!]  \centering
\includegraphics[width=3.5in]{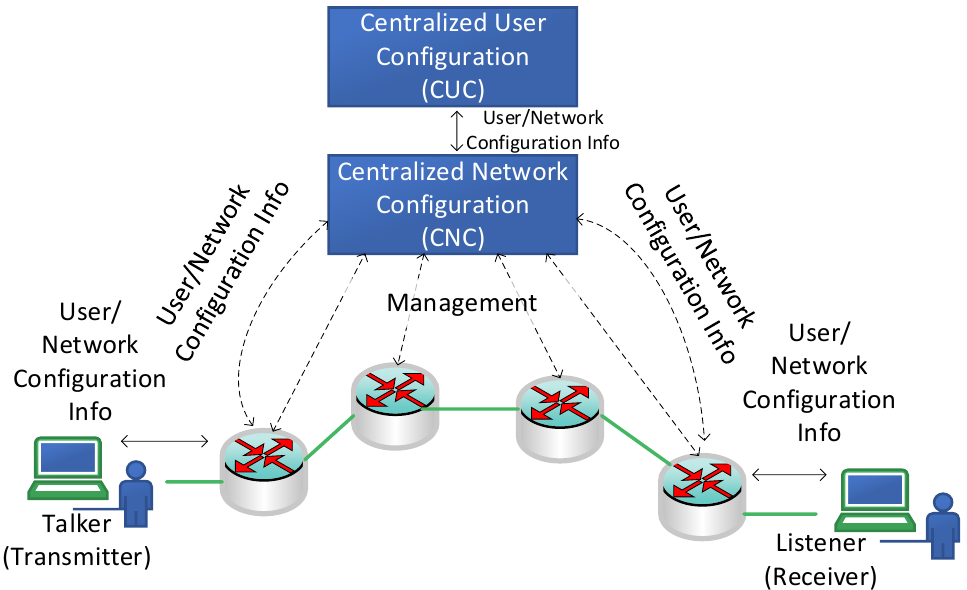} \vspace{-0.5cm}
\caption{Illustration of Centralized Network Configuration (CNC): End
  stations interact with the network entities via the User-Network
  Interface (UNI). The CNC receives the requests, e.g., flow
  reservation requests, and provides corresponding management
  functions.  An optional CUC provides delay-optimized configuration,
  e.g., for closed-loop IACS applications.  The solid arrows represent
  the protocol, e.g., YANG or TLV, that is used as the UNI for
  exchanging configuration information between Talkers/Listeners
  (users) and Bridges (network).  The dashed arrows represent the
  protocol, e.g., YANG or TLV, that transfers configuration
  information between edge bridges and the CNC.} \label{fig_tsn_Qcc}
\end{figure}

Additionally, IEEE 802.1Qcc~\cite{IEEE8021Qcc} provides a set of tools
to manage and control the network globally.  In particular,
IEEE~802.1Qcc enhances the existing SRP with a User Network Interface
(UNI) which is supplemented by a Centralized Network Configuration
(CNC) node, as shown in Fig.~\ref{fig_tsn_Qcc}. The UNI provides a
common method of requesting layer 2 services. Furthermore, the CNC
interacts with the UNI to provide a centralized means for performing
resource reservation, scheduling, and other types of configuration via
a remote management protocol, such as NETCONF~\cite{enns2006netconf}
or RESTCONF~\cite{bierman2017restconf}; hence, 802.1Qcc is compatible
with the IETF YANG/NETCONF data modeling language.

For a fully centralized network, an optional Centralized User
Configuration (CUC) node communicates with the CNC via a standard
Application Programming Interface (API), and can be used to discover
end stations, retrieve end station capabilities and user requirements,
and configure delay-optimized TSN features in end stations (mainly for
closed-loop IACS applications).  The interactions with higher level
reservation protocols, e.g., RSVP, are seamless, similar to how the
AVB Transport Protocol IEEE 1722.1~\cite{IEEE1722.1} leverages the
existing SRP.

802.1Qcc~\cite{IEEE8021Qcc} still supports the fully distributed
configuration model of the original SRP protocol, i.e., allows for
centrally managed systems to coexist with decentralized ad-hoc
systems. In addition, 802.1Qcc supports a ``hybrid'' configuration
model, allowing a migration path for legacy AVB devices. This hybrid
configuration management scheme when coupled with IEEE 802.1Qca Path
Control and Reservation (PCR) (see Section~\ref{IEEE802.1Qca:sec}) and
the TSN shapers can provide deterministic end-to-end delay and zero
congestion loss.

\subsubsection{IEEE 802.1CS Link-Local Reservation Protocol (LRP)} \label{tsn:std:lrp}
To effectively achieve tight bounds on latency and zero
  congestion loss, traffic streams need to utilize effective admission
  control policies and secure resource registration mechanisms, such
  as the SRP~\cite{IEEE8021Qat} and the SRP enhancements and
  management standard~\cite{IEEE8021Qcc}. While the
MRP~\cite[Section~10]{IEEE8021Q} provides efficient methods for
registering streams; the database holding the stream state
information, is limited to about 1500~bytes. As more traffic streams
coexist and the network scale increases, MRP slows significantly as
the database proportionally increases which results in frequent cyclic
exchanges through the MAD between all bridge neighbors.

The Link-Local Reservation Protocol (LRP)~\cite{IEEE8021CS} has been
introduced by the 802.1 TSN TG to efficiently replicate an MRP
database between two ends of a point-to-point link and to
incrementally replicate changes as bridges report new network
developments or conditions. Additionally, the LRP provides a purging
process that deletes replicated databases when the source of such
databases remains unresponsive or the data gets stale. Furthermore,
the LRP is optimized to efficiently handle databases on the order of
1~Mbyte.
\begin{figure}[t!]  \centering
\includegraphics[width=3.5in]{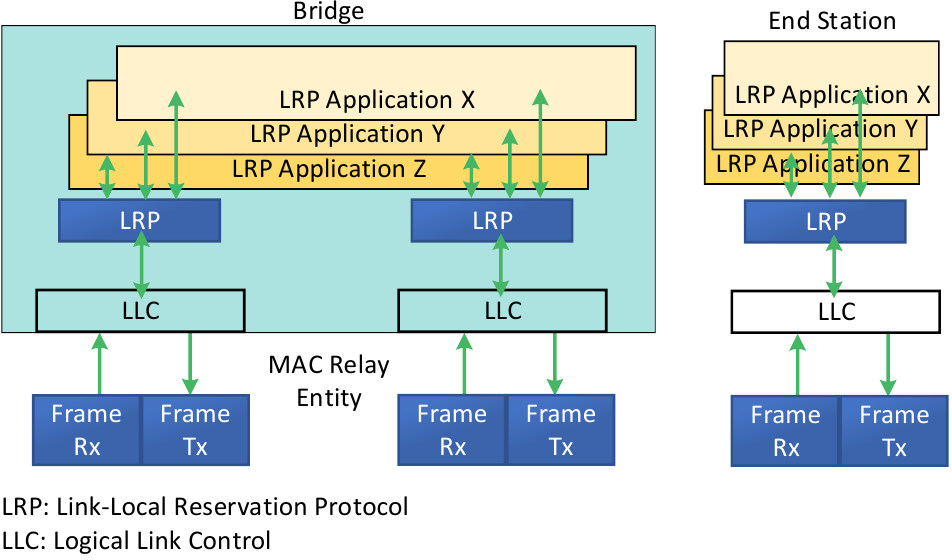} \vspace{-0.5cm}
\caption{Illustration of LRP Architecture: A Link-Local Reservation
  Protocol (LRP) instance (illustrated by the blue LRP box) interacts
  with each application and provides a generic transport service for
  multiple registered LRP applications, which are represented by
  yellow colored boxes near the top of the
  illustration.}  \label{fig_tsn_LRP}
\end{figure}

While MRP is considered application specific, i.e., the MRP operations
are defined by each registered application, LRP is an application
neutral transport protocol. Fig~\ref{fig_tsn_LRP} illustrates the LRP
protocol architecture operating within bridges or end points.

\subsubsection{Resource Allocation Protocol (RAP)---Towards a Distributed TSN Control Model}  \label{RAP:sec}
Although the SRP and the related MSRP (MSRPv1~\cite{IEEE8021Qcc}) were
designed for distributed stream configuration (including registration,
reservation, and provisioning), SRP is generally restricted to A/V
applications with a limited number of Stream Reservation (SR) classes,
e.g., classes A and B for the Credit Based Shaper (CBS), see
Section~\ref{802.1Qav:sec}.  SRP guarantees the QoS characterized by
each stream through the reservation in conjunction with shaper
mechanisms, see Section~\ref{tsn:flow_ctrl:sec}. IEEE 802.1Qcc pushed
for more centralized configuration models, where all the newly
established TSN features, e.g., shaping, preemption, and redundancy,
are supported through the CNC configuration model. Any distributed
model is currently restricted to CBS.

The Resource Allocation Protocol (RAP)~\cite{chen2017rap} leverages the LRP
to propagate TSN stream configuration frames that include resource
reservation and registration information in a manner similar to MSRP. The
MSRP (and MSRPv1) is geared towards AVB systems, while RAP is defined
for TSN enabled systems for distributed stream configuration. The RAP
promises to improve scalability (through LRP), to support all TSN
features, to improve performance under high utilization, and to enhance
diagnostic capabilities.

\subsubsection{Summary and Lessons Learned}
Flow management allows distributed (legacy SRP and RAP) as well as
centralized (802.1Qcc and 802.1CS) provisioning and management
of network resources, effectively creating protected channels over
shared heterogeneous networks. Moreover, flow management offers users
and administrators Operations, Administration, Maintenance (OAM)
functions to monitor, report, and configure (802.1Qcp and 802.1Qcc)
network conditions. This allows for fine-grained support of network
services while enforcing long term allocations of network resources
with flexible resource control through adaptive and automatic
reconfigurations.

However, both centralized and distributed flow management models have
specific deployment advantages and disadvantages.
For example, a centralized entity presents a single
point of failure, whereas, distributed schemes incur extensive control
plane overheads. A centralized scheme can benefit from SDN
implementation and management but could result in new infrastructure
cost for the operators.  Nevertheless, the choice of deployments can
be based on the relative performance levels among centralized and
distributed nodes, as well as the use of existing infrastructures and the
deployment of new
infrastructures. Future research needs to thoroughly examine these
tradeoffs.

Another important future research direction is to examine predictive
models that estimate the resource reservation requirements in bridges.
Estimations may help in effectively managing queues and scheduling
while efficiently utilizing the network resources.

\subsection{Flow Control}   \label{tsn:flow_ctrl:sec}
Flow control specifies how frames belonging to a prescribed traffic
class are handled within TSN enabled bridges.

\subsubsection {IEEE 802.1Qav Forwarding and Queuing of
  Time-Sensitive Streams} \label{802.1Qav:sec} IEEE 802.1Qav specifies
Forwarding and Queuing of Time Sensitive Streams (FQTSS), which has
been incorporated into 802.1Q.  IEEE 802.1Qav serves as a major
enhancement to the forwarding and queuing operation in traditional
Ethernet networks.  IEEE 802.1Qav specifies bridge operations that
provide guarantees for time-sensitive (i.e., bounded latency and
jitter), lossless real-time audio/video (A/V)
traffic~\cite{IEEE8021Qav}.  The IEEE 802.1Qav
standard~\cite[Section~34]{IEEE8021Qav, IEEE8021Q}, details flow
control operations, such as per priority ingress metering and
timing-aware queue draining algorithms.

IEEE 802.1Qav was developed to limit the amount of A/V traffic
buffering at the downstream receiving bridges and/or end stations.
Increasing proportions of bursty multimedia traffic can lead to
extensive buffering of multimedia traffic, potentially resulting in
buffer overflows and packet drops. Packet drops may trigger
retransmissions, which increase delays, rendering the re-transmitted
packets obsolete and diminishing the Quality of Experience (QoE).

IEEE 802.1Qav limits the amount of buffering required in the receiving
station through the Stream Reservation Protocol
(SRP)~\cite{IEEE8021Qat} in conjunction with a credit-based shaper
(CBS). The CBS spaces out the A/V frames to reduce bursting and
bunching. This spacing out of A/V frames protects best-effort traffic
as the maximum AVB stream burst is limited.  The spacing out of A/V
frames also protects the AVB traffic by limiting the back-to-back AVB
stream bursts, which can interfere and cause congestion in the
downstream bridge.

\begin{figure}[t!]  \centering
\includegraphics[width=2.5in]{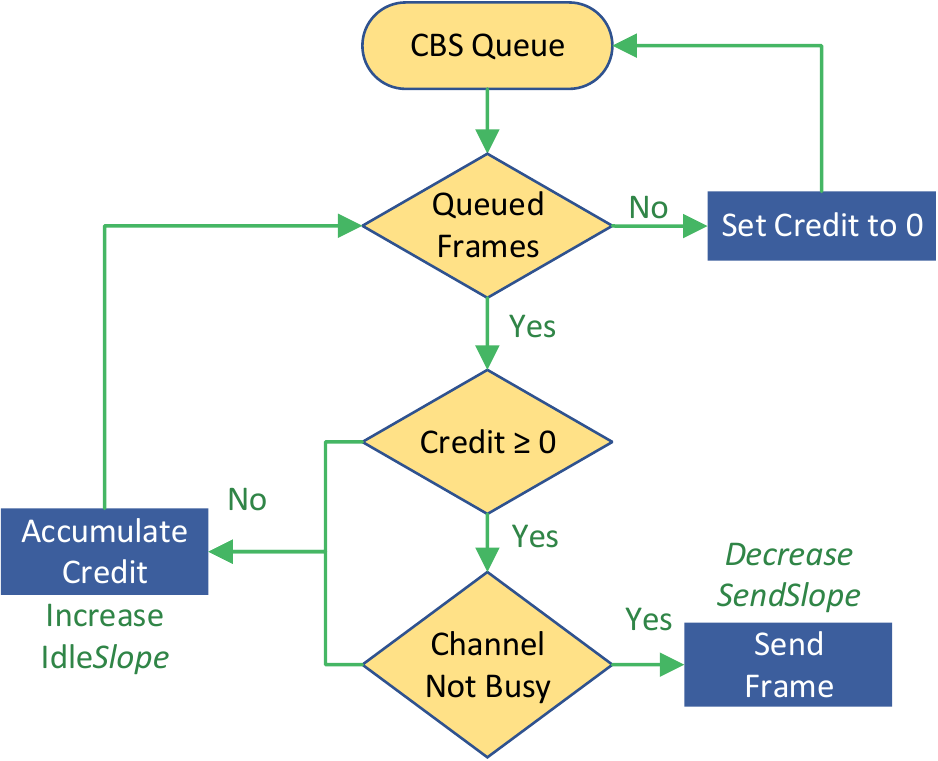}   \vspace{-0.1cm}
\caption{Flow-chart illustration of the Credit-Based Shaper (CBS)
  operation for a given queue. A queue is permitted to transmit if both
  credits are greater than or equal to zero, and the channel is
  vacant.}  \label{fig_tsn_CBS}
\end{figure}
The CBS shaper separates a queue into two traffic classes, class A
(tight delay bound) and class B (loose delay bound).  Each class queue
operates according to the throttling mechanism illustrated in
Fig.~\ref{fig_tsn_CBS}. When no frame is available in the queue, the
credit for the queue is set to zero. A queue is eligible for
transmission if the credit is non-negative.
The credit is increased by
\textit{idleSlope} when there is at least one frame in the queue, and
decreased by \textit{sendSlope} when a frame is transmitted.
The \textit{idleSlope} is the actual bandwidth reserved (in bits
  per second) for the specific queue and traffic class within a
  bridge~\cite[Section~34]{IEEE8021Q}, while the \textit{sendSlope} is
  the port transmit rate (in bits per second) that the underlaying MAC
  service supports. Furthermore, two key limiting parameters are
  defined: $i)$ \textit{hiCredit} and $ii)$ \textit{loCredit}, which
  are functions of the maximum frame size (in the case of
  \textit{loCredit}) and maximum interference size (in the case of
  \textit{hiCredit}), the \textit{idleSlope}/\textit{sendSlope}
  (respectively), and the maximum port transmit rate. Further details can
  be found in~\cite[Annex~L]{IEEE8021Qav}.
The CBS throttles each shaped traffic class to not exceed their
preconfigured bandwidth limits (75\% of maximum bandwidth due to
bandwidth intensive applications, e.g., audio and video~\cite[Section~34.3.1]{IEEE8021Qav}).  The CBS in
combination with the SRP is intended to bound delays to under
250~$\mu$s per bridge~\cite{IEEE8021Qat}.  Overall, the IEEE 802.1Qav
Ethernet AVB standard guarantees worst-case latencies under 2~ms for
class A and under 50~ms for class B up to seven network
hops~\cite{IEEE8021Qav}.

However, some key CBS disadvantages are that the average delay is
increased and that the delay can be up to 250~$\mu$s per hop, which
may be too high for industrial control applications~\cite{Specht2016}.
Also, CBS struggles to maintain delay guarantees at high link
utilizations.

In order to address the CBS shortcomings, the TSN TG has introduced
other standards, e.g., IEEE~802.1Qbv, 802.1Qch, and 802.1Qcr,
which are reviewed in the following subsections.
Also, addressing the CBS shortcomings is an active research area,
see Section~\ref{tsn:res:flow_ctrl:sec}.

\subsubsection{IEEE 802.1Qbv Enhancements to Traffic Scheduling: Time-Aware Shaper (TAS)}   \label{802.1Qbv:sec}
\begin{figure}[t!] 	\centering
\includegraphics[width=3.5in]{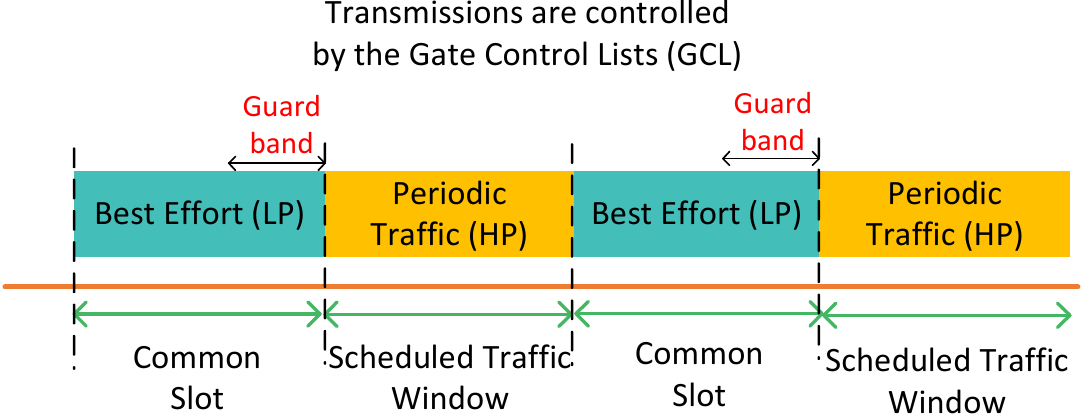}
\vspace{-0.5cm}
\caption{IEEE 802.1Qbv Time-Aware Shaper (TAS)~\cite{IEEE8021Qbv}:
  Scheduled traffic is sent over synchronized Time-Division
  Multiplexing ``windows'' within the Ethernet traffic. Yellow marked
  frames are time-sensitive high priority (HP) traffic that have guaranteed reserved
  resources across the network, while the blue frames correspond to
  best-effort low priority (LP) traffic.}  \label{fig_tsn_periodicFrames}
\end{figure}

\begin{figure}[t!] 	\centering
\includegraphics[width=3.5in]{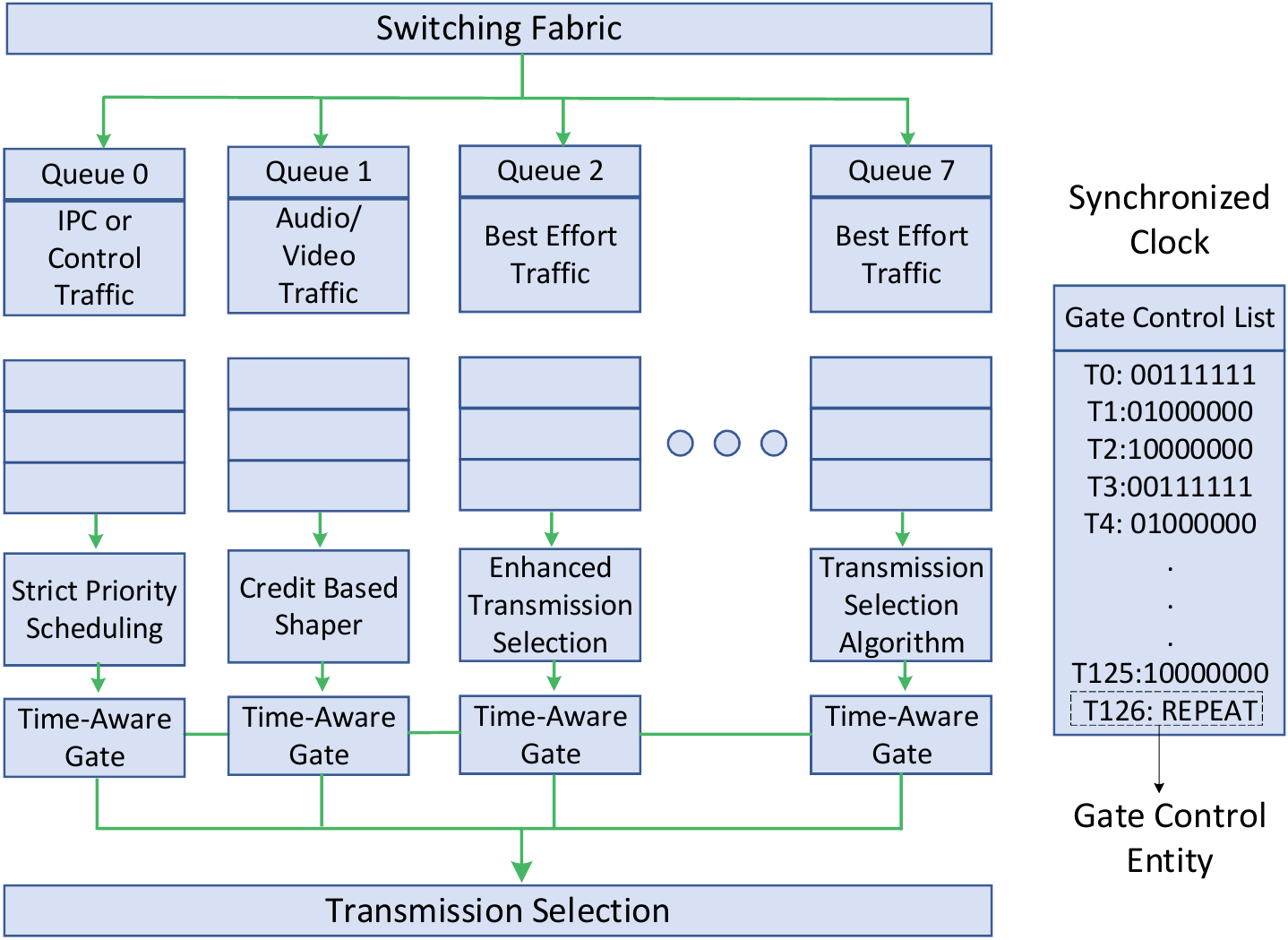}  \vspace{-0.5cm}
\caption{IEEE 802.1Qbv: Illustration of egress hardware queue with 8 software
  queues, each with its unique transmission selection algorithm. The
  transmissions are controlled by the Gate Controlled List (GCL) with
  multiple Gate Control Entries (GCEs) that determine which software queues
  are open.  For instance, in time interval T0, the gates for queues 2
  through 7 are open, and the transmission selection at the bottom
  arbitrates access to the medium~\cite[Section~8.6.8]{IEEE8021Q}. In
  time interval T1, the gate opens for AV traffic from Queue~1, and a
  credit based shaper (CBS) regulates the frame transmissions from
  Queue~1.  In time interval T2, the gate opens for Queue~0 and strict
  priority scheduling selects the frames to transmit from
  Queue~0.}  \label{fig_tsn_frameScheduling}
\end{figure}
As a response to the IEEE 802.1Qav shortcomings, the TSN task group
proposed a new traffic shaper, namely the IEEE~802.1Qbv Time-Aware
Shaper (TAS)~\cite{IEEE8021Qbv} along with the IEEE 802.1Qbu frame preemption technique~\cite{IEEE8021Qbu}
to provide fine-grained QoS~\cite{sim2018des}. The TAS and frame preemption mechanisms are
suitable for traffic with deterministic end-to-end ULL requirements,
e.g., for critical control or Interprocess Communication (IPC)
traffic, with sub-microseconds latency requirements.  In particular,
the TAS schedules critical traffic streams in
time-triggered windows, which are also referred to as protected
traffic windows or as time-aware traffic windows.  Thus, TAS follows
the TDMA paradigm, similar to Flexible Time-Triggered Ethernet
(FTT-E)~\cite{pedreiras2005ftt, meyer2013extending}, whereby each
window has an allotted transmission time as shown in
Fig.~\ref{fig_tsn_periodicFrames}.
In order to prevent lower priority traffic, e.g., best effort
traffic, from interfering with the scheduled traffic transmissions,
scheduled traffic windows are preceded by a so-called guard band.
\begin{figure}[t!]  \centering
\includegraphics[width=3.5in]{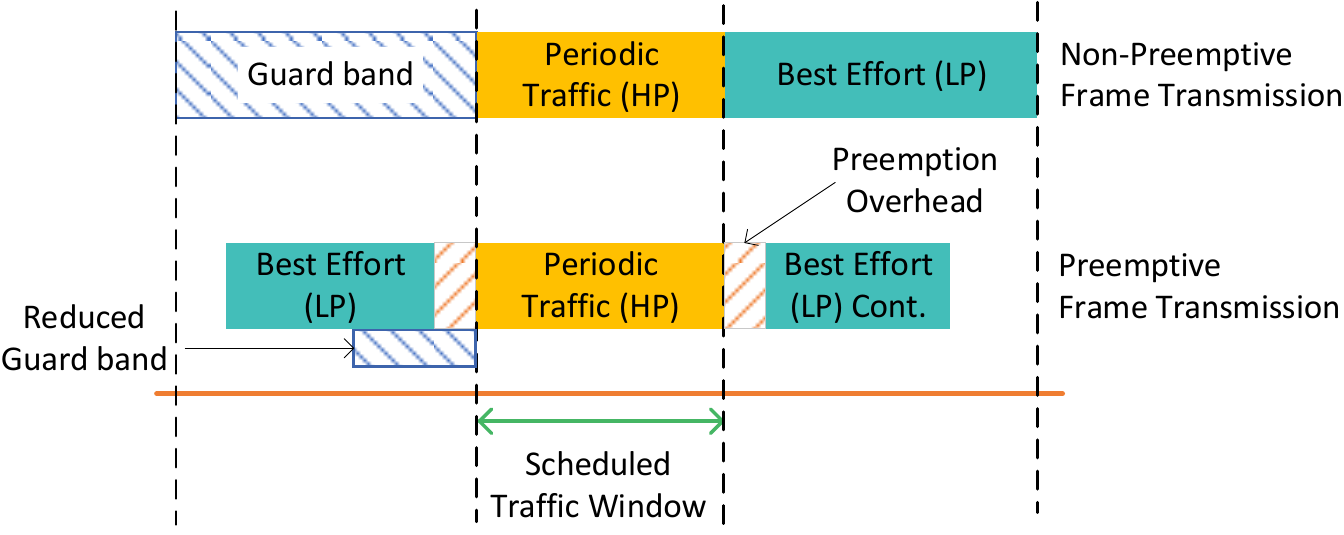} \vspace{-0.6cm}
\caption{The IEEE 802.1Qbv transmission selection prevents low priority
  (best effort) frames from starting transmission if the transmission
  cannot be completed by the start of the scheduled traffic
  window. This transmission selection essentially enforces a guard
  band (sized as a maximum size frame) to protect the scheduled
  traffic window. With preemption (IEEE 802.3br, IEEE 802.1Qbu) the
  guard band can be reduced to the smallest Ethernet frame fragment.}
	\label{fig_tsn_PreemptionExample}
\end{figure}

TAS is applicable for ULL requirements but needs to have all
time-triggered windows synchronized, i.e., all bridges from sender to
receiver must be synchronized in time.  TAS utilizes a gate driver
mechanism that opens/closes according to a known and agreed upon time
schedule, as shown in Fig.~\ref{fig_tsn_frameScheduling}, for each
port in a bridge. In particular, the Gate Control List (GCL) in
Fig.~\ref{fig_tsn_frameScheduling} represents Gate Control Entries
(GCEs), i.e., a 1 or 0 for open or close for each queue, respectively.
The frames of a given egress queue are eligible for transmission
according to the GCL, which is synchronized in time through the
802.1AS time reference.  The GCL is executed in periodically repeating
cycle times, whereby the each cycle time contains one GCL execution.
Within a cycle time, the time period during which a gate is open is
referred to as the time-aware traffic window.  Frames are transmitted
according to the GCL and transmission selection decisions, as
illustrated in Fig.~\ref{fig_tsn_frameScheduling}.  Each individual
software queue has its own transmission selection algorithm, e.g.,
strict priority queuing (which is the default).  Overall, the IEEE
802.1Qbv transmission selection at the bottom of
Fig.~\ref{fig_tsn_frameScheduling} transmits a frame from a given
queue with an open gate if: $(i)$ The queue contains a frame ready for
transmission, $(ii)$ higher priority traffic class queues with an open
gate do \textit{not} have a frame to transmit, and $(iii)$ the frame
transmission can be completed before the gate closes for the given
queue.  Note that these transmission selection conditions ensure that
low priority traffic is allowed to \textit{start} transmission only if
the transmission will be completed by the start of the scheduled
traffic window for high priority traffic.  Thus, this transmission
selection effectively enforces a ``guard band'' to prevent low
priority traffic from interfering with high priority traffic, as
illustrated in Fig~\ref{fig_tsn_PreemptionExample}.

One critical TAS shortcoming is that some delay is incurred due to
additional sampling delay, i.e., due the waiting time until the next
time-triggered window commences.  This sampling delay arises when
unsynchronized data is passed from an end-point to the network.  Task
and message scheduling in end-nodes would need to be coupled with the
TAS gate scheduling in the networks in order to achieve the lowest
latencies.  Moreover, synchronizing TSN bridges, frame selections, and
transmission times across the network is nontrivial in moderately
sized networks, and requires a fully managed network.  Also, the
efficient use of bandwidth with TAS needs to be thoroughly examined.
Overall, TAS has high configuration complexity.  Future research needs
to carefully examine the scalability to large networks, runtime
reconfiguration, and the integration of independently developed
sub-systems.

\subsubsection{IEEE 802.3br and 802.1Qbu Interspersing Express Traffic
  (IET) and Frame Preemption}  \label{802.1Qbu:sec}
\begin{figure}[t!] 	\centering
\includegraphics[width=2in]{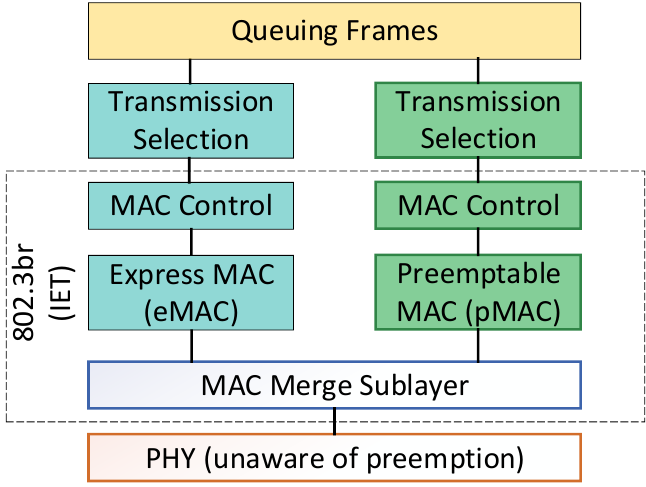} \vspace{-0.2cm}
\caption{Illustration of the layering for the Ethernet MAC Merge
  Sublayer: The MAC Merge Sublayer provides a Reconciliation Sublayer
  (RS) service for pMAC and eMAC frames.  The RS service supports two
  main ways to hold the transmission of a pMAC frame in the presence of an
  eMAC frame: By preempting (interrupting) the pMAC frame
  transmission, or by preventing the start of the pMAC frame
  transmission. }  \label{fig_tsn_802.3IET}
\end{figure}
To address the ULL latency requirements and the inverted priority
problem, i.e., the problem that an ongoing transmission of a low
priority frame prevents the transmission of high priority frames,
the 802.1 TG along with the 802.3 TG introduced frame preemption
(802.1Qbu and 802.3br)~\cite{IEEE8021Qbu, IEEE8023br}.  Frame
preemption separates a given bridge egress port into two MAC service
interfaces, namely preemptable MAC (pMAC) and express MAC (eMAC), as
illustrated in Fig.~\ref{fig_tsn_802.3IET}. A frame preemption status
table maps frames to either pMAC or eMAC; by default all frames are
mapped to eMAC. Preemptable frames that are in transit, i.e., they are
holding on to the resource (transmission medium), can be preempted by
express frames. After the transmission of an express frame has
completed, the transmission of the preempted frame can resume.

With preemption, the guard band in Fig.~\ref{fig_tsn_periodicFrames}
can be reduced to the transmission time of the shortest low priority
frame fragment. Thus, in the worst case, the transmission of the low
priority frame fragment can be completed before starting the
transmission of the next high priority frame.  The transmission of the
leftover fragmented frame can then be resumed to completion.  Note
that this preemption occurs only at the link-level, and any fragmented
frame is reassembled at the MAC interfaces. Hence the switches process
internally only complete frames. That is, any frame fragments
transmitted over a physical link to the next bridge are re-assembled
in the link layer interface; specifically, the MAC merge sublayer (see
Fig.~\ref{fig_tsn_802.3IET}) in the link layer of the next bridge, and
the next bridge then only processes complete frames.  Each preemption
operation causes some computational overhead due to the encapsulation
processing by the bridge to suspend the current fragment and to
transition the operational context to the express traffic frame and
vice versa, which is illustrated in
Fig.~\ref{fig_tsn_PreemptionExample}.  Note that this overhead occurs
only in layer 2 in the link interface.

\subsubsection{IEEE 802.1Qch Cyclic Queuing and Forwarding (CQF)}
\label{tsn_std_CQF}
While the IEEE~802.1Qav FQTSS with CBS works well for soft real-time
constraints, e.g., A/V traffic, the existing FQTSS has still several
shortcomings, including, $i)$ pathological topologies can result in
increased delay, and $ii)$ worst-case delays are topology dependent,
and not only hop count dependent, thus buffer requirements in switches
are topology dependent. The TSN TG introduced Cyclic Queuing and
Forwarding (CQF)~\cite{IEEE8021Qch}, also known as the Peristaltic
Shaper (PS), as a method to synchronize enqueue and dequeue
operations. The synchronized operations effectively allow LAN bridges
to synchronize their frame transmissions in a cyclic manner, achieving
zero congestion loss and bounded latency, independently of the network
topology.

\begin{figure}[t!]  \centering
\includegraphics[width=3.5in]{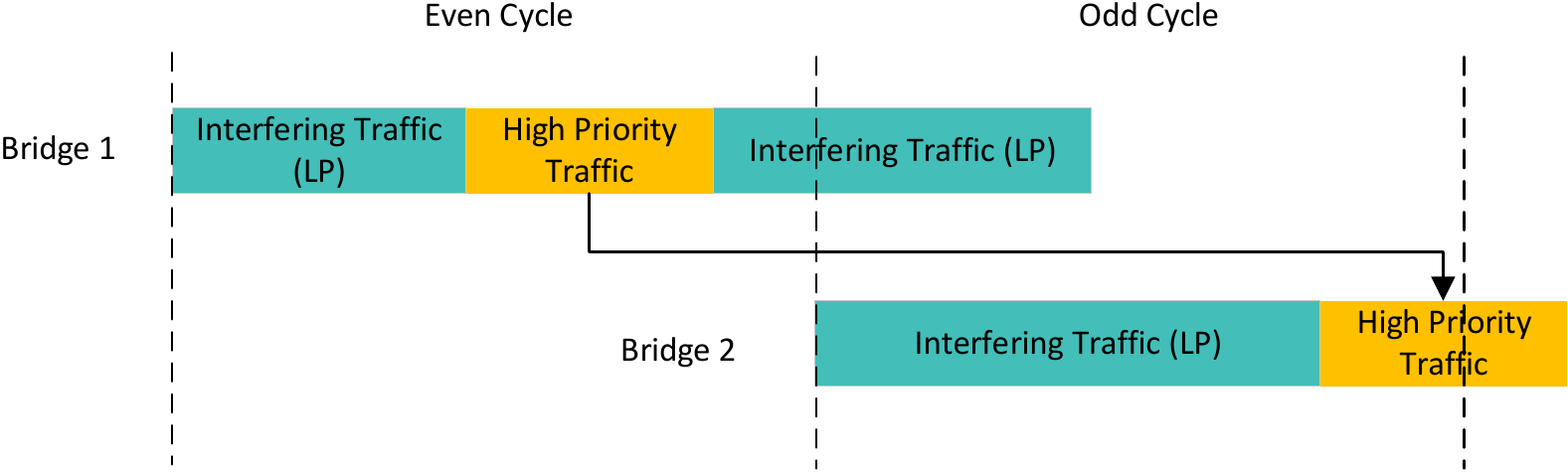} \vspace{-0.5cm}
\caption{Illustration of Cyclic Queuing and Forwarding (CQF) without
  preemption for a linear network: Each High Priority (HP) traffic
  frame scheduled on a cycle (even or odd) is scheduled to be received
  at the next bridge in the next cycle, whereby the worst-case HP
  frame delay can be two cycle times.  In the illustrated example, the
  HP traffic is delayed due to low priority interfering traffic, but still
  meets the two cycle time delay bound.}   \label{fig_tsn_CQF}
\end{figure}

Suppose that all bridges have synchronized time, i.e., all bridges are
802.1AS enabled bridges, and suppose for simplicity of the discussions
that wire lengths and propagation times are negligible.  Then, time
sensitive streams are scheduled (enqueued and dequeued) at each time
interval or cycle time with a worst-case deterministic delay of two
times the cycle time between the sender (talker) and the downstream
intermediate receiver, as illustrated in Fig.~\ref{fig_tsn_CQF}. In
essence, the network transit latency of a frame is completely
characterized by the cycle time and the number of hops. Therefore, the
frame latency is completely independent of the topology parameters and
other non-TSN traffic.

\begin{figure}[t!] 	\centering
\includegraphics[width=3.5in]{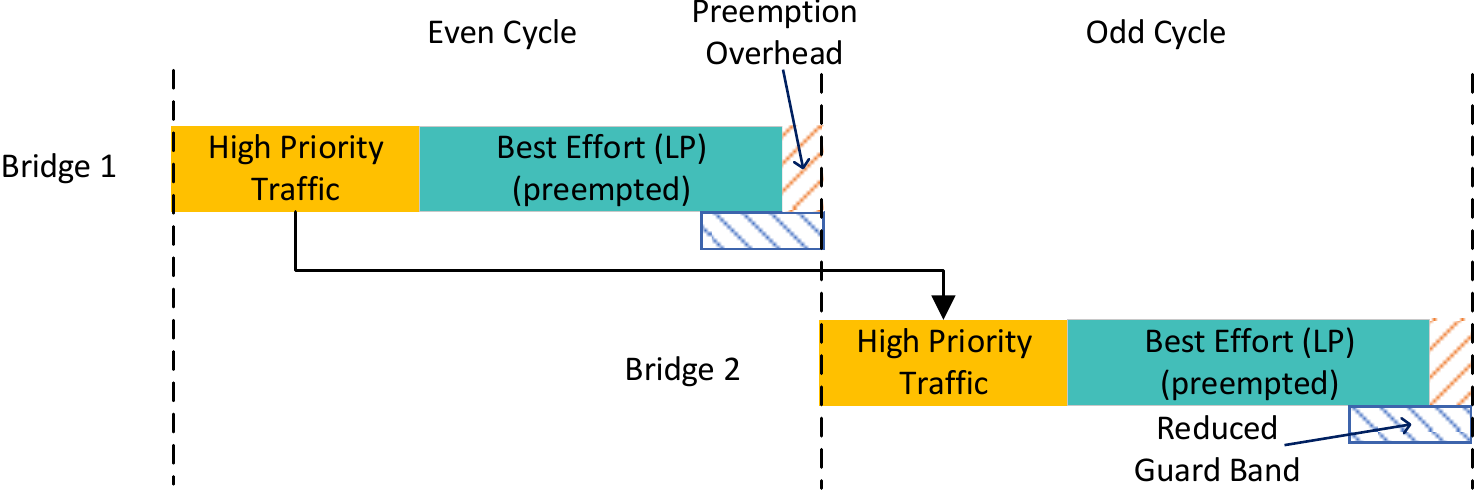} \vspace{-0.5cm}
\caption{Illustration of CQF with preemption for a linear network: A
  Guard Band (GB) before the start of the cycle prevents any
  interfering (LP) traffic from affecting the High Priority (HP)
  traffic. The CQF without preemption in Fig.~\ref{fig_tsn_CQF} did
  not prevent the LP traffic from interfering with HP traffic, while
  the CQF with preemption prevented the LP traffic from interfering
  with HP traffic.  Thus, preemption can improve the performance for
  HP traffic.}  \label{fig_tsn_CQFpreempt}
\end{figure}
CQF can be combined with frame preemption specified in IEEE~802.3Qbu,
to reduce the cycle time from the transmission time of a full size
frame to the transmission time of a minimum size frame fragment (plus
all the TSN traffic), as illustrated in
Fig.~\ref{fig_tsn_CQFpreempt}. Note however that for CQF to work
correctly, all frames must be kept to their allotted cycles, i.e., all
transmitted frames must be received during the expected cycle at the
receiving downstream intermediate
bridge~\cite{IEEE8021Qch}. Therefore, the cycle times, the alignment
of the cycle times among the bridges in the network, and the timing of
the first and last transmissions within a cycle need to be carefully
considered in order to ensure that the desired latency bounds are
achieved. To this end, CQF in conjunction with IEEE~802.1Qci ingress
policing and the IEEE~802.1Qbv TAS ensures that all frames are kept
within a deterministic delay and guaranteed to be transmitted within
their allotted cycle time.

\subsubsection{IEEE 802.1Qcr Asynchronous Traffic Shaping (ATS)}
While CQF and TAS provide ULL for critical traffic, they depend on
network-wide coordinated time and, importantly, due to the enforced
packet transmission at forced periodic cycles, they utilize network
bandwidth inefficiently~\cite{Specht2016}. To overcome these shortcomings,
the TSN TG has proposed the IEEE 802.1Qcr Asynchronous Traffic Shaper
(ATS)~\cite{IEEE8021Qcr}, which is based on the urgency-based
scheduler (UBS)~\cite{Specht2016,spe2017syn}. The ATS aims to smoothen traffic
patterns by reshaping TSN streams per hop, implementing per-flow
queues, and prioritizing urgent traffic over relaxed traffic.  The ATS
operates asynchronously, i.e., bridges and end points need not be
synchronized in time. Thus, ATS can utilize the bandwidth efficiently
even when operating under high link utilization with mixed traffic
loads, i.e., both periodic and sporadic traffic.

The UBS is based on the Rate-Controlled Service
Disciplines (RCSDs)~\cite{zhang1994rate}. RCSDs are a non-work
conserving class of packet service disciplines which includes
Rate-Controlled Static Priority~\cite{zhang1993rate} and Rate-Controlled
Earliest Deadline First~\cite{georgiadis1996efficient}. The RCSD packet
scheduling consist of two components: the rate controller implements the
rate-control policies, and the scheduler  implements the packet scheduling
according to some scheduling policy, e.g., Static-Priority,
First-Come-First-Serve, or Earliest Due-Date First. By separating the rate
controller and scheduler, the RCSD effectively decouples the bandwidth for
each stream from its delay bound, i.e., allocating a prescribed amount of
bandwidth to an individual stream is independent of the delay
bound. Hence, RCSD can support low delay and low bandwidth streams.

UBS adds a few improvements to RCSDs~\cite{zhang1994rate}, namely: 1)
UBS provides low and predictable worst-case delays even at high link
utilization, 2) low implementation complexity due to the separation of
per-flow queues from per-flow states where flow state information,
such as Head-of-Queue frame and time stamp, is stored, and 3)
independence from the global reference time synchronization;
specifically, individual flow delays are analyzed at each hop, i.e.,
per-hop delay calculation, and end-to-end delays are calculated based
on the network topology and by the closed-form composition of the
per-hop delays calculated initially.

The fundamental aim of the RCSD is to individually control frame
selection and transmission at each hop between the transmitter and
receiver, i.e., per hop shaping. As pointed out by Specht et
al.~\cite{Specht2016}, the RCSD has multiple scalability problems,
including dynamic reordering of packets within separate queues
according to the packets' eligibility times, i.e., priority queue
implementation with non-constant complex data structures, such as
heaps. Specialized calender queues have been proposed to achieve
constant complexity~\cite{Specht2016}. However, calender queues
require RCSD capable switches to have large memory pools, are
difficult to control as the network size scales up, and are ideal only
for some specific applications with special properties.  Therefore,
Specht et al.~\cite{Specht2016} utilize the RCSD concept with
the outlined improvements and have proposed a novel UBS
solution as the core of the ATS standard.

\subsubsection{Summary and Lessons Learned}
Flow control mainly enforces rules to efficiently forward and
appropriately queue frames according to their associated traffic
classes. All existing flow controls follow similar principles, namely,
certain privileges are associated with TSN flows while non-TSN flows
are delayed. Nearly all existing schedulers and shapers enforce fair
transmission opportunities according to each flow's traffic class. The
transmission selection algorithm selects the appropriate stream within
a given traffic class according to the network and traffic
conditions. Flow control collaborates with flow management, see
Section~\ref{tsn:flow_mgt:sec}, and flow integrity, see
Section~\ref{tsn:flow_int:sec}, to ensure adequate resources are
available for TSN streams.

Overall, we can classify real-time TSN systems into event-triggered
systems and time-triggered systems. For example, IEEE 802.1Qbv is a
time-triggered shaper, while IEEE 802.1Qcr is an event-triggered
shaper.  An interesting future research direction is to explore
whether both types of shapers can be combined.  That is,
would it be efficient to dynamically change a flow's priority,
individually or collectively, and to reshape flows based on neighbor
network conditions while each flow is shaped by a centralized computed
schedule incorporating time slots at each egress's port? For example,
a stream initially sent with a certain high priority can be downgraded
to low priority based on downstream network conditions while adhering
to each bridge's time-aware scheduler and gating mechanism.

Also, it will be interesting to investigate whether IEEE~802.1Qbv can
be replaced with an event-triggered shaper that guarantees an upper
bound on latency, but not generally a deterministic latency. Changing
TAS into an event-triggered shaper can lead to more flexible and
easily computed schedules since certain events, e.g., incoming frames
or network changes, can require schedule changes at runtime.

\subsection{Flow Integrity}  \label{tsn:flow_int:sec}
To accomplish the goals of deterministic ultra-low latency, jitter,
and packet loss, TSN streams need to deliver their frames regardless of
the dynamic network conditions, including physical breakage and link
failures.  Several techniques have been standardized to enable flow
integrity.

\subsubsection{IEEE 802.1CB Frame Replication and Elimination for
  Reliability (FRER)}
\begin{figure}[t!]  \centering
\includegraphics[width=3.5in]{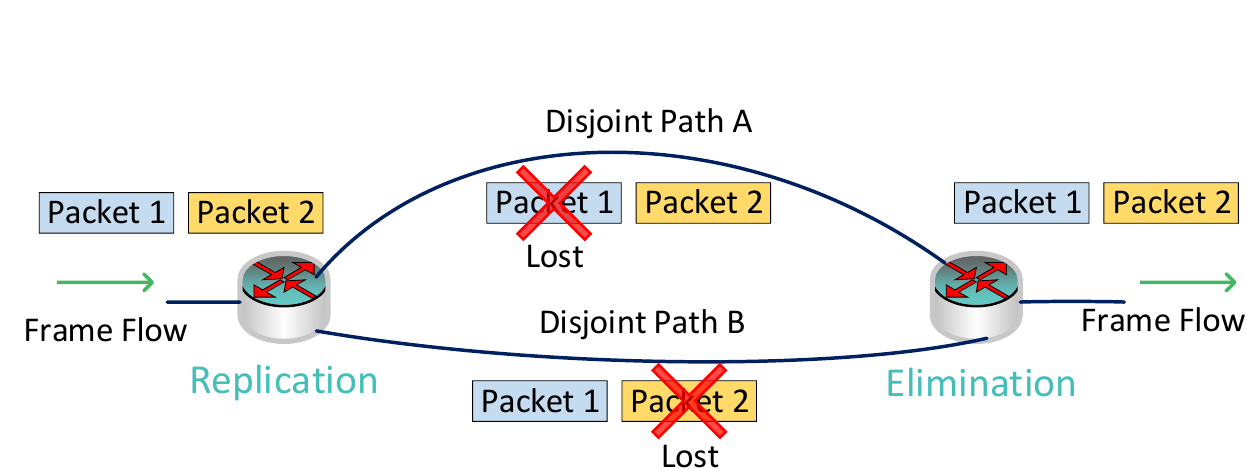} \vspace{-0.1cm}
\caption{Illustration of FRER operation: The first bridge replicates
  the frame and transmits the duplicated frames on two disjoint
  paths. The FRER operation can be started and ended at any bridge
  between the sender and receiver.}
	\label{fig_tsn_FRER}
\end{figure}

IEEE 802.1CB Frame Replication and Elimination
for Reliability (FRER)~\cite{IEEE8021CB}, is a stand-alone standard
that ensures robust and reliable communication using proactive
measures for applications that are intolerant to packet losses, such
as control applications.  802.1CB FRER minimizes the impact of
congestion and faults, such as cable breakages, by sending duplicate
copies of critical traffic across disjoint network paths, as shown in
Fig.~\ref{fig_tsn_FRER}. If both frames reach their destination, the
duplicate copy is eliminated. If one copy fails to reach its
destination, the duplicate message can still be received, effectively
providing seamless proactive redundancy at the cost of additional
network resources.

In order to minimize network congestion, the packet replication can be
selected based on traffic class and the path information acquired
through the TSN stream identification (\textit{stream\_handle}), plus
a sequence generation function.  The sequence generation function
generates identification numbers for replicated frames to determine
which frames to discard and which frames to pass on so as to ensure correct
frame recovery and merging.  The frame redundancy information is
carried in a Redundancy Tag~\cite{IEEE8021CB}.  Frame sequence numbers
and timing information are also needed to limit the memory
needed for duplicate frame detection and elimination.  For example,
FRER may only be employed for critical traffic, while best effort and
other loss-tolerant traffic is transmitted normally.  FRER is
compatible with industrial fault-tolerance architectures, e.g., High
Availability and Seamless Redundancy (HSR)~\cite{kirrmann2009hsr} and
the Parallel Redundancy Protocol (PRP)~\cite{kirrmann2007iec}.  We
note that frame duplication, routing, and elimination are non-trivial
tasks that will likely require centralized management. Hence, such
protocols can be combined with other standards, e.g., 802.1Qcc and
802.1Qca, to ensure seamless redundancy and fast recovery in time-sensitive
networks.

\subsubsection{IEEE 802.1Qca Path Control and Reservation (PCR)}
\label{IEEE802.1Qca:sec}

IEEE 802.1Qca Path Control and Reservation (PCR) is based on and
specifies TLV extensions to the IETF Link State Protocol (LSP), the
Intermediate Station to Intermediate Station (IS-IS)
protocol~\cite{oran1990intermediate}. IEEE 802.1Qca allows the IS-IS
protocol to control bridged networks beyond the capabilities of
shortest path routing
(ISIS-SPB)~\cite[Section~28]{fedyk2012extensions, IEEE8021Q},
configuring multiple paths through the network~\cite{IEEE8021Qca,
  farkas2016path}. IEEE 802.1Qca PCR aims to integrate control
protocols required to provide explicit forwarding path control, i.e.,
predefined protected path set-up in advance for each stream, bandwidth
reservation, data flow redundancy (both protection and restoration),
and distribution of control parameters for flow synchronization and
flow control messages~\cite{IEEE8021Qca}.

In general, 802.1Qca specifies bridging on explicit paths (EPs)
for unicast and multicast frame transmission, and protocols to
determine multiple active topologies, e.g., Shortest Path, Equal Cost
Tree (ECT), Internal Spanning Tree (IST), Multiple Spanning Tree
Instance (MSTI), and Explicit Tree (ET), in a bridged
network. Explicit forwarding paths, as opposed to hop-by-hop
forwarding, mitigate disruptions caused by the reconvergence of bridging
protocols. PCR has similar goals and evolved from spanning tree
protocols, e.g., the Rapid Spanning Tree Protocol
(RSTP)~\cite[Section~13.4]{IEEE8021Q}, the Multiple Spanning Tree
Protocol (MSTP)~\cite[Section~13.5]{IEEE8021Q}, and the Shortest Path
Bridging (SPB)~\cite[Section~27]{IEEE8021Q}.

\begin{figure}[t!] 	\centering
\includegraphics[width=3in]{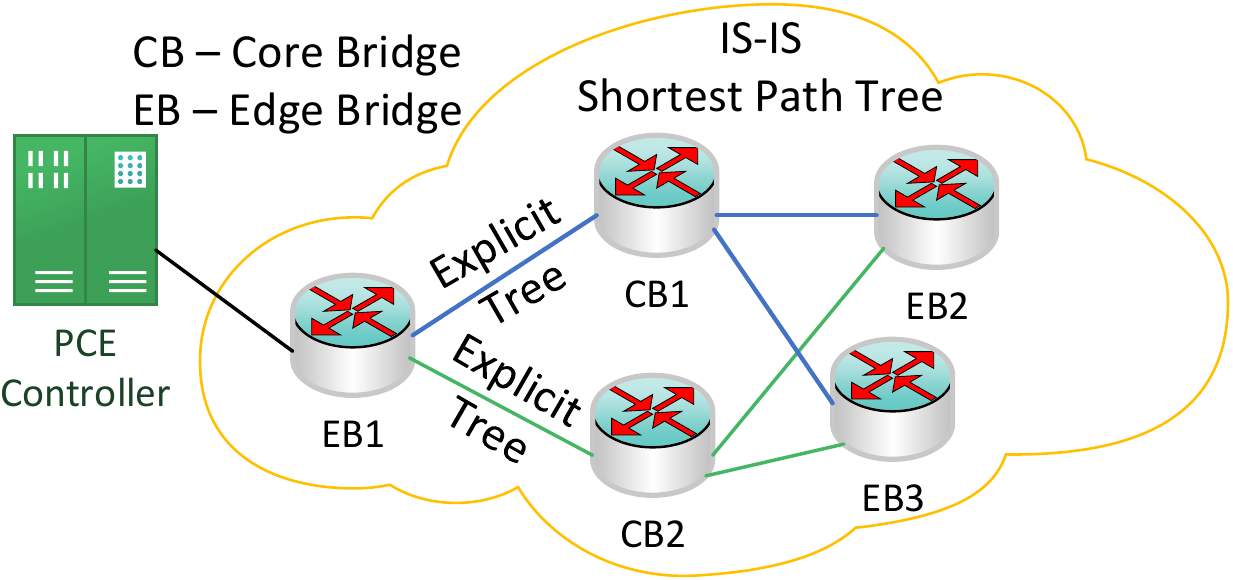} \vspace{-0.2cm}
\caption{Illustration of Explicit Paths (EPs): A control plane PCE SDN
  controller installs computed Explicit Tree (ET) paths via the IS-IS
  data plane. Two computed ET paths are shown represented by the green
  and blue lines.}	\label{fig_tsn_explicitTree}
\end{figure}

The IEEE 802.1Qca standard is based on Shortest Path Bridging
(SPB)~\cite[Section~27]{IEEE8021Q} and incorporates a Software Defined
Networking (SDN) hybrid approach~\cite{IEEE8021Qca}.  In the hybrid
approach, the IS-IS protocol in the data plane handles basic
functions, e.g., topology discovery and default path computation,
while the SDN controller~\cite{ban2018dis} in the control plane manages the Explicit
Paths (EPs), as shown in Fig.~\ref{fig_tsn_explicitTree}.  In
particular, the controller utilizes dedicated path computation server
nodes called Path Computation Elements (PCEs)~\cite{farrel2006rfc},
defined by the IETF PCE WG~\cite{farrel2006rfc}, to manage the EPs. A
PCE interacts with the IS-IS protocol to handle and install requests
for the network and can interact with the SRP protocol, see
Section~\ref{tsn:flow_mgt:sec}, to reserve resources along the
EPs. Additionally, the PCEs can manage redundancy on the EPs, thus
providing protection on top of the EPs by utilizing alternate paths,
e.g., Loop Free Alternates (LFAs)~\cite{IEEE8021Qca}, that reroute in
a few milliseconds.

\subsubsection{IEEE 802.1Qci Per-Stream Filtering and Policing (PSFP)}
\begin{figure}[t!]	\centering
\includegraphics[width=0.7in]{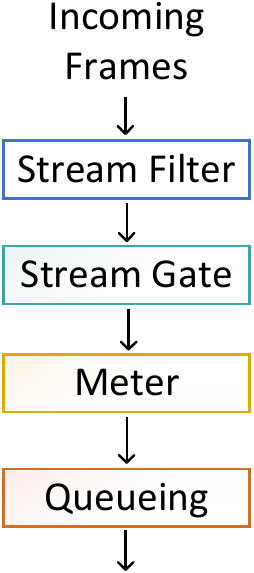}
\vspace{-0.1cm}
\caption{Illustration of PSFP flow: The flow is first filtered
  according to per-flow policies. Then, a gating mechanism
  regulates the flow. Finally, flow metering ensures bandwidth
  limitations before a frame is queued for forwarding.}
	\label{fig_tsn_PSFPflow}
\end{figure}

The IEEE 802.1Qci per-stream filtering and policing (PSFP)
  standard~\cite{IEEE8021Qci}, also known as ingress policing/gating
  standard, filters and polices individual traffic
  streams based on rule matching. IEEE 802.1Qci prevents traffic
  overload conditions, that are caused, for instance, by erroneous
  delivery due to equipment malfunction and Denial of Service (DoS)
  attacks, from affecting intermediate bridge ports and the receiving
  end station, i.e., improves network robustness.
IEEE 802.1Qci may
be used to protect against software bugs on end points or bridges, but
also against hostile devices and attacks.
IEEE 802.1Qci
  specifies filtering on a per flow (stream) basis by identifying
  individual streams with a \textit{StreamID}, which utilizes the
  802.1CB stream handler method~\cite{IEEE8021CB}. The identified
  individual streams can then be aggregated, processed, and finally
  queued to an input gate.  As illustrated in
  Fig.~\ref{fig_tsn_PSFPflow}, each gate performs three functions.

The PSFP stream filter performs per-flow filtering by matching frames
with permitted stream IDs and priority levels, and then applies policy
actions.  The PSFP stream gate coordinates all streams such that all
frames proceed in an orderly and deterministic fashion, i.e., similar
to the 802.1Qch signaling process, see Section~\ref{tsn_std_CQF}.  The
PSFP flow metering enforces predefined bandwidth profiles for
streams. The metering may, for instance, enforce prescribed maximum
information rates and burst sizes.

\subsubsection{Summary and Lessons Learned}
Flow integrity provides path redundancy, multi-path selection, as well
as queue filtering and policing. Flow integrity also prevents
unauthorized or mismanaged and rogue streams on bridged LAN networks.

In general, as network devices improve in terms of hardware
performance, they can be equipped with more state information within
the core network.  The increased state information allows for fine
granular QoS management at the expense of control messages for
efficient control dissemination in the network.  Future research needs
to carefully examine the tradeoffs between disseminating more
extensive control messages and the resulting QoS management
improvements.

\subsection{Discussion on TSN Standardization}
The IEEE TSN TG has standardized deterministic networking for Layer 2
Ethernet based bridging LANs. These standards have been revised and
continue to be updated to reflect the convergence of the industrial
and consumer markets. Overall, the TSN standards guarantee the
required QoS requirements for data transmission and provide sufficient
measures to enable end-to-end functional communication safety in the
network. Essentially, the TSN standardization provides the recommended
practices for enabling low latency, jitter, and data loss, as well as
redundancy and reservation.  In addition, the TSN standardization
provides mechanisms for bandwidth limitation, dynamic reconfiguration,
centralized management, and strict timing features.

Timing measurement and sub-microsecond time synchronization as
basis for TSN standard mechanisms can be achieved with IEEE 802.1AS
and the updated revised version 802.1AS-REV.  Essentially, all gPTP
network entities contribute to distributing and correcting delay
measurement timing information based on the source GM.  802.1AS-REV
provides, among others, GM redundancy for fast convergence.

Several flow management standards, including IEEE 802.1CB (FRER),
802.1Qca (PCR), 802.1Qci (PSFP), 802.1Qcc (Enhanced SRP and
centralized Management), 802.1CS (LRP) and RAP have been published or
are in progress to enable redundancy, path reservation, bandwidth
limitation, dynamic reconfiguration, as well as overall flow integrity
and management.  Although standard Ethernet provides redundancy
features, e.g., through spanning tree protocols, the convergence time
in the event of a failure is too slow for real-time IACS
applications. Therefore, FRER is used to proactively enable seamless
data redundancy at the cost of additional bandwidth
consumption. Moreover, PCR in combination with FRER and 802.1Qcc
enables fast recovery, efficient path redundancy, and dynamic runtime
flow management. Furthermore, PSFP manages, controls, and prevents
rogue flows from deteriorating the network performance. Since SRP and
the related signaling protocols are fully distributed mechanisms
targeted towards AVB applications, the SRP and MRP protocols are not
scalable to large networks with real-time IACS applications due to a
limited state information database for the registered flows, see
Section~\ref{tsn:std:lrp}. Therefore, LRP in conjunction with RAP as
the signaling protocol features a decentralized approach to support
resource reservations for scalable TSN enabled networks.

To achieve low latency, several flow control standards have been
released, including IEEE 802.1Qbv (TAS), 802.1Qch (CQF), and IEEE
802.1Qcr (ATS).  For TAS, IEEE 802.1Qbu frame preemption can ensure
that the transmission channel is free for the next express traffic
transmission.  CQF can coordinate ingress and egress operations to
reduce the TAS configuration complexity, albeit at the expense of
higher delays. Finally, ATS has been proposed to provide deterministic
operations independently of the reference time synchronization and low
delays for high link utilization.  The efficient dynamic configuration
of these flow control standard mechanisms, including IEEE~802.1Qbv, is
an open challenge that requires extensive future standardization and
research efforts.

The TSN mechanisms (and similarly the DetNet mechanisms) do not
  explicitly define mechanisms to specifically reduce packet
  jitter. The various TSN mechanisms for ensuring very short
  deterministic packet delays implicitly achieve very low packet
  jitter. Moreover, resource reservation and admission control can
  further reduce end-to-end jitter by limiting interfering traffic,
  which is typically the main cause of jitter. Additionally, CQF can
  coordinate ingress and egress operations, which can cause jitter, to
  reduce delays to sub-microsecond levels or to bound delays to within
  a few microseconds, effectively eliminating jitter caused by the
  physical properties of links and switching
  fabrics~\cite{nay2018inc}. However, while it is very
  unlikely that high jitters occur in a TSN network, in the event of
  high jitter, the TSN standards do not actively delay or throttle
  flows to compensate for the high jitter condition. Such specific
  jitter control operations are an open issue for potential future TSN
  standards development.

The TSN standardization has so far excluded the specific consideration
of security and privacy.  The IEEE 802.1 Security TG has addressed
security and privacy in general IEEE 802.1 networks, i.e.,
functionalities to support secure communication between network
entities, i.e., end stations and bridges. The TG has detailed a number
of standards and amendments, including 802.1X Port-based Network
Access Control (PNAC)~\cite{IEEE8021X,IEEE8021Xbx} , 802.1AE MAC
Security (MACsec)~\cite{IEEE8021AE, IEEE8021AEbn, IEEE8021AEbw,
  IEEE8021AEcg}, and 802.1AR Security Device Identity
(DevID)~\cite{IEEE8021AR}, that focus on providing authentication,
authorization, data integrity, and confidentiality.  Specifically,
PNAC utilizes industry standard authentication and authorization
protocols enabling robust network access control and the establishment
of a secure infrastructure. Furthermore, PNAC specifies the MACsec Key
Agreement (MKA)~\cite{IEEE8021Xbx} protocol.  MACsec specifies the use
of cryptographic cipher suites, e.g., Galois/Counter Mode of Advanced
Encryption Standard cipher with 128-bit key (GCM-AES-128), that allow
for connectionless user data confidentiality, frame data integrity,
and data origin authentication, essentially providing a set of
protocols that ensures protection for data traversing Ethernet LANs.
For instance, DevID is a unique per-device identifier that
cryptographically binds a device to the DevID. Thus, 802.1 LAN devices
can be authenticated and appropriate policies for transmission and
reception of data and control protocols to and from devices can be
applied. The IEEE 802.1 Security TG is working on a couple of
amendments to address privacy concerns and to include a YANG model
allowing configuration and status reporting for PNAC in 802.1 LANs.
The integration of the security protocols and standards with TSN
enabled networks needs to be addressed in future research and
standardization.  For instance, the impact of the security stack
overhead on TSN flows and the impact of the security overhead on OT
related applications running over Ethernet LANs need to be investigated.
Thus, there are ample research opportunities for testing and
benchmarking to ensure the efficient integration of legacy security
protocols with TSN.

The important area of networks for industrial applications often
employs cut-through switching techniques. An interesting future
research direction is to investigate how networking with cut-through
switching compares with networking based on the TSN standards (tool
sets).

More broadly, even though many standards and recommended practices
addressing deterministic networking have been published, significant
testing and benchmarking is needed to provide assurances to the
industry and consumer markets.

\begin{figure*}[t!]
	\footnotesize
	\setlength{\unitlength}{0.10in}
	\centering
	\begin{picture}(33,33)
	\put(8,33){\textbf{TSN Research Studies, Sec.~\ref{tsn:res:sec}}}
	\put(-14,30){\line(1,0){59}}
	\put(-14,30){\vector(0,-1){2}}
	\put(-17,27){\makebox(0,0)[lt]{\shortstack[l]{			
				\textbf{Flow Synchronization},\\
				\ Sec.~\ref{tsn:res:flow_sync:sec} \\ \\
				Clock Precision~\cite{gutierrez2017synchronization} \\
				Freq. Sync.~\cite{Li2017arch} \\
				Timing Accuracy~\cite{Noseworthy2016,shr2018pre}
	}}}
	\put(0,30){\vector(0,-1){2}}
	\put(-3,27){\makebox(0,0)[lt]{\shortstack[l]{			
				\textbf{Flow Management},\\
				\ Sec.~\ref{tsn:res:flow_mgt:sec} \\ \\
				Resource Resrv.~\cite{Park2016} \\
                                Reconfiguration~\cite{raagaard2017fog}\\
                                Bandw. Alloc.~\cite{ko2015research} \\
				Routing~\cite{Arif2016} \\
				SDN TSN~\cite{Nayak2015,Thiele2016}
	}}}
	\put(15,30){\line(0,1){2}}
	\put(23,30){\vector(0,-1){2}}
	\put(20,27){\makebox(0,0)[lt]{\shortstack[l]{	
				\textbf{Flow Control},\\
				\ Sec.~\ref{tsn:res:flow_ctrl:sec}
	}}}
	\put(12,23){\line(1,0){23}}
	\put(23,24){\line(0,-1){1}}
	\put(12,23){\vector(0,-1){2}}
	\put(8,20){\makebox(0,0)[lt]{\shortstack[l]{
				\underline{Shaping} \\
				\ Traf. Shap. Ana..~\cite{Thangamuthu2015} \\
				\ Shaping Overhead~\cite{Farzaneh2016ontology}
	}}}
	\put(23,23){\vector(0,-1){2}}
	\put(21,20){\makebox(0,0)[lt]{\shortstack[l]{
	      \underline{Scheduling} \\
              	TTEthern. vs.~TSN~\cite{craciunasoverview} \\
				Ctrl. Traf. Sch.~\cite{Bello2014} \\
				Opti. Sch.~\cite{durr2016no, craciunas2016scheduling,far2017gra} \\
				Urgency Sch.~\cite{Specht2016,spe2017syn} \\
				Joint Rout. \& Sch.~\cite{Pop2016,smi2017opt,mah2018sta} \\
				Traf. Sch. Impact~\cite{Smirnov2017, park2016simulation}
	}}}
	\put(35,23){\vector(0,-1){2}}
	\put(33,20){\makebox(0,0)[lt]{\shortstack[l]{
			\underline{Preemption} \\
			\ Preemption~\cite{Lee2016time} \\
			\ Pre. effect on non-CDT~\cite{thiele2016preempt} \\
			\ Preempt. M/G/1 Analy.~\cite{Zhou2017}
	}}}
	\put(45,30){\vector(0,-1){2}}
	\put(42,27){\makebox(0,0)[lt]{\shortstack[l]{
				\textbf{Flow Integrity},\\
				\ Sec.~\ref{tsn:res:flow_int:sec} \\ \\
				Fault Tolerance.~\cite{Kehrer2014,alv2017tow}
	}}}		
	\end{picture}
	\vspace{-3cm}	
	\caption{Classification of TSN research studies.}
	\label{TSN:res:fig}
\end{figure*}
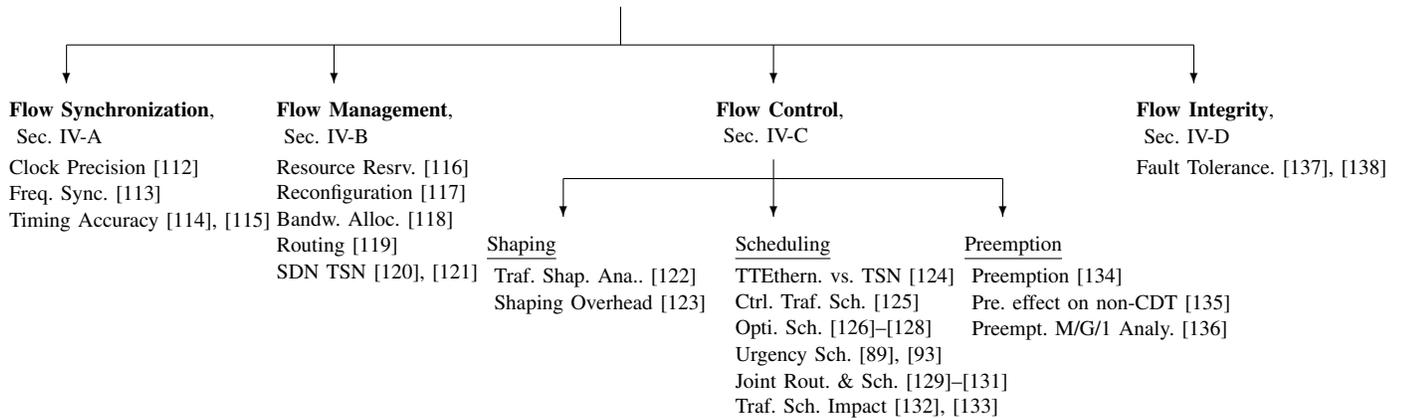

\section{TSN Research Studies} \label{tsn:res:sec}
This section surveys the existing research studies towards achieving
ULL in the context of the TSN standards. The TSN standards provide
tool sets to enable TSN characteristics, such as flow synchronization
and flow control (see Sec.~\ref{tsn:std:sec}), in conventional
networks. Based on the application requirements, various TSN standard
tools can be independently and selectively adopted on network segments
to enable TSN characteristics.  Similar to the organization of the
review of TSN standards in Fig.~\ref{ctl_class:fig}, we organize the
survey of TSN related research studies in Fig.~\ref{TSN:res:fig}
according to the same classification as the TSN standards in
Fig.~\ref{ctl_class:fig}.  To date there have been no specific
research studies on the TSN flow concept; therefore, we omit the flow
concept category in Fig.~\ref{TSN:res:fig}.

\subsection{Flow Synchronization} \label{tsn:res:flow_sync:sec}

\subsubsection{Clock Precision}

Most existing time synchronization implementations are limited to
clock precisions on the order of
sub-microseconds~\cite{sommer2013race}. The global sharing of the
timing information across the network elements allows the clocks in
the network elements to be precisely synchronized relative to each
other (see Section~\ref{tsn:flow_sync:sec}).  The challenges
associated with network wide clock synchronization are not limited to
one particular network attribute. Rather, a wide set of network
attributes, including hardware capabilities, such as clock stability,
and isolation from environmental impacts e.g., temperature, and
software implementations, e.g., for designing an effective closed-loop
system to track and correct the timing drifts, influence the
synchronization quality in the network as a whole.  As a result, most
current deployments rely on sub-microseconds clock precision
techniques.  However, future trends in network applications require a
tighter clock synchronization on to the order of sub-nanoseconds in
Ethernet networks. For instance, the control system of the CERN Large
Hadron Collider (LHC) communication network has to operate with
sub-nanosecond precision to share timing and perform time-trigger
actions~\cite{cussans2017trigger}.

Gutierrez et al.~\cite{gutierrez2017synchronization} have analytically
evaluated the synchronization process and the quality of the timing
error estimation in large scale networks based on the IEEE 802.1AS TSN
synchronization standard.  In particular, Gutierrez et al.~focused on
the clock synchronization quality with a small margin of error between
each node for a large network consisting of a few thousand nodes with
maximum distances between the grandmaster clock and synchronized node
clocks spanning up to 100 hops.  The study of the protocol behavior
included various network aspects, such as clock granularity,
network topology, PHY jitter, and clock drift.  The results from
probabilistic analytical modeling and simulation evaluations
indicate that implementation specific aspects, such as PHY jitter and
clock granularity, have a significant impact on the clock precision with
deviations reaching 0.625~$\mu$s in the TSN synchronization process.
Therefore, it is critical to ensure that the physical properties of
the clock within each node are accurate so as to ensure the overall
quality of the synchronization process in TSN networks that adopt IEEE
802.1AS.

\subsubsection{Frequency Synchronization}
\begin{figure}[t!] \centering
\includegraphics[width=3in]{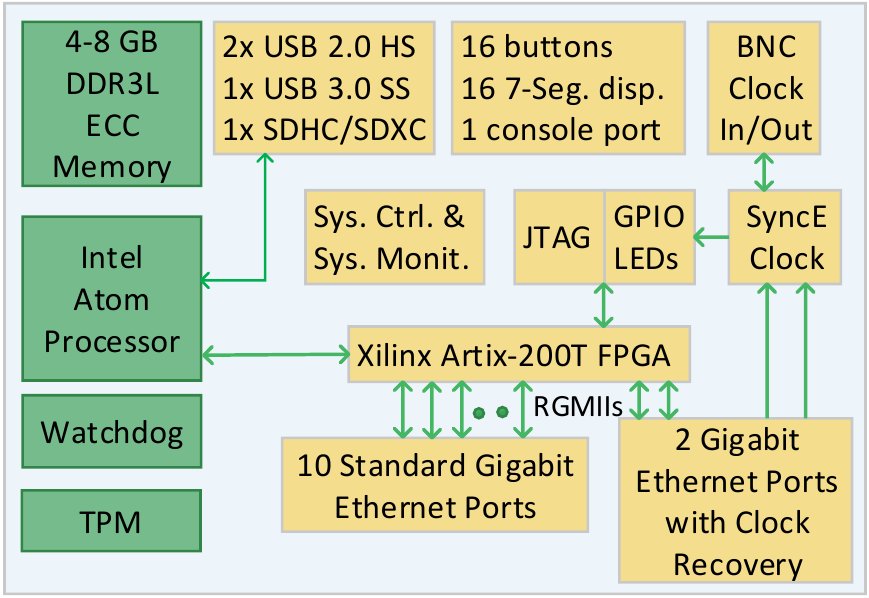} \vspace{-0.1cm}
\caption{Illustration of frequency synchronization design supporting TSN with
  clock recovery and network wide synchronization~\cite{Li2017arch}.}
	\label{fig_system_sync}
\end{figure}
Li et al.~\cite{Li2017arch} have introduced a novel networking device
architecture that provides ULL switching and routing based on
synchronization.  Their design integrates a state-of-the-art FPGA with
a standard x86-64 processor (which supports both 32 and 64 bit
operation) to support TSN functions.  The system provides frequency
synchronization over standard Ethernet to the entire network.
Frequency synchronization enables distribution of timing information
with low-jitter across the network.  In the frequency synchronization
design illustrated in Fig.~\ref{fig_system_sync}, datapaths are
enabled with one or more synchronous modules supported by clock
synchronization.  These datapaths are allocated resources in terms of
bit rate and packet rate based on the worst-case traffic load.  This
design exploiting hardware synchronization capabilities achieves
cut-through latencies of 2 to 2.5~$\mu$s for twelve Gigabit Ethernet
ports at full line rate packet processing~\cite{Li2017arch}.  The
constituents of the observed latency were identified as pipeline
delay, arbitration delay, aggregation delay, backpressure cycles,
cross-clock domain synchronization cycles, datapath width adaptation
cycles, and head-of-line blocking cycles.  Emphasizing the importance
of the hardware implementation of the frequency synchronization
process, Li et al.~\cite{Li2017arch} suggest that their novel hardware
implementation and timing distribution process based on frequency
synchronization across networks can be easily extended to other custom
designs.

\subsubsection{Timing Accuracy}
Although TSN protocols offer very accurate timing information for the
inter clock alignment, the validity and accuracy of the received
timing information can still be uncertain. That is, typically the
timing information received from the grand master is blindly followed
by the clock alignment process, which can potentially result in
out-of-sync clocks if the received timing information is not accurate.
The detection of erroneous timing information by the receiving node
can potentially help time critical network applications to re-trigger
the verification, calibration, and re-synchronization process.
Moreover, nodes can use this information to alert network applications
to request a new path or to terminate critical operations that require
timing precision. Therefore, timing accuracy is an essential aspect in
TSN networks.

The time-error is the relative clock difference between the slave and
the grand master. The time-error can still exist even if the slave
node applies the timing corrections based on timing error estimation.
The timing accuracy represents the overall quality of the timing
distribution throughout the network.  The timing accuracy at a node
can be estimated in two ways: $i)$ by receiving the timing information
from another source and periodically comparing to check the accuracy,
and $ii)$ keeping track of the node's self error and (ingress and
egress) port latencies to predict the inaccuracy in the received
timing.  Noseworthy et al.~\cite{Noseworthy2016} have specifically
addressed the timing inaccuracy of a Precision-Time Protocol (PTP)
node with the help of an auxiliary node. The proposed network-based
system monitors and measures the timing errors and port latencies to
track the self errors independently of the PTP protocol and network
application.  Such a system can share the information with other nodes
so that the other nodes can estimate the timing errors.  In addition
to the timing error of a PTP node, the ingress and egress delays in
the PTP nodes for a specific TSN flow have been estimated and used in
the process of clock reference maintenance.
A PTP extension to wireless networks has
been investigated in~\cite{shr2018pre} while related measurement
techniques have been examined in~\cite{kovacshazy2016towards,kov2018low}.

\subsubsection{Summary and Lessons Learnt}
An important aspect of timing and synchronization in TSN networks is
to estimate the relative timing difference between two nodes.  Timing
differences may arise because of clock errors, synchronization errors,
as well as tracking and estimation
errors~\cite{levesque2016survey}. Clock errors are caused by the
timing drifts resulting from hardware imperfections. Synchronization
errors are caused by false timing information and wrong interpretation
of timing information. Tracking and estimation errors can, for
instance, arise due to sleep states for power savings.  In deep-sleep
states, only a minimal set of sub-systems is kept alive. Moreover, the
clock system is typically switched from high resolution and high
precision to low resolution and low precision, which may incur large
clock drifts.  The repeated switching of the clocking system may
accumulate significant synchronizing errors that need to be corrected
by external sources.  In order to achieve high-order precision in the
clock implementation for TSN applications, all aspects of the clock
errors must be considered to mitigate the effects arising from
incorrect local timing.

The clock synchronization in the network requires significant
bandwidth, i.e., imposes a significant overhead in the network.  The
synchronization data needs to be propagated throughout the network in
a deterministic fashion. Hence, the synchronization traffic interferes
with the scheduled and regular traffic. Therefore, the design of TSN
networks requires careful consideration of the overhead resulting from
the synchronization process and efforts to reduce the overhead.  On
the other hand, the effectiveness of the protocol that facilitates the
synchronization process is limited by the node capability to preserve
a synchronized local clock.  If the local clock skew is high compared
to the frequency of the synchronization process, then the local clock
will often have the wrong timing. Therefore, the future design of
synchronization protocols and the frequency of synchronization should
be based on the node characteristics.

\subsection{Flow Management}  \label{tsn:res:flow_mgt:sec}

\subsubsection{Resource Reservation}
\begin{figure}[t!] \centering
\includegraphics[width=2in]{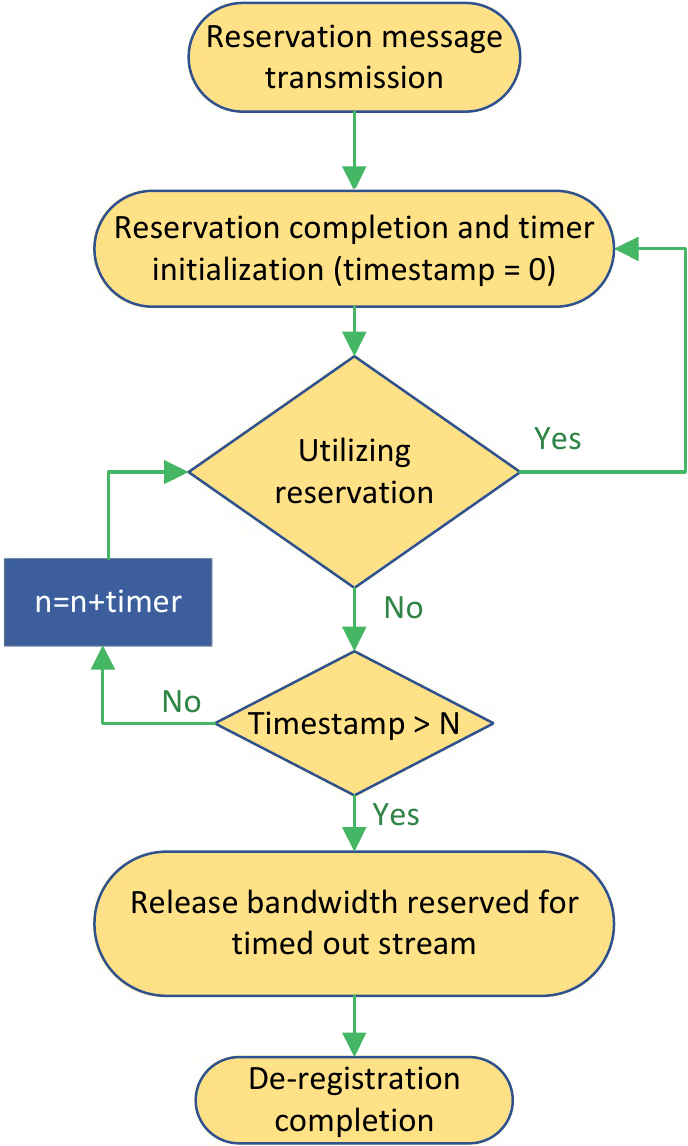} \vspace{-0.2cm}
\caption{The automatic flow de-registration process monitors the
  network for transmission activity and removes the resource
  reservations when a flow is idle for more than a threshold
  duration~\cite{Park2016}.}   \label{fig_de_registration}
\end{figure}

A resource reservation process is typically applied across the network
elements so as to ensure that there are sufficient resources for
processing TSN flow frames with priority. The TSN IEEE 802.1 Qat
protocol defines the resource reservation mechanism in TSN networks,
see Section~\ref{802.1Qat:sec}.  Park et al.~\cite{Park2016} revealed
that the TSN IEEE 802.1 Qat standard lacks effective procedures for
terminating reserved resources. The existing standardized resource
release mechanism involves signaling among TSN nodes to establish a
distributed management process, such that the connection reservations
are torn-down and the resources released when the TSN flow is no
longer needed.  Similarly, when there is a renewed need for the TSN
flow, the connection with its resource reservation is re-initiated
based on the flow's traffic requirements.  For networks with a few
nodes and short end-to-end delays, the management process has
relatively low signaling complexity and does not significantly impact
the TSN flows.  However, Park et al.~\cite{Park2016} found that the
numbers of nodes that are typical for in-vehicle networks result in a
pronounced increase of the overall control message exchanges for the
tear-down and re-initiation of connections.

Therefore, Park et al.~\cite{Park2016} have proposed an automatic
de-registration to tear down reservations.
All participating nodes run the algorithm to de-register the reserved
resources in a synchronized manner across the entire network based on
the network wide synchronization capability in TSN networks.
Figure~\ref{fig_de_registration} presents the flow chart of the
automatic de-registration process: A timer is initialized to track the
idle times for a specific TSN flow. Once the timer meets a predefined
threshold, the resource reservations of the flow are automatically
torn-down by all the participating nodes. The de-registration process
is simultaneously performed throughout the network based on the
synchronized timers. The downside of such an automatic de-registration
process is the overhead for the re-activation process of the resource
reservation for TSN flows which were deactivated due to a short period
of inactivity.  Thus, for highly bursty traffic, the automatic
de-registration process may negatively impact the overall network
performance since the idle times between traffic bursts may trigger
the automatic de-registration.

Raagaard et al.~\cite{raagaard2017fog} have examined GCL reconfiguration
in the context of CNC and CUC (see Section~\ref{802.1Qat:sec}).  The
actual underlying scheduling mechanism is an elementary greedy
earliest deadline first heuristic. That is, flows with earlier
deadlines are scheduled first. A weakness of the approach appears to
be the long reconfiguration time. Despite the algorithmic simplicity,
reconfigurations take between several seconds to up to a
minute. Dynamic runtime management and reconfiguration of the IEEE 802.1Qbv GCL
schedules thus continue to pose a significant challenge and are an
important topic for future
research~\cite{bha2018cen,cra2018dem,gav2018sch,gut2017sel,hei2017sel,oli2018iee,pop2018ena,sch2018dyn}.

\subsubsection{Bandwidth Allocation}
Bandwidth allocation reserves the physical transmission resources
required to meet the delay requirements of an end-to-end flow.
A specific bandwidth allocation challenge in TSN arises
from the multiple traffic classes, such as the different priority
levels for scheduled traffic and best-effort non-scheduled
traffic.

Ko et al.~\cite{ko2015research} have developed a simulation model to
study the impact of the Maximum Transmit Unit (MTU) size of TSN
traffic packets on the performance for scheduled traffic within a
specific bandwidth allocation framework.
Specifically, Ko et
  al. have examined bandwidth allocations for the scheduled traffic
  based on TSN definitions.  Ko et al. assume that 75\% of the
  bandwidth is allocated to the different QoS traffic classes, while
  the remaining 25\% of the bandwidth are allocated to best-effort
  traffic.  In particular, two classes of QoS traffic were considered,
  namely scheduled traffic and audio/video traffic.  Bandwidth is
  allocated such that the total bandwidth allocated to scheduled and
  audio/video is always 75\%, i.e., the allocation ratio between QoS
  traffic and best-effort traffic is maintained constant (75\% to
  25\%).  The study varies the bandwidth ratio between the scheduled
  traffic and the audio/video traffic.  The bandwidth allocation for
  the scheduled traffic was varied by varying its MTU size.
The
  simulations for a specific in-vehicle network scenario found that an
  MTU size of 109 bytes (corresponding to a bandwidth allocation of
  7\% to scheduled traffic), optimally allocated bandwidth to the
  scheduled traffic, which achieved an average end-to-end latency of
  97.6~$\mu$s.

\subsubsection{Routing}
In contrast to routing mechanisms in conventional networks, Arif et
al.~\cite{Arif2016} have proposed a computationally efficient
optimization method to evaluate the routing paths for a TSN end-to-end
connection.  The proposed solution considers an optimality criterion
that minimizes the routing path delays which effectively reduces the
end-the-end latency of the TSN flows across the network. The proposed
approach also considers multipath jitter, as well as the probability
of loop occurrence while evaluating the end-to-end routing path of the
TSN flow. The main purpose of the routing is to load balance the TSN
flows in the network nodes and thus to reduce the routing path delays.

\subsubsection{Software Defined Networking for TSN}
\label{sec:tsn:res:mgt:sdn}
\begin{figure}[t!] \centering
\includegraphics[width=3.5in]{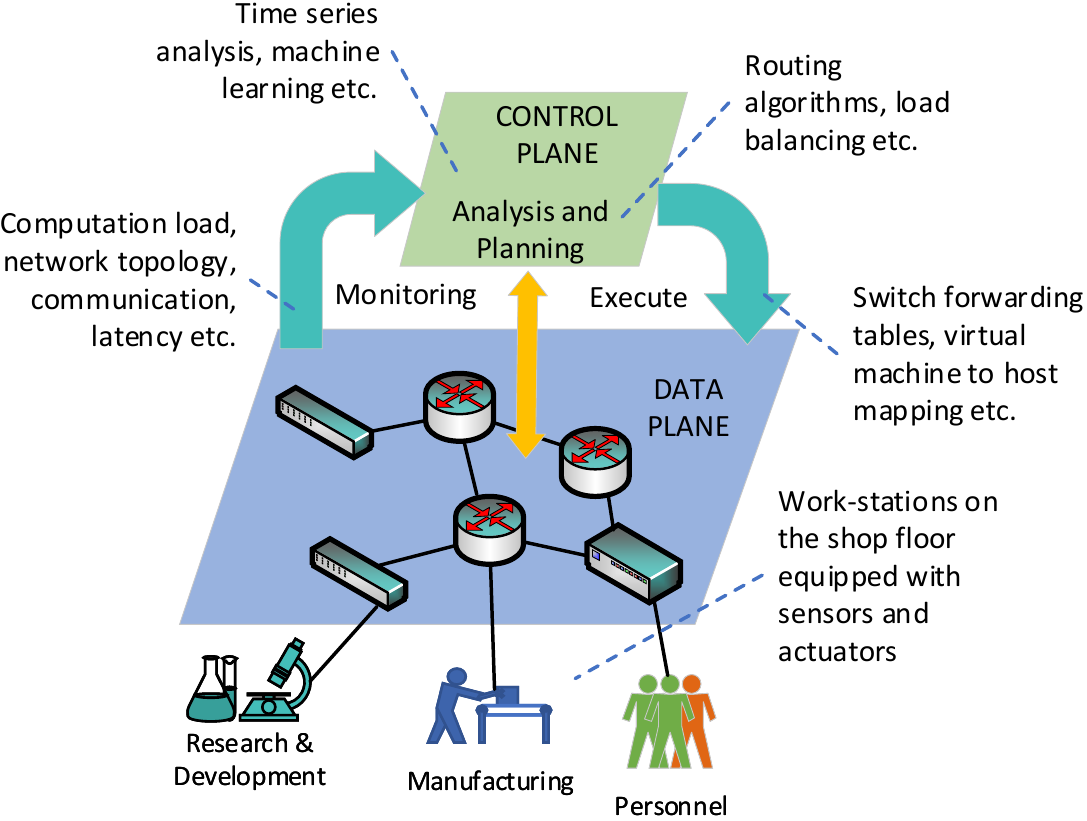} \vspace{-0.5cm}
\caption{Software Defined Networking (SDN) based Time Sensitive Networking
  (TSN) in industrial network setting: Monitoring sensors from various
  factory locations deliver information to the centralized
  controller. The centralized controller applies the time sensitive
  networking rules across the industrial networks to support critical
  connectivity paths~\cite{Nayak2015}.}
	\label{fig_tsn_sdn}
\end{figure}
The centralized computation and management of routing of an end-to-end
TSN flow follows similar principles as the central control in the SDN
paradigm. A formal adoption of the SDN paradigm in TSN networks has
been presented by Nayak et al.~\cite{Nayak2015}.  Nayak et
al.~employed SDN principles to evaluate the routing of TSN flows and
to apply the evaluated routes to the network nodes.  As shown in
Fig.~\ref{fig_tsn_sdn}, the proposed SDN controller implements four
main management functions, namely monitor, analyze, plan, and execute
to establish and control the TSN flows.  Nayak et al.~have conducted
delay and flexibility simulation evaluations of several routing
mechanisms with the SDN approach and without the SDN approach to
quantify the benefits offered by SDN.  Based on simulations, Nayak et
al. have proposed the adoption of SDN to existing processes for the
network management of time-sensitive applications.

While SDN inherently provides management flexibility~\cite{ami2018hyb,hua2017sur,tro2016sur}, the actual
deployment characteristics of SDN for TSN still need to be carefully
characterized.  Towards this goal, Thiele et al.~\cite{Thiele2016}
have presented the challenges in adapting SDN for TSN
networks. Specifically, Thiele et al.~have performed a timing analysis
of an end-to-end TSN flow in the SDN framework to verify the
limitations of SDN, such as overhead, scalability, and control plane
delay in meeting the TSN requirements for in-vehicle networks.  Thiele
et al. used a compositional performance analysis framework to model
the SDN network performance for TSN. The SDN deployment requires a
centralized controller for the global management of the TSN network
from flow establishment to tear-down. The placement of the controller
among the TSN nodes can be challenging since the control signalling
communication between nodes and controller can span across the entire
network.  Each TSN flow establishment process requires the exchange of
control messages between a TSN node and the controller. As the numbers
of TSN nodes and flows increase, the overhead due to control message
exchanges could increase, affecting the overall TSN performance. Moreover, the flow setup process requires the TSN node to request
the flow rules from the SDN controller which can increase the flow
setup time as compared to a static non-SDN scenario. Therefore, to
determine the feasibility of SDN for in-vehicle TSN networks
an analytical formulation was verified through simulations.  The
simulation results demonstrate that the worst-case SDN network
configuration delay is 50~ms, which is typically tolerable
for admission control and fault recovery in conventional Ethernet
networks. A related SDN based control plane architecture has recently
been proposed in~\cite{sch2018inv}.

\subsubsection{Summary and Lessons Learnt}
In addition to dynamic flow establishment based on current network
characteristics, flow management ensures that TSN networks preserve
the time-sensitive characteristics, such as low end-to-end delay, when
the network characteristics, such as topology and number of nodes,
change. The adaptability of the network to changes in network
characteristics is an important network design aspect that needs to be
examined in detail in future research. This future research needs to
address the control plane as well as the data plane.

Currently, IEEE~802.1Qcc has centralized management, but does not
preclude distributed management.  The TSN TG has started the process
of chartering a project to standardize RAP, see Section~\ref{RAP:sec},
which uses distributed management.  Generally, centralized management
can reduce the traffic overhead and reduce the management complexity.
The detailed investigation of the tradeoffs between centralized and
distributed management is an important direction for future research.

The static allocation of link resources to a TSN flow
can result in low network efficiency. Dynamic link resource allocation
provides more efficiency and flexibility.  More specifically, a flow
management technique can be implemented to statistically multiplex
several flows sharing common network resources, while the worst-case
flow performance is still bounded by a maximum prescribed value.  A
pitfall that needs to be carefully addressed is the network complexity
in developing and deploying flow management techniques in actual
networks.  SDN may be a promising technology for the management of
dynamic resource allocation in TSN networks. SDN also provides an
inherent platform to design advanced TSN flows management mechanisms,
such as admission control and security mechanisms.

\subsection{Flow Control}  \label{tsn:res:flow_ctrl:sec}
The overall temporal characteristics of a TSN flow are dictated by the
flow control mechanisms that are applied in the intermediate nodes.
The flow control mechanisms implemented at each TSN node directly
impact the process of frame traversal through each node that a
particular flow is defined to pass through.  A variety of flow control
mechanisms are employed in the intermediate nodes before an enqueued
frame is scheduled for transmission over the physical link.
The most critical flow control mechanisms in TSN nodes are traffic
shaping as well as scheduling and preemption.

Traffic shaping limits the traffic rate to a maximum allowed rate,
whereby all traffic exceeding the maximum allowed rate is buffered and
scheduled for transmissions at an available opportunity.  (In
contrast, traffic policing simply drops the exceeding traffic.)  The
downside of traffic shaping is queuing delay, while the downside of
policing is that excess frame dropping can affect the TCP transmission
windows at the sender, reducing the overall network throughput.

\subsubsection{Traffic Shaping}  \label{sec:traffic:shaping:analysis}
Control-Data Traffic (CDT) is the TSN traffic class for transmissions
of control traffic with the shortest possible delay. In addition to
the CDT class, TSN distinguishes traffic class A and class
B. Collectively, these traffic classes are shaped by the traffic
shapers in the TSN nodes to meet the delay requirements. The traffic
shapers ensure that $i)$ the CDT is allocated resources with strict
priority, $ii)$ the TSN traffic is isolated from the regular traffic,
and $iii)$ the wait times for enqueued frames are bounded. Towards
these goals, various traffic shaping methods have been standardized, see
Section~\ref{tsn:flow_ctrl:sec}, in order to satisfy the requirements
of the flows based on their traffic classes.

\paragraph{Shaping Analysis}
Thangamuthu et al.~\cite{Thangamuthu2015} have conducted a detailed
comparison of the standard TSN traffic shaping methods.  In
particular, Thangamuthu et al. have compared the burst limiting shaper
(BLS, a variation of CBS, which was considered in research
but not incorporated into the TSN standard),
the time aware shaper (TAS), and the peristaltic shaper (PS),
see Section~\ref{tsn:flow_ctrl:sec}. The simulations show that for
typical 100~Mbps Ethernet network deployments the in-vehicle delay
requirements are met for most applications, except for applications
with strict delay requirements. Therefore, additional ULL mechanisms
are recommended, in addition to the traffic shaping, to satisfy strict
application requirements.  Complementarily, Thiele et
al.~\cite{thiele2015timeaware,Thiele2016bustlimit,mig2018ins} have conducted a
formal timing analysis and worst-case latency analysis of the
different shapers for an automotive Ethernet topology,
while an avionics context has been considered in~\cite{he2017imp}.
Moreover, general latency and backlog bounds have recently been
  derived in~\cite{cao2018ind,fin2018inc,fin2018net,jia2018bas,moh2018end,zha2018tim,zha2018wor }.
  As alternative to CBS and TAS shaping, a pre-shaping approach at
  the senders has been explored in~\cite{nav2018pre}.
  A complementary analysis of the ATS shaper has bee conducted
  in~\cite{zho2018ana}.
  Pre-shaping
  has been found to be effective for a low number of hops. However,
  the pre-shaping effectiveness decreases with increasing hop count.
  Also, pre-shaping does not protect the shaped traffic flows from
other unshaped or misbehaving flows in the network.
The wireless fronthaul context, see Section~\ref{CPRI:sec},
has been considered in~\cite{wan2015per}.

\paragraph{Traffic Shaping Overhead}
Traffic shaping, in particular the TAS can significantly impact the
configuration overhead throughout the network, especially for
temporary (short lived) TSN flows. Typically, the TSN flows that
originate from plug-and-play devices attached to the TSN network are
temporary in nature. The transmission schedule for TAS gate control
must be evaluated and maintained at each traversed TSN node
corresponding to each temporary flow. The schedule information at each
node is generated and managed as a network configuration.  These
network configurations must be applied across the network to establish
an end-to-end TSN flow.  The temporary TSN flows resulting from
plug-and-play connections can create a deluge of management traffic
overhead.

To address this overhead issue, Farzaneh et
al.~\cite{Farzaneh2016ontology} have presented an ontology based
automatic configuration mechanism. Application management service and
TSN management service entities coordinate the connection
establishment and tear-down procedures, managing the control plane
actions for the TSN network. A TSN knowledge database is implemented
to track and manage new, existing, and previous connections. For each
connection, QoS requirements, assignments, and source details, such as
port, related topics and devices are identified and analyzed to build
an ontology of TSN flows corresponding to an application and device.
Thus, whenever the plug-and-play event for a specific device occurs in
the network, the TSN configurations are automatically
retrieved and applied, lowering the overhead compared to the
conventional connection management scheme.
Although the automatic configuration mechanism is similar to the
principles of SDN, Farzaneh et al. have discussed the process based
automatic configuration mechanism independently of SDN. Nevertheless,
the ontology based automatic configuration mechanism can be easily
adapted to SDN by implementing the proposed application management
service and TSN management functions as an SDN application.

\subsubsection{Scheduling}
\paragraph{TTEthernet vs. TSN}
Craciunas et al.~\cite{craciunasoverview} have presented an overview
of scheduling mechanisms for Time-Triggered Ethernet
(TTEthernet)~\cite{cra2016com,kop2005tim,ste2009tte} and TSN.  In the
TTEthernet switch, the incoming frames for an outgoing egress port are
temporarily stored in a buffer, and wait for the scheduler to assign a
transmission-slot based on the precomputed schedule.  In contrast, the
incoming frames in TSN are directly inserted into priority queues, and
these priority queues are served based on prescribed schedules.  The
fundamental difference between TTEthernet and TSN is the scheduling
procedure, whereby the TTEthernet buffer is served based on global
static scheduling information, i.e., a \textit{tt-network-schedule}
assigned to meet the end-to-end delay requirements. In contrast, TSN
employs a dynamic schedule local to each node for control frame
transmissions from priority queues.  TSN switches may be synchronized
to network timing and can preempt an ongoing lower priority
transmission, which is not possible in a TTEthernet switch. Thus, the
deployment of TSN switches as opposed to TTEthernet switches can
improve support for delay critical applications. However, the
implementation cost and complexity (due to synchronization) of TSN is
typically higher than for TTEthernet.

\paragraph{Control Traffic Scheduling}
Bello et al.~\cite{Bello2014} have presented an overview of TSN
standards and examined the scheduling of control traffic flows in
intra-vehicular Ethernet networks.  More specifically, Bello et
al.~focused on the IEEE 802.1Qbv standard for scheduled traffic.
Bello et al. have implemented the scheduled traffic mechanism for
automotive connectivity applications by utilizing the time-sensitive
properties of TSN.  In particular, flow prioritization has been used
to prioritize the control traffic flows over regular data flows.  The
traffic flows are separated into multiple priority queues and
scheduling procedures are applied across the queues.  Bello et
al.~\cite{Bello2014} developed a simulation model for an automotive
network to study the behaviors of TSN supported network modules. The
simulation evaluations indicated significant latency reductions by up
to 50\% for the control traffic flows, i.e., the scheduled traffic
flows, compared to non-scheduled traffic.  A limitation of the Bello
et al.~\cite{Bello2014} study is that it considered only the
automotive network domain and did not consider the wider applicability
and potential of TSN.

\paragraph{Optimization Based Scheduling}
An important shortcoming of the IEEE 802.1Qbv standard, which defines
the transmission of scheduled traffic in TSN, is that there are no
specific definitions of algorithms to determine the transmission
schedule of frames on a link.  In addition, the IEEE 802.1Qbv standard
enforces a time spacing, i.e., guard bands, between the
scheduled traffic types. The guard bands isolate scheduled traffic
belonging to a specific class from other traffic classes, including
the best-effort traffic class.  A critical pitfall in the IEEE
802.1Qbv standard is that as the number traffic
classes increases, there can potentially be a large number of guard
band occurrences during the traffic transmissions over the link.
Traffic schedules with frequent guard bands waste bandwidth and can
contribute to latency increases.  Hence, an important future work
direction is to develop traffic transmission schedules with reduced
numbers of guard band occurrences in order to prevent wasted
bandwidth and to keep latencies low.

D\"urr et al.~\cite{durr2016no} have modeled TSN scheduling as a
no-wait job-shop scheduling problem~\cite{yamada1997job}.  D\"urr et
al.~then have adapted the
Tabu search
  algorithm~\cite{bat1994rea,glover1989tabu,mac1999mod} to
efficiently compute optimal TSN transmission schedules while reducing
the occurrences of guard bands. The simulations evaluations indicate
that the proposed algorithm can compute the near-optimal schedules for
more than 1500 flows on contemporary computing systems while reducing
the guard band occurrences by 24\% and reducing the overall end-to-end
latency for TSN flows.  With the minimal duration of guard bands, see
Section~\ref{sec:tsn:res:preemp}, the receivers have to be actively
synchronized for the correct reception of TSN frames.  The existence
of guard bands in the traffic flows provides an inherent secondary
synchronization for the receivers. However, it should be noted that
the implementation of such optimization algorithms can increase the
network node complexity as well as protocol operations, increasing the
overall operational cost of the device.
These scheduling
  principles have been further developed in~\cite{nay2018inc} towards
  the incremental addition of new flows.

Craciunas et al.~\cite{craciunas2016scheduling} have examined the
scheduling of real-time traffic, whereby the transmission schedules
are computed through optimization methods.  The constraints for the
optimization problem formulation are based on the generalized TSN
network configuration in terms of the characteristics of the Ethernet
frames, physical links, frame transmissions, end-to-end requirements,
and flow isolation.  While considering a comprehensive set of
parameters, the optimization problem is modeled to compute
transmission schedules in online fashion (i.e., is frame arrival event
driven) to achieve low latency and bounded jitter.  While a complex
optimization problem can provide a near optimal solution, it is also
important to consider the required computation times Addressing the
complexity aspect, Craciunas et al. have proposed several extensions
to the optimization process and outlined the implications for the
computation time.  Craciunas et al.~\cite{craciunas2016scheduling}
have conducted simulation evaluations for various network loads and
configurations.  The simulation results indicate that an optimization
process can be scalable while achieving the desired level of
scheduling benefits, i.e., bounded latency and jitter for an
end-to-end connection carrying real-time traffic.
Craciunas et al.~have further developed this optimal scheduling problem
in~\cite{craciunas2017formal,ste2018tra}
A related scheduling approach based on a graphical model
  has recently been examined by Farzaneh et al.~\cite{far2017gra},
  while a recent study by Kentis et al.~\cite{ken2017eff} has
  examined the impact of port congestion on the scheduling.

\paragraph{Joint Routing and Scheduling}
TSN frame transmissions out of the queues can be controlled through
gating (see Section~\ref{802.1Qbv:sec}), whereby a predefined event
triggers the gate to transmit a frame from a queue according to a
prescribed scheduling policy.  With event triggering, the frame
transmissions follow the predefined time triggered pattern, resulting
in so-called time triggered traffic~\cite{ein2018int,kop2005tim,
  rumpf2014software, meyer2013extending}.
Pop et al.~\cite{Pop2016} have designed a joint routing and scheduling
optimization that evaluates the time trigger events to minimize the
worst-case end-to-end frame delay.
The time trigger schedule is based on an
optimization problem formulated with integer linear programming.  The
proposed optimization problem comprehensively considers the network
topology as well as time trigger flows and AVB flows.
The time trigger flows follow the shortest route, while AVB flows
follow a greedy randomized adaptive search approach.  Simulation
evaluations indicate that the compute time to evaluate the time
triggered scheduling and AVB routing optimization is acceptable as
compared to the timing of the frame flows.
A limitation of the approach by Pop et al. is that the optimizations
are not scalable and flexible when there are changes in the properties
of network infrastructures, e.g., topology changes.  When there are
such network infrastructure changes, then the entire optimization
process must be reconfigured.
The recent related study by
  Smirnov et al.~\cite{smi2017opt} has focused on mixed criticality
  levels while the study by Mahfouzi et al.~\cite{mah2018sta} has
  focused on the stability aspects of joint routing and scheduling.

\paragraph{Impact of Traffic Scheduling}
Although TSN networks provide a pathway to achieve ULL through
enhancements to the existing Ethernet standards, the benefits are
limited to TSN flows as opposed to best-effort traffic. That is, in
case of mixed transmissions, where the TSN defined transmissions are
multiplexed with non-scheduled best effort traffic transmissions,
there are no guarantees for the effective behavior of the
non-scheduled best-effort traffic.  If there are requirements for the
non-scheduled traffic, such as a hard deadline for frame delivery in
an end-to-end connection, the application can be severely affected due
to the interference from the scheduled TSN traffic. The behavior
characterization of non-scheduled traffic can be challenging and
unpredictable due to the interference from scheduled TSN traffic.
Therefore, Smirnov et al.~\cite{Smirnov2017} have provided a timing
analysis to study the uncertainty of critical non-scheduled traffic in
presence of scheduled TSN traffic interference. The challenge in the
characterization of scheduled interference is to consider all possible
traffic scenarios, such as all possible scheduling types, resulting in
long computation times.  Smirnov et al. propose an approach to
integrate the analysis of worst-case scheduled interference with
traditional end-to-end timing analysis approaches to reduce the
computation times.  Such an integrated approach can estimate an upper
bound on the scheduled interference for various scheduling types, and
the evaluations show significant computation time reductions.

A complementary study by Park et al.~\cite{park2016simulation} has
investigated the performance of scheduled traffic as opposed to the
non-scheduled traffic. Park et al.~preformed extensive simulations
focusing on TSN to verify whether the end-to-end flow requirements are
impacted by increasing numbers of TSN nodes in the presence of
non-scheduled traffic. The simulations employed the general network
wide synchronous event-triggered method for frame transmissions in TSN
networks.  The simulations for an in-vehicle network based on the event
triggered scheduling for various traffic types show that the delay
requirements of control traffic can be successfully met for up to
three hops. However, the scheduled traffic needs to be transferred
within at most five hops to meet the typical 100~$\mu$s delay
requirement for critical control data in in-vehicle networks.

At a given TSN node, the events to trigger an action that is then
utilized for traffic scheduling can either be generated by a
processing unit within the TSN node or by an external control entity.
With the development and proliferation of SDN, future research can
develop various event generation techniques based on the centralized
SDN control and management.  The generated events can trigger various
TSN specified actions, such as frame transmissions, frame dropping, or
frame preemption, enabling new applications for SDN control and
management. To the best of our knowledge, event triggering methods
based on SDN have not yet been investigated in detail, presenting an
interesting direction for future research.  However, SDN based
management of TSN has already proposed and we discuss the
applicability of SDN for managing TSN flows in
Sec.~\ref{sec:tsn:res:mgt:sdn}.

While scheduled TSN transmissions provide low latency for prioritized
traffic, lower-priority traffic which is also TSN scheduled can be
significantly affected by higher priority traffic.  In order to
advance the understanding of the behaviors of traffic shapers on
low-priority TSN traffic, Maixum et al.~\cite{maxim2017delay} have
analyzed the delay of Ethernet frames that are scheduled according to
a hierarchical CBS or TAS in TSN switches.  The evaluations by
Maixum et al. indicate that the traffic scheduling for higher priority
TSN flows can potentially result in traffic burstiness for lower
priority TSN flows, increasing the overall delay for the lower
priority traffic.  This is because, long bursts of higher priority
traffic starve the scheduling opportunities for lower priority frames,
leading to the accumulation of low priority traffic.  In addition to
the static scheduling order, Maixum et al. have also studied the
effects of changing the scheduling orders in terms of end-to-end delay
for both higher and lower priority levels. The formal worst-case delay
analysis and simulation results indicate that low priority traffic is
severely affected by the scheduled higher priority traffic.
Simulations of an automotive use-case indicate a worst-case delay for
the prioritized traffic of 261~$\mu$s, while the worst-case delay for
low priority traffic is 358~$\mu$s.

\subsubsection{Preemption} \label{sec:tsn:res:preemp}
\paragraph{Preemption Mechanism}
Lee et al.~\cite{Lee2016time} have examined the preemption mechanism
(see Section~\ref{802.1Qbu:sec}) in conjunction with the TSN timing
and synchronization characteristics to estimate the transmission
properties of CDT and non-CDT frames.  In particular, Lee et al. have
proposed to insert a special preemption buffer into the transmission
selection module that operates across all the different queues at the
bottom in Fig.~\ref{fig_tsn_frameScheduling} to aid with the
preemption mechanism.  Lee et al.~have then analyzed the timing
dynamics of the preemption.  Lee et al.~note that in actual
deployments there are likely timing synchronization errors which
impact the frame boundary calculations.  Therefore, a minimum safety
margin that avoids collisions should be maintained while implementing
the preemption mechanism.  Lee et al.~\cite{Lee2016time} advocate for
a safety margin size of 20~bytes, accounting 5~bytes for an error
margin and 15~bytes for synchronization errors. The simulation
evaluations justify the impact of the synchronization errors on the
safety margin duration and end-to-end delay. Related preliminary
preemption analyses have been conducted in~\cite{jia2013per}.

\paragraph{Preemption Effect on Non-CDT}
Preemption prioritizes CDT frame transmissions over the transmission
of regular Ethernet frames.  Thus, preemption of non-CDT traffic can
negatively impact the end-to-end characteristics of non-CDT
traffic. In addition, low priority CDT frames can be preempted by
high priority CDT frames.  Hence, the preemption process can impact
the end-to-end delay differently for the different priority levels
even within the CDT traffic.  Thiele et al.~\cite{thiele2016preempt}
have formulated an analytical model to investigate the implications of
preemption on the end-to-end delay characteristics of CDT
and non-CDT traffic.  Thiele et al. have compared standard Ethernet
with preemption (IEEE~802.1Q + IEEE~802.3br) and TSN Ethernet with
time triggered scheduling and preemption (IEEE~802.1Qbv + IEEE
802.3br) with the baseline of standard Ethernet (IEEE~802.1Q) without
preemption.  The worst-case end-to-end latency of CDT with preemption
was on average 60\% lower than for 802.1Q without preemption.  Due to
the CDT prioritization, the worst-case latency of non-CDT traffic
increased up to 6\% as compared to the baseline (802.1Q) due to the
overhead resulting from the preemption process.  Hence, the impact of
preemption of non-CDT traffic is relatively minor as compared
to the performance improvements for CDT traffic.  Additionally, the
latency performance of standard Ethernet with preemption is comparable
to that of Ethernet TSN with preemption.  Therefore, Thiele et
al.~\cite{thiele2016preempt} suggest that standard Ethernet with
preemption could be an alternative to TSN for CDT traffic.  Standard
Ethernet would be much easier to deploy and manage than TSN, as TSN
requires the design and maintenance of the IEEE~802.Qbv gate
scheduling processes along with time synchronization across the
network.

\paragraph{Preemption Analysis and Hardware Implementation}
Zhou et al.~\cite{Zhou2017} conducted a performance analysis of frame
transmission preemption.  In particular, Zhou et al. adapted a
standard M/G/1 queueing model to estimate the long run average delay
of preemptable and non-preemptable frame traffic and evaluated the
frame traffic through simulations.  The numerical results from the
adapted M/G/1 queueing model and the simulations indicate that
preemption is very effective in reducing the frame delays for express
non-preemptable traffic relative to preemptable traffic; the average
frame delays of the express traffic are one to over three orders of
magnitude shorter than for preemptable traffic.  Zhou et al. have also
provided the VHDL design layout of the transmit unit and receive unit
for frame preemption for an FPGA based hardware implementation.

\subsubsection{Summary and Lessons Learnt}
Flow control mechanisms ensure that intermediate nodes support
the end-to-end behavior of a TSN flow.  Traffic shaping controls the
frame transmission over the egress port in a TSN switch. Each traffic
shaper strives to transmit a frame from a priority queue within the
shortest possible deadline while minimizing the impact on the
transmissions from other queues.  A finer resolution of priority
levels, i.e., a higher number of priority levels provides increasingly
fine control over frame transmissions from multiple queues. As a
limiting scenario, an independent queue can be implemented for each
individual flow in a TSN node.  However, such fine-grained
prioritization would require extensive computation and memory
resources in each TSN node.  To overcome this, virtual queues can be
implemented by marking the frames in a single queue, eliminating the
need for a number of queues equal to the number of TSN flows. Each
marked frame can be scheduled based on the marking value.  As low
priority flows can potentially face long delays due to resource
starvation from the scheduling of high priority flows, dynamic (i.e.,
changeable) priority values can be assigned to virtual queues.
Dynamic priorities can prevent prolonged delays for flows that
were initially assigned low priority.  The priority levels can be
dynamically changed based on the wait time or the total transit delay
of a frame compared to a predefined threshold.  Advanced dynamic
priority techniques, such as priority inversion, could be implemented
such that the worst-case delay of low priority traffic is kept
within prescribed limits.

\subsection{Flow Integrity}  \label{tsn:res:flow_int:sec}
\subsubsection{Fault Tolerance}
The AVB task group was mainly introduced to add real-time capabilities
to the best effort Ethernet service.  Industrial control networks
expect more reliable and stricter QoS services as compared to best
effort Ethernet network service.  Fault tolerance is a critical part
of industrial networks. The general principle for enabling fault
tolerance in a network is to introduce redundancy.

Following this general principle, TSN provides fault tolerance through
redundancy mechanisms, such as frame replication and elimination as
well as path control and reservations, see
Section~\ref{tsn:flow_int:sec}.  Kehrer et al.~\cite{Kehrer2014} have
conducted research on possible fault-tolerance techniques for TSN
networks. The main challenges associated with fault tolerance
mechanisms in TSN networks are the restoration processes for the
end-to-end link failures while preserving the network topology, i.e.,
without causing any significant break in continuous network
connectivity.  To address this, Kehrer et al. have compared two
approaches: $i)$ decoupled stream reservation and
redundancy~\cite{kle2012fau}, and $ii)$ harmonized stream reservation
and redundancy (which corresponds to IEEE~802.1CB).

In the decoupling approach, the stream reservation protocol registers
and reserves the streams independently of the redundancy requirements.
This decoupled approach allows for arbitrary redundancy protocols to
be utilized.  In contrast, the harmonized approach integrates
establishment of the reservation and the redundancy requirements.
More specifically, the IEEE 802.1Qca stream reservation protocol is
coupled with the IEEE~802.1CB frame duplication.

The main pitfall to avoid is to understand the application
requirements in terms of flexibility before choosing the redundancy
approach. Specific industrial automation networks may have peculiar
reliability requirements that may be more flexibly met with the
decoupled approach.  On the other hand, the decoupled approach has a
higher protocol overhead and requires more network bandwidth due to
the distributed and independent mechanisms along with the lack of
coordination between stream reservation and redundancy, as opposed to
the integrated approach.
A related fault tolerance approach
  based on redundant packet transmissions has been examined
  in~\cite{alv2017tow} while a mixing of temporal and spatial redundancy
  has been proposed in~\cite{alv2018mix}.

\subsubsection{Summary and Lessons Learnt}
Failure recovery and fault tolerance are key aspects of reliable
network design. However, to date there has been only very scant
research to address the critical challenges of resource reservation
for fault tolerance while considering ULL requirements. Future
research has to investigate the wide range of tradeoffs and
optimizations that arise with reliability through frame
replication. For instance, high priority flows could have reservations
of dedicated resources, while low priority flows could share a common
reserved resource. The dedicated resources would enable the
instantaneous recovery of the high priority TSN flows; albeit, at the
expense of a slight reduction of the overall network efficiency due to
the redundancy.  In the event of failure for a low priority traffic
flow, the connection could be reestablished with a new flow path
considering that the flows can tolerate delays on the order of the
connection reestablishment time.  Centralized SDN management can also
provide the flexibility of dynamic path computation and resource
reallocation in the event of failures.  Therefore, the area of flow
integrity requires immediate research attention to design and evaluate
the performance of efficient recovery processes based on priority
levels.

\subsection{General TSN Research Studies}
TSN is being widely adopted in critical small-scale closed automotive
and industrial networks to establish reliable ULL end-to-end
connections.  However, a key TSN limitations is exactly this focus on
closed networks, e.g., in-vehicle networks and small-scale robotic
networks.  The network applications running in robots and in
in-vehicle networks often involve significant interactions with
external non-TSN networks.  Robotic and vehicular network applications
require a tight integration with mobility handling procedures by the
external network. If advanced network features, such as mobility, are
not properly supported in the external network, then the TSN benefits
are fundamentally limited to small-scale closed networks. Therefore,
smooth interoperability between TSN and different external networks is
essential for TSN operation in heterogeneous network scenarios.
Ideally, the connectivity between TSN and non-TSN networks should be
able to accommodate similar characteristics as TSN to ensure the
overall end-to-end connection requirements in heterogeneous
deployments.

\subsubsection{V2X Communication}
Juho et al.~\cite{Juho2017} have proposed iTSN, a new methodology for
interconnecting multiple TSN networks for large-scale applications.
The iTSN methodology utilizes wireless protocols, such as IEEE
802.11p, for the inter-networking between different TSN networks.  In
particular, the sharing of global timing and synchronization
information across the interconnected network is important for
establishing a common timing platform to support TSN characteristics
in the external networks. The iTSN network uses the IEEE 802.11p WAVE
short message protocol to share the timing information between
different TSN networks.  Critical rapid alert messages can be
prioritized not only within a given TSN network, but also across
multiple interconnecting networks.  Thus, the iTSN methodology
enables, for instance, vehicular networks to transmit safety critical
messages to control nodes, e.g., Road Side Units
(RSUs)~\cite{mac2018v2x}, with delays on the order of microseconds in
a heterogeneous deployment.  Through the adoption of such reliable
inter-connectivity techniques, the vehicle braking safety distance can
be achieved in much shorter (microseconds) time spans than the
currently feasible range of milliseconds. Overall, TSN and an
interconnecting technique, such as iTSN, can create a communication
platform for safe autonomous driving systems.

\subsubsection{Network Modeling}
Although TSN standards have received significant attention
in networks for automotive driving, a major
challenge in network deployment is managing the complexity. As
automotive driving technology progresses, more requirements are
imposed on the existing in-vehicle network infrastructure.  As the
number of sensors increase in an in-vehicle network, the increasing
connectivities and bandwidth requirements of the sensors should be
correspondingly accommodated in the network planning. However, the
dynamic changes in the network requirements for an in-vehicle control
system could require a more extensive network infrastructure,
resulting in higher expenditures.  Considering the complexities of
automotive networks, Farzaneh et al.~\cite{Farzaneh2016} have proposed
a framework to analyze the impact of adding new sensors to an existing
infrastructure that supports critical applications. In particular, the
network configuration that fulfills all the requirements, including
newly added sensors, must be dynamically evaluated and implemented.
Towards this end, the Farzaneh et al.~\cite{Farzaneh2016} framework
involves a design and verification tool based on a Logic Programming
(LP) method to support the reconfiguration and design verification
processes for an in-vehicle TSN network.  The proposed framework
consists of comprehensive logical facts and rules from which a user
can query the database with the requirements to obtain configurations
that satisfy the requirements. A key characteristic of the
proposed approach is that the network modeling process considers the
most accurate logical facts and rules of the TSN applications and
requirements to obtain an efficient configuration and verification
process.

\subsubsection{TSN Simulation Framework}
Heise et al.~\cite{Heise2016} have presented the TSimNet simulation
framework to facilitate the development and verification of TSN
networks.  TSimNet was primarily implemented to verify industrial
use-cases in TSN networks. The simulation framework is based on
OMNeT++, whereby the non time-based features, such as policy
enforcement and preemption
are implemented in a modular fashion to increase the flexibility of
designing new network mechanism suitable for industrial networks. For
instance, the initial evaluation of the simulation framework for frame
preemption mechanisms indicates that the end-to-end latency can be
increased if the network is not configured in an optimized way for
critical functions, such scheduling and traffic shaping. Heise et
al.~have evaluated the computational cost of the TSimNet framework for
various network function simulations, such as policing, recovery, and
preemption in terms of CPU and memory requirements.  The simulation
framework also features Application Programming Interfaces (APIs) for TSN
mechanisms that do not require time synchronization, such as stream
forwarding, per-stream filtering, as well as frame replication and
recovery.  APIs can be invoked by the simulation framework through a
profile notification. The basic framework modules also include the
TSimNet Switch Model, which can identify streams based on MAC, VLAN,
and/or IP addresses, while the TSimNet Host Model implements complex
functions, such as ingress and egress policy, as well as traffic
shaping.
Related simulation evaluations with OMNeT++ have been
  reported in~\cite{nsaibi2017formal}, while a TSN simulation model
  based the OPNET simulation framework has been presented
  in~\cite{pah2018eva}.

\subsubsection{Hardware and Software Design}
Hardware and software component designs to support TSN
functions, such as scheduling, preemption, and time-triggered event
generation in TSN nodes require significant engineering and
development efforts.  Hardware implementations are highly efficient in
terms of computational resource utilization and execution latency but
result in rigid architectures that are difficult to adapt to new
application requirements.  On the other hand, software implementations
can flexibly adapt to new application requirements, but can overload
CPUs due to the softwarization of network functions, such as
time-triggered scheduling and hardware virtualization.

Gross et al.~\cite{gross2014hardware} have presented a TSN node
architecture design where the time-sensitive and computationally
intensive network functions are implemented in dedicated hardware
modules to reduce the CPU load.  The proposed hardware/software
co-design approach flexibly allocates network function to be executed
completely in hardware, completely in software, or in both hardware
and software based on the dynamic load.  The flexible allocation is
limited to network functions that independently scale with the timing
requirements, such as the synchronization protocol. More specifically,
Gross et al. have considered time-triggered transmissions, frame
reception and timestamping, and clock synchronization. The hardware
modules can produce the time-triggered events nearly jitter free,
implement frame reception and time-stamping in real-time, and
synchronize clocks with a high degree of precision. Thus, the hardware
modules improve the overall TSN node performance compared to a
software-only implementation. The performance evaluations from a
prototype implementation based on a Virtex-6 FPGA showed a significant
reduction in the CPU load compared to a software-only implementation.
Additionally, the precision of the time-triggered event generation in
the hardware implementation was improved by a factor of ten compared
to software triggered events.

\subsubsection{Summary and Lessons Learnt}
The general aspects of TSN that determine the overall success of TSN
designs and implementations are the inter-interoperability with
heterogeneous network architectures, such as LANs, WANs, and core
networks.  Most of the research on TSN to date has focused on
in-vehicle networks which are independent and isolated from
external networks. Another limitation of the TSN research field is the
lack of a simulation framework that encompasses large-scale
heterogeneous network architectures. Valid use cases that include both
localized and external network interactions, such as automotive
driving, should be created and considered in benchmark evaluations.
Currently, the general use-case in most TSN research studies is an
in-vehicle network supporting on-board sensor connectivity and
audio/video transmission for infotainment.  Future custom TSN
simulation frameworks should be based on networks that support
next-generation applications with localized and external network
interactions, such as automotive driving. Similarly, the SDN based TSN
management could exploit hierarchical controller designs to extend
the management from localized networks, such as in-vehicle networks,
to external networks, such as vehicle-to-any (V2X) networks.

\subsection{Discussion on TSN Research Studies}
The TSN network infrastructure and protocols have to support bounded
end-to-end delay and reliability, to support basic features related to
critical applications of IoT, medical, automotive driving, and smart
homes. TSN based solutions for addressing the requirements of these
applications result in complex network infrastructures supporting
various protocols. Hence, simplified TSN network management mechanisms
are essential to reduce the complexity while achieving the critical
needs of the ULL applications.

The deterministic TSN network behavior has so far been generally
applied to a closed network, i.e., a network spanning only the scope
of a particular application, for instance, in-vehicle networks.
However, the connectivity to external networks, such as cellular and
WLAN networks, enhances the capabilities of TSN networks. For
instance, in automotive driving, the application requirements can be
controlled by weather data from the cloud or by sharing information
with a neighboring TSN in-vehicle network.  Therefore, reliable,
secure, and low-latency communication between multiple TSN networks is
essential to support a wide range of future applications.  The lack of
TSN standards for connecting and communicating with external TSN and
non-TSN networks is impeding the research activities in
inter-operating networks and needs to be urgently addressed.  In
summary, we identify the following main future design requirements for
TSN research:
\begin{itemize}
\item[i)] Support for a wide range of applications spanning from
  time-sensitive to delay tolerant applications with flow level
  scheduling capabilities.
\item[ii)] Connectivity between multiple closed TSN architectures.
\item[iii)] Flexible and dynamic priority allocations to ensure bounded
	end-to-end latency for lower priority traffic.
\item[iv)] Adoption of SDN for the centralized management of TSN functions
	with a global network perspective.
\item[v)] Efficient timing information sharing and accurate clock design
	through self-estimation and correction of local clock skewness.
\item[vi)] Computationally efficient hardware and software designs.
\end{itemize}

Generally, TDM can enforce a deterministic (100\%) latency bound, but
the TDM average delay is typically somewhat higher than the
statistical multiplexing average delay (and TDM has low utilization
for bursty traffic).  With proper admission control, statistical
multiplexing can provide statistical guarantees for latency bounds~\cite{che2018pro},
e.g., the probability for exceeding the delay bound can be very low,
e.g., less than $10^{-4}$ probability that the delay bound is
violated. These rare occurrences of violating the delay bound ``buy''
usually much higher utilization (throughput) than TDM and lower
average delay (for bursty data
traffic)~\cite{kni1999adm,lie1996exa,rei1999gua,rei2002fra,zha1997cal}.
An interesting future research direction is to examine the tradeoffs
between deterministic and probabilistic delay bound assurances in
detail for ULL traffic served with TSN mechanisms.

\section{DetNet Standardization} \label{detnet:std:sec}
In this section, we present a detailed overview of the current
standardization of the IETF Deterministic Networking (DetNet) WG.  The
IETF DetNet WG collaborates with the IEEE 802.1 TSN TG to define a
common architecture for layers 2 and 3, whereby the TSN TG focuses on
layer 2 bridged networks and the DetNet WG focuses on layer 3 routed
segments. Similar to the TSN goals, DetNet aims to support
deterministic worst-case bounds on latency, packet delay variation
(jitter), and extremely low/zero packet loss.  Moreover, both TSN and
DetNet strive for high reliability and redundancy over disjoint paths
targeted towards IACS real-time applications.

The charter of the DetNet WG is to specify an overall architecture
that standardizes the data plane and the Operations, Administration,
and Maintenance (OAM) for layer 3 ULL support.  This charter includes
the time synchronization, management, control, and security operations
that enable multi-hop routing.  Moreover, the DetNet charter includes
the various forms of dynamic network configuration (automated and
distributed as well as centralized and distributed) and the multi-path
forwarding. In general, DetNet focuses on extending the TSN data and
control plane into the layer 3 domain, thus expanding the scope of TSN
beyond LANs.

Since the DetNet WG has only been established recently (started in
October 2014, and became a WG in October 2015), no IETF RFCs exist
yet. However, at the time of writing this article, several IETF
drafts have become available and will be covered in the following
subsections to provide a comprehensive overview of the ongoing IETF
DetNet standardizations.

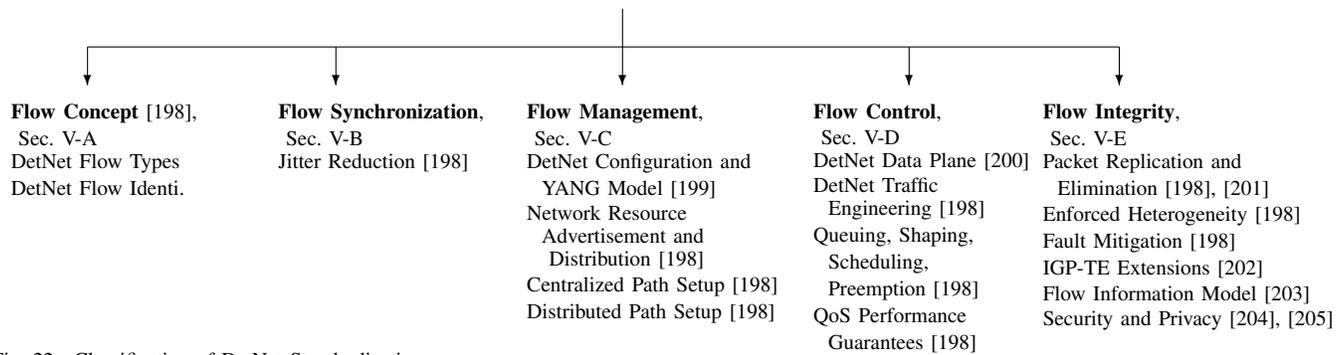
\begin{figure*}[t!]
	\footnotesize
	\setlength{\unitlength}{0.10in}
	\centering
	\begin{picture}(33,33)
	\put(8,33){\textbf{DetNet Standardizations, Sec.~\ref{detnet:std:sec}}}
	\put(-14,30){\vector(0,-1){2}}
	\put(-14,30){\line(1,0){54}}
	\put(-18,27){\makebox(0,0)[lt]{\shortstack[l]{			
		\textbf{Flow Concept}~\cite{finn2017deterministic}, \\
				\ Sec.~\ref{detnet:flow_def:sec} \\
				DetNet Flow Types \\
				DetNet Flow Identi.
	}}}
	\put(-1,30){\vector(0,-1){2}}
	\put(-4,27){\makebox(0,0)[lt]{\shortstack[l]{			
				\textbf{Flow Synchronization},\\
				\ Sec.~\ref{detnet:flow_sync:sec} \\
				Jitter Reduction~\cite{finn2017deterministic}
	}}}
	\put(14,30){\line(0,1){2}}
    \put(14,30){\vector(0,-1){2}}				
    \put(9,27){\makebox(0,0)[lt]{\shortstack[l]{			
			\textbf{Flow Management},\\
			\ Sec.~\ref{detnet:flow_mgt:sec} \\
			DetNet Configuration and \\
			\ \ YANG Model~\cite{yang2017deterministic} \\
			Network Resource \\
			\ \ Advertisement and \\
			\ \ \ Distribution~\cite{finn2017deterministic} \\
			Centralized Path Setup~\cite{finn2017deterministic} \\
			Distributed Path Setup~\cite{finn2017deterministic}
    }}}
	\put(29,30){\vector(0,-1){2}}
	\put(24,27){\makebox(0,0)[lt]{\shortstack[l]{
				\textbf{Flow Control},\\
				\ Sec.~\ref{detnet:flow_ctrl:sec} \\
				DetNet Data Plane~\cite{dataPlane2017deterministic} \\
				DetNet Traffic \\
				\ \ Engineering~\cite{finn2017deterministic} \\
				Queuing, Shaping, \\
				\ \ Scheduling, \\
				\ \ Preemption~\cite{finn2017deterministic} \\
				QoS Performance \\
				\ \ Guarantees~\cite{finn2017deterministic}
	}}}
	\put(40,30){\vector(0,-1){2}}
	\put(36,27){\makebox(0,0)[lt]{\shortstack[l]{	
				\textbf{Flow Integrity},\\
				\ Sec.~\ref{detnet:flow_int:sec} \\
				Packet Replication and \\
				\ \ Elimination~\cite{finn2017deterministic, redundancy2017deterministic} \\
				Enforced Heterogeneity~\cite{finn2017deterministic} \\
				Fault Mitigation~\cite{finn2017deterministic} \\
				IGP-TE Extensions~\cite{information2017deterministic} \\
				Flow Information Model~\cite{informationModel2017deterministic} \\
				Security and Privacy~\cite{security2017deterministic, usecases2017deterministic}
	}}}
	\end{picture}
	\vspace{-4cm}	
	\caption{Classification of DetNet Standardization. }
	\label{detnet_fig}
\end{figure*}

\subsection{Flow Concept}   \label{detnet:flow_def:sec}
Similar to the TSN TG, DetNet flows are specified by their QoS
classes. DetNet defines each flow's QoS by 1) the maximum and minimum
end-to-end latency, and 2) the packet loss probability
requirements~\cite{finn2017deterministic}. DetNet strives to transport
unicast and multicast ULL data flows for real-time applications with
extremely low packet loss. In essence, DetNet emulates point-to-point
links over a packet switched network, where each link can be shared
between multiple DetNet and non-DetNet flows, each with varying flow
requirements and properties. A key aspect of DetNet flow control and
management is ensuring that non-DetNet flows have no influence on
DetNet flows. Maintaining each DetNet flow's QoS is achieved through
the mechanisms surveyed in this section.

\subsubsection{DetNet Flow Types}
Before introducing the DetNet flow types, we first give a brief
overview of two main layers of the DetNet architecture stack model.
The DetNet Transport Layer has an option to provide congestion
protection (see Section~\ref{detnet:flow_ctrl:sec}).  The DetNet
Service Layer provides service protection, e.g., through flow
duplication (see Section~\ref{detnet:flow_int:sec}).  Four main DetNet
flow types have been identified~\cite{finn2017deterministic}:
\begin{enumerate}
	\item \textbf{App-flow:} The native data flow between the source and
	  destination end systems within a DetNet enabled network.
	\item \textbf{DetNet-t-flow:} The specific data flow format bound to
	the transport layer within a DetNet enabled network. The
	DetNet-t-flow contains the specific data attributes that provide
	features for congestion protection.
	\item \textbf{DetNet-s-flow:} The specific data flow format bound to
	the service layer within a DetNet enabled network. The DetNet-s-flow
	contains the specific data attributes that provide features for
	replication and elimination functions supporting service protection.
	\item \textbf{DetNet-st-flow:} The specific data flow format that is
	bound to both the transport and service layers within a DetNet
	enabled network. The DetNet-st-flow signals the appropriate
	forwarding function utilizing both the service and transport layer
	attributes.
\end{enumerate}

\subsubsection{DetNet Flow Identification}
In contrast to a conventional strictly layered network architecture,
DetNet nodes intentionally violate ``layering norms'' so that lower
layers can detect and become aware of higher layer flow types.  This
awareness of the higher layer flow types helps to provide specific
queuing, shaping, and forwarding services as flows are transported
across multiple technology domains. However, violating the layering
norms creates new layering and re-layering complexities. Therefore,
DetNet must provide a way to easily and correctly identify flows and
their associated types.  DetNet is architected to allow nodes within
the network data plane to distinguish DetNet flows based on the flow
ID and DetNet Control Word (CW), i.e., sequencing information,
appended in the packet header, whereby the CW is used for replication
and elimination purposes.

To achieve accurate flow detection and identification, the flow
attribute mapping between layers and across technology domains has to
be standardized.  For each forwarding of a DetNet flow between
different technology domains, the relay node (i.e., router) needs to
acquire upper layer information related to the flow type and
corresponding attributes. For example, when a DetNet flow is forwarded
between two Label Switching Routers (LSRs) that interconnect different
Layer 2 bridged domains, then at each domain boundary, the higher
layer flow information is passed down to the node for correct
forwarding. Three main forwarding methods are considered in DetNet: 1)
IP routing, 2) MPLS label switching, and 3) Ethernet bridging. For
forwarding across technology domains, each DetNet App-flow packet is
appended or encapsulated with multiple flow-IDs (IP, MPLS, or
Ethernet). This enables DetNet routing and forwarding between
different and disparate IP and non-IP networks, essentially providing
network interoperability.

\subsection{Flow Synchronization}   \label{detnet:flow_sync:sec}
The main objective of DetNet is to expand the TSN capabilities to
layer 3 routing segments. DetNet relies heavily on the services of the
IEEE TSN TG mechanisms. Flow synchronization with respect to the
DetNet flow architectural model has not been specifically addressed
in~\cite{finn2017deterministic}. Therefore, it is likely that DetNet
will ensure timing synchronization between DetNet capable network
entities (bridges, routers, and end systems) through various existing
synchronization techniques and profiles, e.g., IEEE 802.1AS and IEEE
1588v2.

Applications in the mission critical latency traffic class require
extremely low delay variations (jitter). High jitter can lead to
packet loss downstream and in the worst-case, loss of human life in
factory networks.  DetNet strives to support minimal jitter by
bounding the minimum and maximum latency~\cite{finn2017deterministic},
which is challenging in large scale packet switched networks.  DetNet
specifies jitter reduction through two main principles: 1)
sub-microsecond time synchronization between network entities, and 2)
time-of-execution fields embedded within the application
packets~\cite{finn2017deterministic}.  While no specific
specifications regarding time synchronization for DetNet network
devices exist, the DetNet WG have overall hinted at using other
Standardization Development Organization's (SDO), e.g., IEEE TSN's
802.1AS methods, see Section~\ref{tsn:flow_sync:sec}.

\subsection{Flow Management}  \label{detnet:flow_mgt:sec}
Flow management describes and specifies the mechanisms
for discovering and configuring node capabilities.

\subsubsection{DetNet Configuration and YANG Model}
In order for DetNet to enable seamless configuration and
reconfiguration across various DetNet enabled network entities, a
uniform and scalable configuration model needs to be defined. The
Internet draft~\cite{yang2017deterministic} defines distributed,
centralized, and hybrid configuration models, related attributes, and
the YANG model for DetNet.

\paragraph{DetNet Configuration Model}
Three configuration models have been
introduced~\cite{yang2017deterministic}: fully distributed, fully
centralized, and hybrid. For a fully distributed configuration model,
UNI information is sent over a DetNet UNI protocol, i.e., sent using
the flow information model discussed in
Section~\ref{detnet:flow_int:sec}. A distributed DetNet control plane
propagates the UNI and configuration information to each data plane
entity.  In the centralized configuration model, the CUC sends the UNI
information to the CNC, similar to the IEEE 802.1Qcc centralized
configuration model, see Section~\ref{tsn:flow_mgt:sec}.  For the
hybrid configuration approach, a combination of distributed and
centralized protocols within the control planes are used to coordinate
configuration information. The fully distributed and hybrid
configuration models are not covered in~\cite{yang2017deterministic}
and are left for future work.

\paragraph{DetNet Configuration Attributes} Depending on the
configuration model and control plane associated protocols (i.e., IGP
and RSVP-TE, or CNC and CUC), different configuration parameters or
attributes are used.
The following main attributes have been defined for the
centralized configuration model~\cite{yang2017deterministic}:

\begin{enumerate}
\item \textbf{DetNet Topology Attributes} specify topology
  related attributes, such as the node type, whether it is
  Packet Replication and Elimination Function (PREF) capable
  or not, and the queueing management algorithm.
\item \textbf{DetNet Path Configuration Attributes} specify the
  networked path related attributes, such as the constraints (required min/max
  latency), and explicit routes using a PCE (with
  PREF).
\item \textbf{DetNet Flow Configuration Attributes} specify the DetNet
  flow attributes, such as the flow ID, priority, traffic
  specification, and encapsulation method.
\item \textbf{DetNet Status Attributes} specify the flow status
  feedback attributes, such as the flow performance (delay, loss,
  policing/filtering), and the PREF status.
\end{enumerate}

\textbf{DetNet YANG Model:} Similar to IEEE 802.1Qcp
(see Section~\ref{IEEE802.1Qcp:sec}), a DetNet YANG
model has been defined~\cite{yang2017deterministic} for the
centralized configuration model to convey network configuration
parameters.

\subsubsection{Network Resource Advertisement and Distribution}
To supplement the DetNet Congestion Protection mechanisms (which are
defined for DetNet as flow control mechanisms, including shaping,
scheduling, and preemption), and to accurately provision network
resources for DetNet flows, i.e., admission control, each node (or
central controller in a centralized setup) needs to share and alert nearby
networks of its (end system and/or transit node)
capabilities~\cite{finn2017deterministic} including:
\begin{enumerate}
	\item System capabilities, e.g., shaping and queuing algorithm used,
	buffer information, and worst-case forwarding delay
	\item Dynamic state of the node's DetNet resources
	\item Neighbor nodes and the properties of their relationships, i.e.,
	the properties of the links connecting them, e.g., length and bandwidth.
\end{enumerate}

How this information is carried over the control plane and the
implementation specification is not available nor standardized
yet. However, with this information, PCE's automatic path installation
(distributed or centralized) can handle each DetNet flow's QoS
requirement assuming that enough resources are available,
which
  is enforced by admission control mechanisms similar to the TSN SRP
  (MRP) protocols (see Section~\ref{802.1Qat:sec}).

\subsubsection{Centralized Path Setup}
Similar to IEEE TSN's centralized management model (802.1Qcc, see
Section~\ref{802.1Qat:sec}), DetNet's centralized path setup leverages
PCEs and packet based IP or non-IP network information dissemination
to enable global and per-flow optimization across the DetNet enabled
network.  The DetNet WG~\cite{finn2017deterministic} has addressed
several related key issues, such as the installation of the paths
corresponding to the received path computation (whether by the Network
Management Entity (NME) or end systems), and how a path is set up,
i.e., through direct interactions between the forwarding devices and
the PCEs, or by installing the path on one end of the path through
source-routing or explicit-routing information~\cite{finn2017deterministic}.

\subsubsection{Distributed Path Setup}
The DetNet WG has developed initial design specifications for
a distributed path setup (similar to the 802.1Qat, 802.1Qca, and
MRP signaling protocols) utilizing Interior-Gateway Protocol Traffic
Engineering (IGP-TE) signaling protocols, defined in
Section~\ref{detnet:flow_ctrl:sec}, e.g., MPLS-TE, RSVP-TE, OSPF-TE,
and ISIS-TE~\cite{finn2017deterministic}. A key issue
is how the interactions and integration between layer 2 sub-network
peer protocols for TE and path installation will be defined, since
significant work has been accomplished by the IEEE 802.1 TSN TG
regarding distributed and centralized protocols on path and multi-path
setup and signaling protocols.

\subsubsection{Summary and Lessons Learned}
Before controlling a DetNet flow, the node's capabilities need to be
distributed to the PCE in the control plane. To efficiently
disseminate the node capability information, a configuration and YANG
model need to be standardized to allow for dynamic
reconfiguration, management, and status collection in large scale
IP/non-IP based networks.

Additionally, as networks under the control of DetNet related services
and mechanisms may become saturated with flows, effective admission
control mechanism, e.g., similar to the admission control mechanisms
researched within the IETF IntServ
framework~\cite{kni1999adm,lie1996exa,rei1999gua,rei2002fra,zha1997cal},
must be researched to operate within the DetNet framework. Based on
the admission control, network resources must be managed such that ULL
applications/traffic that is marked with higher priorities than other
traffic can be allocated the appropriate resources.

\subsection{Flow Control}  \label{detnet:flow_ctrl:sec}
While most control functions for DetNet flows follow the same
principles used for IEEE TSN TG deterministic flows, key integration
mechanisms and several differences are outlined as follows.

\subsubsection{DetNet Data Plane}
To better understand how DetNet services operate, we first provide a
brief overview of the DetNet data plane. A DetNet capable network is
composed of interconnected end systems, edge nodes, and relay
nodes~\cite{finn2017deterministic}. Transit nodes (e.g., routers or
bridges) are used to interconnect DetNet-aware nodes, but are not
DetNet-aware themselves. Transit nodes view linked DetNet
nodes as end points.
DetNet is divided into two main layers: 1) the DetNet service
  layer, and 2) the DetNet transport layer.  The DetNet service layer is the
  layer responsible for specific DetNet services, such as congestion
  and service protection, while the DetNet transport layer is
  responsible for optionally providing congestion protection for
  DetNet flows over paths provided by the underlying
  network~\cite{finn2017deterministic}. More specifically, the service
  layer can apply specific services, such as packet sequencing, flow
  replication/duplicate elimination, and packet encoding, while the
  transport layer can apply congestion protection mechanisms (through
  the underlaying subnetworks, e.g., MPLS TE, IEEE 802.1~TSN, and OTN)
  and explicit routes.
DetNets can have several hierarchical DetNet topologies where each
lower layer services the higher layers.  Furthermore, DetNet nodes
(end systems and intermediary nodes) are inter-connected to form
sub-networks. These sub-networks, e.g., Layer 2 networks, can support
DetNet traffic through compatible services, e.g., IEEE 802.1 TSN or
point-to-point Optical Transport Network (OTN) service in 5G
systems~\cite{finn2017deterministic}.

There are currently various protocol and technology options under
consideration for DetNet service and transport layer protocols.
Table~\ref{tab_tsn_DetNEtprotocolTable} provides an overview of these
protocol candidates for the DetNet service and transport layers,
including a brief description of each protocol and the latency impact
on a DetNet flow. Although no official solution has emerged yet for
the DetNet data plane encapsulation at the network layer, a couple of
proposals exist to tackle this problem. According to Korhonen et
al.~\cite{dataPlane2017deterministic}, two of the most prominent
deployment candidates for the data plane protocols are either a
UDP/TCP service layer over a native-IP (IPv6) transport layer or a
PseudoWire-based (PW)~\cite{bryant2005pseudo} service layer over an
MPLS Packet Switched Network (PSN) transport layer.
\begin{table*}[t]  \centering
\caption{Candidate Protocols for DetNet Service and Transport Layers.
 A prominent deployment candidate is a UDP service layer over an IP transport layer.}
	\label{tab_tsn_DetNEtprotocolTable}
	\footnotesize
	\def\arraystretch{1.5}
	\begin{tabular}{c|p{0.2\linewidth}|p{0.53\linewidth}|p{0.1\linewidth}}
		\multicolumn{1}{c}{\textbf{Layer}}
		&\multicolumn{1}{c}{\textbf{Candidate Protocol}}
		&\multicolumn{1}{c}{\textbf{Description}}
		&\multicolumn{1}{c}{\textbf{Latency Imp.}} \\ \hline
		\multirow{6}{*}{\textbf{Service}}
		& PseudoWire (PW)
		& Emulates networking services across packet switched networks (PSNs), delivers bare minimum
		network service functionality on physical infrastructure with
		some degree of
		fidelity.
		& \multicolumn{1}{c}{Moderate}              \\
		& User Datagram Prot.
		(UDP)
		& Connection-less transmission of packets with low
		overhead, though no feedback services provided.
		& \multicolumn{1}{c}{Low}              \\
		& Generic Rout. Encap.
		(GRE)
		& Tunneling protocol that encapsulates arbitrary network layer
		protocol over another network layer protocol, e.g., IPv6 over
		IPv4.
		& \multicolumn{1}{c}{Moderate}               \\
		\hline
		\multirow{10}{*}{\textbf{Transport}}
		& Internet Prot. Ver. 4 (IPv4)
		& Connection-less protocol for use in PSNs supporting
		best-effort services.
		& \multicolumn{1}{c}{Moderate}               \\
		& Internet Prot. Ver. 6 (IPv6)
		& Similar to IPv4 but with a larger address space, includes a few
		improvements and simplifications.
		& \multicolumn{1}{c}{Moderate}              \\
		& Multi-Prot. Label Swit. Label Swit. Path (MPLS LSP)
		& Routing prot. that forwards labeled
		packets that define the source-destination paths without routing
		table look-ups. Instead, at each hop, the
		label is used as an index and a new label is generated and sent
		along the packet to the next hop.
		& \multicolumn{1}{c}{Moderate}               \\
		& Bit Ind. Explicit Rep. (BIER)
		& An alternative multicast forwarding technique that
		does not use per-flow forwarding entries.
		Instead, a BIER header is used to identify the packet's egress
		nodes in the BIER domain. A bit
		string that is set at each ingress node is used, and the flow is
		replicated at each egress node represented by the bit string.
		& \multicolumn{1}{c}{High}               \\
		& BIER-Traffic Engin. (BIER-TE)
		& Operates similarly to BIER but does not require an Interior Gateway Protocol.
		TE by explicit hop-by-hop forwarding and
		loose hop forwarding~\cite{ietf-bier-te-arch-00} of packets is supported.
		& \multicolumn{1}{c}{High}               \\ \hline
	\end{tabular}
\end{table*}
While many options exist for DetNet data encapsulation, it is
imperative to test and discern the corresponding performance overhead
for each proposed DetNet node's packet manipulation technique.

\subsubsection{DetNet Traffic Engineering}
The IETF Traffic Engineering Architecture and Signaling (TEAS) WG
considers Traffic Engineering (IE) architectures for packet and
non-packet networks~\cite{lee2015requirements}, essentially allowing
network operators to control traffic traversing their networks. Since
DetNet operates with explicit paths, the DetNet WG has drafted a TE
architectural design for DetNet utilizing similar methodology as the
Software Defined Networking (SDN) paradigm. The DetNet WG defines
three main planes~\cite{finn2017deterministic}: 1) the (user)
application plane, 2) the control plane, and 3) the network plane.
The network plane conforms with the specification of the Internet
Research Task Force (IRTF) RFC 7426~\cite{haleplidis2015software} that
details the structure and architecture of the SDN networking
paradigm. This DetNet SDN approach shares similarities with the IEEE
TSN's 802.1Qcc management scheme (see Section~\ref{802.1Qat:sec}) and
centralized SDN approach.

\paragraph{Application Plane} The collection of applications and
services that define the network behavior constitute the application
plane. For example, network services, such as network topology
discovery, network provisioning, and path reservation, are all part of
network applications that can be utilized through the application
plane and can be accessed by a user-application interface or by other
services through the service interface~\cite{finn2017deterministic}.
Moreover, the DetNet WG has defined a user agent application for
passing DetNet service requests from the application plane via an
abstraction Flow Management Entity (FME) to the network plane. The
management interface handles the negotiation of flows between end
systems, where requested flows are represented by their corresponding
traffic specification (Tspec), i.e., the flow characteristics. The
applications in the application plane communicate via the service
interface with the entities in the control plane

\paragraph{Control Plane} The collection of functions responsible for
controlling (e.g., flow installation and processing in the forwarding
plane) and managing (e.g., monitoring, configuring, and maintaining)
network devices constitute the control plane. The DetNet TE
architecture utilizes the Common Control and Measurement Plane (CCAMP)
standardized by the IETF CCAMP WG, where the aggregate control plane,
i.e., the control and management planes, is distinctly split between
management and measurement entities within the control
plane. Additionally, the control plane leverages PCEs and NMEs. PCEs
are considered the core of the control plane. Given the relevant
information through the network interface, the PCEs compute the
appropriate deterministic path that is installed in the network plane
devices.

\paragraph{Network Plane} The aggregate network plane constitutes the
operational (control), forwarding (data), and parts of the
applications plane aspects under the RFC 7426 standard.
The network plane interconnects all the Network Interface Cards (NICs)
in the end systems and intermediate nodes (i.e., IP hosts and
routers/switches). Additionally, UNIs and
Network-to-Network (NNI) interfaces are used for TE path reservation
purposes. A network interface is used to enable communication between
the network plane and the control plane, whereby the control
plane can describe and install the physical topology and resources in
the network plane.

In general, this DetNet TE architecture envisions a highly
scalable, programmable, and uPnP scheme, where network functionality
and configurations are easily implemented and extended.

\subsubsection{Queuing, Shaping, Scheduling, and Preemption}
While identifying the appropriate data and control plane solutions is
imperative for correct operations in DetNet environments, flow control
principles (e.g., queuing, shaping, scheduling, and preemption) must
be defined to enable DetNet flows to achieve deterministic bounded
latency and packet loss~\cite{finn2017deterministic}.  Flow control
usually involves admission control and network resource reservation,
i.e., bandwidth and buffer space allocation. However, a key aspect of
reservation is to standardize reservations across multi-vendor
networks, such that any latency in one system that differs in another
system is accounted for and handled appropriately.

DetNet flow control will accordingly leverage the IEEE
802.1 TSN queuing and enhanced transmission and traffic
shaping techniques surveyed in Section~\ref{tsn:flow_ctrl:sec}.
These TSN mechanisms include the credit-based shaper
(802.1Q, Section 34), the time-gated or time-aware transmission
selection (802.1Qbv), the cyclic queuing and forwarding or peristaltic
shaper (802.1Qch), the asynchronous traffic shaper (802.1Qcr), and
the preemption within bridges (802.1Qbu and 802.3br). These
techniques (except for packet
preemption) can relatively easily be implemented in DetNet networks
and are a focus of collaboration
between the DetNet WG and the TSN TG.

\subsubsection{QoS Performance Guarantees between Synchronous and Asynchronous DetNet Flows}
DetNet flows, similar to TSN flows,
can be transmitted synchronously or asynchronously.  Each method has
advantages and disadvantages with respect to congestion
protection. Synchronous DetNet flows traverse DetNet nodes that are
closely time synchronized (e.g., better than one microsecond accuracy). The
time synchronized DetNet nodes can transmit DetNet flows belonging to
different traffic classes in a coordinated timely fashion, i.e., based
on repeated periodic schedules that are synchronized between the
DetNet nodes. This synchronized transmission follows the same
principles as the TSN time-aware gated mechanism (802.1Qbv) where
buffers are shared based on the coordinated time among the nodes. A
main disadvantage of synchronous transmission is that there is a
tradeoff between fine-grained time synchronized schedules and the
required network resource allocation~\cite{finn2017deterministic}.

In contrast, asynchronous DetNet flows are relayed
based on the judgment of a given individual node.
More specifically, the node
assumes the worst-case latency interference among the queued DetNet flows
and characterizes flows based on three properties:
\begin{enumerate}
	\item The maximum packet size of each DetNet flow
	\item The observational interval, i.e., the time a DetNet flow is
	occupying the resource
	\item The maximum number of transmissions during the observational interval.
\end{enumerate}

Based on the DetNet packet properties and the various header fields
resulting from the employed protocol stack, the transmission control
limits the DetNet flow's transmission opportunities to a prescribed
number of bit times per observational interval.  DetNet's design goal
of deterministic operation with extremely low packet loss dictates
that each flow must be regulated in terms of consumed
bandwidth. Furthermore, any unused bandwidth can be allocated to
non-DetNet flows, and not to any other DetNet flow since each DetNet
flow has its own resource reservation allowance.

\subsubsection{Summary and Lessons Learned}
DetNet specifies the control parameters and properties that can
integrate with lower layer L2 network transport functionalities.
These specifications enable deterministic bounds on QoS flow
requirements across L3 networks that consist of multiple L2 network
segments. DetNet defines a high-level TE architecture that follows an
SDN approach, where key concepts and functions that control and manage
DetNet flows and the relationships between the planes are defined and
specified. This allows users and operators to easily control, measure,
and manage flows dynamically while introducing fast recovery and
deterministic bounds on QoS parameters.

In contrast to the TSN flow control operations and services which are
contained within a given L2 network segment, we anticipate that the
DetNet flow control operations will have significantly larger scale
and higher complexity. DetNet flow control will pose several
challenges in areas of interoperability, control data overhead, and,
importantly, in guaranteeing QoS metrics across a wide range of L2
network segments.  In addition, there may arise complex contractual
aspects of QoS Service-Level Agreements (SLAs) among owners of
different network segments.

\subsection{Flow Integrity} \label{detnet:flow_int:sec}
DetNet flow integrity follows similar principles and methods used in
IEEE TSN standards and recommended practices. However, some key
differences include terminology, L2/L3 integration, and
security/privacy considerations.

\subsubsection{Packet Replication and Elimination Function}
The Packet Replication and Elimination Function (PREF) shares several
similarities with the TSN TG 802.1CB standard and is derived from
the IETF HSR and PRP mechanisms. PREF operates in the DetNet service
layer with three main functions~\cite{finn2017deterministic}:

\paragraph{Packet Sequencing Information}
Packet sequencing adds sequence numbers or time-stamps to each packet
belonging to a DetNet flow once. The sequence numbers are used to
identify the duplicates if two or more flows converge at a transit or
relay node. Moreover, these sequence numbers can be used to detect
packet loss and/or reordering.

\paragraph{Replication Function}
Flows are replicated at the source, i.e., with explicit source routes,
whereby a DetNet stream is forwarded on two disjoint paths directed to
the same destination.

\paragraph{Elimination Function}
Flow elimination is performed at any node in the path with the intent of
saving network resources for other flows further downstream. However,
most commonly, the elimination point is at the edge of the DetNet
network, near or on the receiver end system.
The receiving port selectively combines the replicated flows and performs
packet-by-packet selection of which to discard based on the
packet sequence number.

PREF is a proactive measure to reduce or even nullify packet
loss. However, the PREF replication mechanism needs at least two
disjoint paths to ensure reliability.  Therefore, in an effort to
enable PREF over networks lacking disjoint paths, Huang et
al.~\cite{redundancy2017deterministic} defined a single-path PREF
function. The single-path PREF function does not replicate the DetNet
flow over multiple paths; instead, it uses the same path as the
original flow.  Therefore, only the terminating or edge node has to
apply PREF on the flow.  The main rationale behind using such a
technique is that if parts of a flow on the same path is corrupted or
lost, then the replicated flow can cross-check and rebuild the
original flow's corrupted or lost packets, essentially performing
error correction and remediation.  Since more packets are sent on the
same link for a single flow than usual, more bandwidth is
needed. Therefore, the technique is mainly used for applications that
require low-rate bursty or constant traffic services, e.g., blockchain
and IoT constrained protocols.

\subsubsection{Enforced Heterogeneity}
Similar to its TSN counterpart, DetNet enforces bandwidth
discrimination between DetNet and non-DetNet flows. The DetNet network
dedicates 75\% of the available bandwidth to DetNet
flows~\cite[Section~3.3.1]{finn2017deterministic}. However, to keep
bandwidth utilization high, any bandwidth that has been reserved for
DetNet flows, but is not utilized can be allocated to non-DetNet flows
(though not to other DetNet flows).  Thus, DetNet's architectural
model ensures proper coexistence between differentiated services and
applications~\cite{finn2017deterministic}. Additionally, DetNet flows
are transmitted in a way that prevents non-DetNet flows from being
starved. Moreover, some flow control properties from
Section~\ref{detnet:flow_ctrl:sec} are employed so as to guarantee the
highest priority non-DetNet flows a bounded worst-case latency at any
given hop.

\subsubsection{Fault Mitigation}
In addition to the flow replication and bandwidth discrimination,
DetNet networks are designed with robustness that reduces the chances of a variety of possible
failures. One of the key mechanisms for reducing any disruption of
DetNet flows is applying filters and policies, similar to IEEE
802.1Qci (PSFP), that detect misbehaving flows and can flag flows that
exceed a prescribed traffic
volume~\cite{finn2017deterministic}. Furthermore, DetNet fault
mitigation mechanisms can take actions according to predefined rules,
such as discarding packets, shutting down interfaces, or entirely
dropping the DetNet flow. The filters and policers prevent
rogue flows from degrading the performance of conformant DetNet flows.

\subsubsection{IGP-TE Extensions for DetNet Networks}
To effectively utilize DetNet techniques, i.e., explicit routes as well as
congestion and resource protection, important network information,
such as node capabilities, available resources, and device performance,
needs to be communicated to and processed at the control
entities~\cite{information2017deterministic}. The DetNet WG utilizes a
PCE where the necessary network information is fed as input, and the
PCE can effectively compute a path that satisfies the QoS requirements
of the DetNet flow. Additionally, some information can be distributed
and collected using already defined TE metric extensions for OSPF and
ISIS.

Key parameters, including the employed congestion control method, the
available DetNet bandwidth, as well as the minimum and maximum queuing
delay are embedded in sub-TLVs~\cite{information2017deterministic}.
Based on these parameters, OSPF and ISIS can accurately compute the
path according to the perceived network topology and status.

\subsubsection{Flow Information Model}
In order to simplify implementations and to enable DetNet services to
operate on Layers 2 and 3, a DetNet flow information model must be
defined to describe the flow characteristics such that nodes within L2 or
L3 provide support flows properly between the sender and receiver end
systems~\cite{informationModel2017deterministic}. Farkas et
al.~\cite{informationModel2017deterministic} have specified a DetNet flow
and service information model based on the data model described in the IEEE
802.1Qcc centralized management and reservation standard
(see Section~\ref{802.1Qat:sec}).

\subsubsection{Security and Privacy Considerations}
While ensuring bounded worst-case latency and zero packet loss are the
main goals of DetNet, security and privacy concerns are also
important~\cite{finn2017deterministic}.  DetNet is envisaged as a
converged network that integrates the IT and OT domains.  Technologies
that once operated in isolation or with very limited Internet
connectivity, e.g., cyber-physical systems (CPSs), such as the power
grid as well industrial and building control, are now
interconnected~\cite{wat2017int}.  The interconnection makes these CPS
applications susceptible to external attacks and threats that are
widespread on consumer IT-based
networks~\cite{security2017deterministic}. Since any potential attack
can be fatal and cause considerable damage, CPS applications present
attractive targets for cyber-attackers.

Mizrahi et al.~\cite{security2017deterministic} have defined a threat
model and analyzed the threat impact and mitigation for the DetNet
architecture and DetNet enabled network.  The attacks that are
associated with several use cases have been detailed
in~\cite{usecases2017deterministic}. Since security models and threat
analysis are outside the scope of this paper, we only briefly note
that the three main DetNet security aspects are $(i)$ protection of
the signaling or control protocol, $(ii)$ authentication and
authorization of the physical controlling systems, and $(iii)$
identification and shaping of DetNet flows and protection from
spoofing and Man-in-the-Middle (MITM) attacks and refer
to~\cite{security2017deterministic} for further details.

\subsubsection{Summary and Lessons Learned}
The integrity and protection of DetNet flows against possible
failures, including intentional and non-intentional failures, is
imperative for the envisaged convergence of the IT and OT domains,
i.e., the linking of CPSs with the consumer/enterprise
systems. Furthermore, the secure information dissemination across
DetNet enabled networks, including access control and authentication,
must be addressed.

Future work should examine whether it would be feasible to ensure
reliability without explicit packet replication. The underlying idea of
replication is to proactively replicate packets for mission-critical
applications, since ULL packets become stale if retransmissions are
used. Therefore, replication is the easiest way to achieve
reliability, albeit at the added cost of bandwidth. State-of-the-art
Ethernet technology has now been standardized to allow up to 400~Gbps
bandwidth. Hence, there should be enough bandwidth for replication for
low to moderate proportions of mission-critical applications.  If
mission-critical applications account for large portions of the
applications, then alternative reliability mechanisms based on
low-latency coding, e.g., low-latency network
coding~\cite{ace2018har,gar2018joi,gab2018catfb,pan2017pac,roc2017blo,shi2016opt,sun2017fee,wun2017net,wun2017cat}, may be required.

\subsection{Discussion on DetNet Standardization}
DetNet strives to extend and integrate L2
techniques and mechanisms with the aim of enabling end-to-end
deterministic flows over bridges and routers, i.e., DetNet L3 nodes
beyond the LAN boundaries.
DetNet is envisioned to run over converged packet switched
networks, in particular IP-based networks. Essentially, the DetNet
architecture provides deterministic properties, e.g. bounded worst-case
latency, jitter, and packet loss, with the goal of IT and OT convergence
requiring L2 and L3 capabilities.

The DetNet WG has so far mainly focused on flow management,
Sec.~\ref{detnet:flow_mgt:sec}, and flow integrity,
Sec.~\ref{detnet:flow_int:sec}.  The DetNet specifications to date
provide correct end-to-end navigation and encapsulation, including the
DetNet data plane and overall DetNet architecture utilizing stable
well-known standards, i.e., IETF RFCs and IEEE standards.  For
instance, DetNet employs PCE for path computation, HSR and PRP for
path redundancy, as well as SDN and centralized approach to the
overall DetNet network.

As DetNet integrates IT and OT, security is an important aspect of the
DetNet architecture and protocols. While previous OT network
topologies and designs have ``air gapped'' security, i.e., completely
isolated OT networks from the outside world, the convergence of IT and
OT will place emphasis on legacy security protocols and consequently
require extensible, flexible, and power efficient security stacks that
can be ported onto OT network components.  Furthermore, with the
emerging ``fog'' computing platforms, i.e., essentially moving IT
(physical datacenters) close to the OT (physical operation points), it
becomes imperative to closely inspect traffic and monitor conditions
since any intrusion can potentially lead to catastrophic situations.

\section{DetNet Research Studies}
\label{detnet:res:sec}
Only very few research studies have examined DetNet aspects. In
particular, the flow control aspect of scheduling, and flow integrity
through replication have been studied, as surveyed in this section.

\subsection{Flow Control: Scheduling}
An important aspect of the deterministic characteristics of the packet
flow is the centralized network wide scheduling. The centralized
network wide scheduling has already been adopted by many low-latency
end-to-end connectivity technologies, such as MultiProtocol Label
Switching (MPLS).  In case of MPLS, the Path Computing Element (PCE)
is a centralized network entity that computes the optimal end-to-end
path based on global topology information.  The PCE also agrees with
the principles of SDN, where all the network control decisions are
centralized. Thus, the PCE can achieve the characteristics of
DetNet. Alternatively, advanced wireless protocols, especially for
industrial applications that require deterministic characteristics,
such as ISA100.11a and wireless HART, already use centralized routing
mechanisms~\cite{nixon2012comparison}.

Adopting wired technologies, such as DetNet, to wireless networks poses
challenges due to the possibility of hidden and exposed
nodes. Additionally, the wireless node mobility makes it more
complicated to track the delay characteristics. As a result, for
wireless technologies supporting DetNet, a promising method for
enabling determinism is by scheduling all transmissions through a
centralized decision entity.
\begin{figure}[t!]	\centering
\includegraphics[width=3.5in]{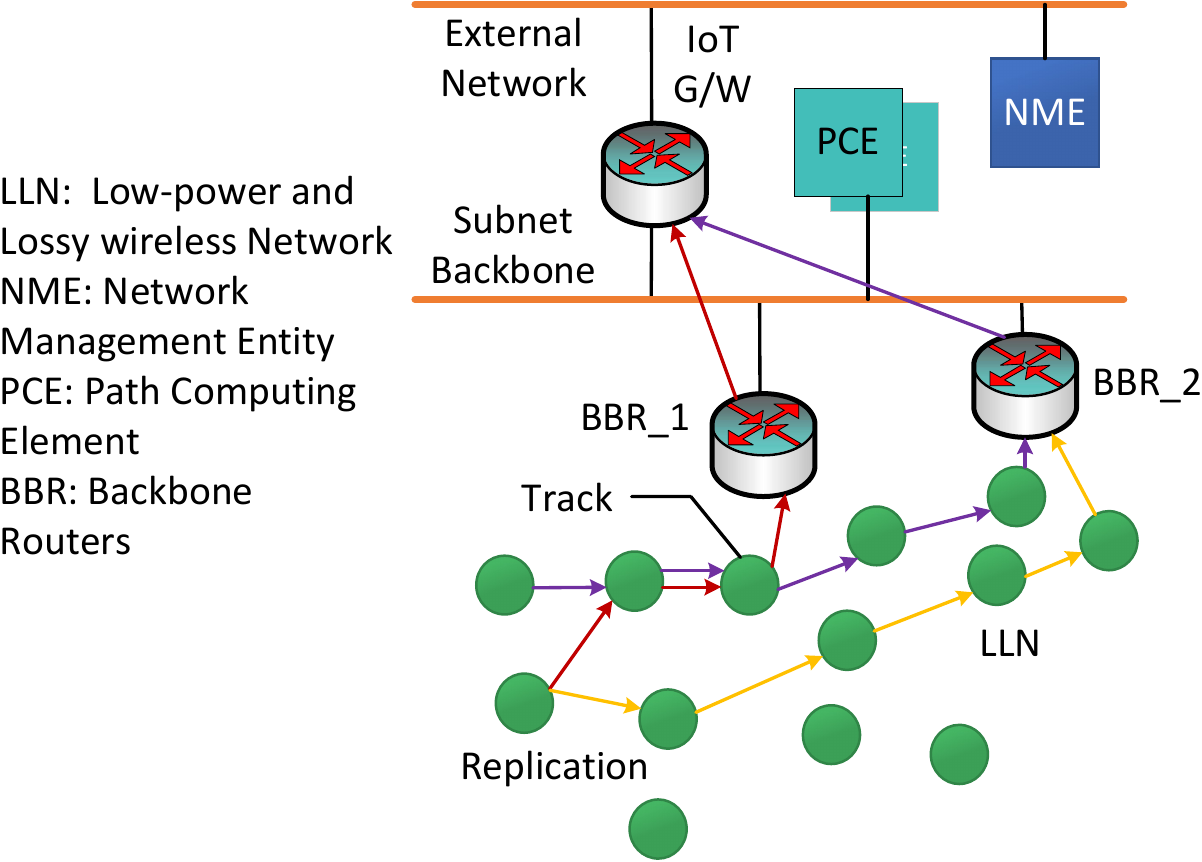} \vspace{-0.5cm}
\caption{IPv6 Time Slotted Channel Hopping (6TiSCH)
  Architecture~\cite{Thubert2015}: Software Defined
  Networking (SDN) based applications for Deterministic Networking
  (DetNet) include the Path Computing Element (PCE) for centralized
  computing of paths supporting frame replication for reliability in
  low-power and lossy networks.}
	\label{fig_6tisch}
\end{figure}
Time Slotted Channel Hopping (TSCH) is a physical layer access
technique where multiple devices access the physical resources in
terms of time and frequency slots~\cite{her2017sch}. However, every
subsequent physical layer access over the same channel hops to a
different frequency slot to achieve independence from interference and
jamming. TSCH has been widely adopted for IoT access
methods~\cite{qiu2018how} because of its simplicity and resilience to
interference~\cite{watteyne2015using}.  Moreover, IoT wireless devices
have widely adopted IPv6 as their default IP layer.  6TiSCH is a
scheduling mechanism~\cite{thubert2015architecture} based on TSCH
supporting IPv6 to achieve DetNet characteristics. Thumbert et
al.~\cite{Thubert2015} have identified the challenges associated with
centralized scheduling in 6TiSCH based on SDN to design end-to-end low
latency connectivity. The Path Computing Element (PCE) in the 6TiSCH
architecture conducts the centralized monitoring and scheduling
management of a TSCH network. The PCE also interacts with the Network
Management Entity (NME) to compute the optimal allocations and to
assign the transmission resources to the devices.  The challenges in
applying DetNet for 6TiSCH include dynamic network topology changes
and the corresponding runtime modifications of the network resources
in response to network topology changes. Additionally, the traffic
classification should be uniformly supported between low power
wireless links and wired networks.

\subsection{Flow Integrity}
Industrial applications require determinism, i.e., a bounded and
deterministic delay value, along with reductions in the end-to-end
packet latency. Towards this end, Armas et al.~\cite{Armas2016} have
examined a path diversity mechanism with packet replication.  Armas et al.~have conducted a comprehensive
performance evaluation to understand the influence of the number of
nodes and the number of replications on the energy consumption and the
end-to-end packet delay. Armas et al.~implemented a centralized
scheduler based on SDN principles in the DetNet architecture framework
to compute the disjoint paths and to apply the flow rules on networks
with up to 80 nodes.  The packet loss over the network was
evaluated through simulations.
The results indicated that with a packet replication factor of one,
where each packet is duplicated once, the packet loss was reduced by
90\% on average, showing the potential of packet replication.  As the
packet replication factor was further increased, the packet loss was
completely eliminated.  For a given network deployment, the complete
packet loss elimination can be achieved with some combination of a
degree of disjoint paths and a packet replication factor; any
additional replication would then waste resources. The energy
consumption almost doubles (is $\sim$1.863 times higher) for a packet
replication factor of 1, while the packet replication factor 4
increases the energy consumption by almost 3 times ($\sim$2.914),
showing significant energy consumption increases due to packet
replication.  In addition to the reliability, the simulation
evaluations have found end-to-end packet latency reductions of up to
40\% with a packet replication factor of one, demonstrating the latency
reduction potential of path diversity.

Pitfalls of packet replication include bandwidth shortages that arise
from the competition between replicated packet traffic and
non-replicated traffic, potentially increasing congestion and delays.
Also, as the number of flows with packet replication increases in the
network, the flow management process becomes extremely difficult in
the event of failures that require the reallocation of resources.
Therefore, addressing the packet replication challenges is a critical
aspect of designing reliability mechanisms.  SDN based flow management
mechanism can potentially optimize the replication factor while
minimizing the bandwidth utilization, consumed energy, and end-to-end
packet latency.

\subsection{Discussion on DetNet Research Studies}
Overall, there has been relatively little DetNet research to date,
leaving a wide scope for future research on architectural and protocol
improvements.  Key future research challenges include the control
plane management, virtualization, and the inter-operation with
external networks. DetNet depends on TSN to support deterministic L2
layer support, and hence requires strict scheduling techniques for
resource sharing over L2 layers.  Moreover, flow synchronization and
flow control (e.g., for traffic shaping) are generally L2 features and
hence DetNet does not address these aspects.  On the other hand, flow
management is a fundamental aspect of DetNet to oversee the management
of end-to-end flow connections. SDN inherently provides a centralized
management platform to manage the end-to-end connections through
continuous monitoring and network reconfigurations to preserve the
deterministic network service characteristics.  SDN can also play an
important role in integrating DetNet with external networks, as well
as in operating in both small scale and large scale wide area
networks.  There has also been a lack of use case definitions in
emerging markets, such as automatic driving and industrial control
networks.

\section{5G Ultra-Low Latency (ULL)}  \label{sec:5g}
5th Generation (5G) cellular technology is a paradigm shift in the
network connectivity as 5G is expected to comprehensively overhaul the
network infrastructure by establishing an end-to-end ultra-reliable
and ultra-low latency connection~\cite{Maier2016, Simsek2016}.  5G is
also expected to improve the network efficiency in terms of network
utilization, control plane overhead, and energy savings.

\begin{figure}[t!]  \centering
\includegraphics[width=3in]{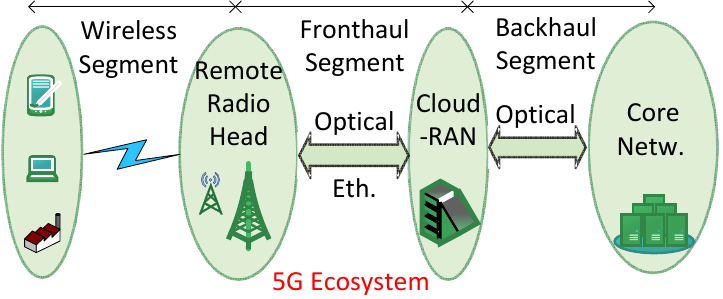}
\caption{The main network segments that constitute the 5G ecosystem
  are the wireless segment, the fronthaul segment,
  as well as backhaul segment with corresponding and core network.
In addition to various research efforts on the wireless segment, a
variety of research efforts have been conducted on the fronthaul as well as
the backhaul and corresponding core network. In
this article we focus mainly on the ULL techniques in the fronthaul
and backhaul network segments.}   \label{fig_5g_ecosystem}
\end{figure}
As illustrated in Fig.~\ref{fig_5g_ecosystem}, the overall 5G
ecosystem can be classified in terms of wireless access, fronthaul, as
well as backhaul segment with corresponding and core network. The
wireless access is responsible for the wireless connectivity between
the devices and the radio nodes. The fronthaul connects the radio
nodes to the radio baseband processing units, while the backhaul
connects the radio baseband processing units to core networks.  The
core network interconnects with the Internet at large, including data
centers, to provide end-to-end services to devices.  A large number of
5G research efforts have been conducted in the wireless access domain;
additionally, many articles have presented overviews of the 5G
advancements~\cite{amj2017ful, bhushan2014network, agiwal2016next,
  hossain20155g, Dutta2017, kak2017cog, Luvisotto2017, She2017,
  Durisi2016, ploder2017cross, mogensen2014centimeter, Pflug2013,
  Beyranvand2017}.

The recent survey on low latency characteristics in 5G by Parvez et
al.~\cite{parvez2018survey} focuses on waveform designs, wireless
protocol optimizations, microwave backhaul architectures, SDN
architectures for backhaul and core networks, and content caching
mechanism for 5G.  To the best of our knowledge, there is no prior
survey that comprehensively covers the ULL aspects across the 5G
network segments from the fronthaul to the core networks focusing on
the transport mechanisms of the user data and the control plane
signalling.  We fill this gap by providing a comprehensive survey of
ULL techniques across the 5G wireless access, fronthaul, as well as
backhaul and core networks in this section.

5G ULL mechanisms are motivated by applications
that require ultra low end-to-end latency.  As discussed by Lema et
al.~\cite{Lema2017}, the business use cases for low latency 5G
networks include health-care and medical applications, driving and transport,
entertainment, and industry automation. Remote health-care and
medical interventions, including robotic tele-surgery, require reliable
communication with ultra-low latency. Assisted and
automatic driving require high data rates for sensor data processing
as well as low latency to ensure quick responses to changing
road conditions. Immersive and integrated
media applications, such as Augmented Reality (AR) and Virtual Reality
(VR) for gaming and entertainment require high data rates for video
transmissions and extremely low latency to avoid jitter in the
video and audio. With these demanding business needs and application
requirements, 5G is expected to continuously evolve to support
ultra and extremely-low latency end-to-end connectivity.

\subsection{5G ULL Standardization} \label{sec:5g:std}
In this section, we identify the key components in 5G standards for
supporting ULL mechanisms. Various standardization organizations
contribute to the development of 5G standards, including the IEEE and
IETF, as well as the Third Generation Partnership Project (3GPP), and
the European Telecommunications Standards Institute (ETSI).  We first
discuss the standards related to the 5G fronthaul interface, and
subsequently we present the 5G architecture components which include
the backhaul.  The fundamental latency limits of 5G standards are
summarized in Table~\ref{tab:5g:latency}.  The 4.9G corresponds to the
optimization efforts for LTE towards 5G, where a drastic more than 10
fold reduction in the latency is achieved. The current standardization
efforts have targeted the total delay for 5G to be 1~ms or lower.

\begin{table}[t] \centering \footnotesize
  \caption{Latency comparison at multiple components of network
    connectivity over 3G (High Speed Packet Access (HSPA)), 4G (LTE),
    4.9G (pre 5G), and 5G~\cite{NokiaLatency}.}
	\label{tab:5g:latency}
	\begin{tabular}{lcccc}
		\hline
		\textbf{Delay Comp. (ms)} & \textbf{3G} & \textbf{4G}
		& \textbf{4.9G} & \textbf{5G} \\ \hline
		DL Trans. & 2 & 1 & 0.14 & 0.125 \\
		UL Trans. & 2 & 1 & 0.14 & 0.125 \\
		Frame alig. & 2 & 1 & 0.14 & 0.125 \\
		Scheduling & 1.3 & 0--18 & Pre-sch. & Pre-sch.\\
		UE proc. & 8 & 4 & 0.5 & 0.250 \\
		eNB proc. & 3 & 2 & 0.5 & 0.250 \\
		Trans.+Core & 2 & 1 & 0.1 & 0.1 \\ \hline
		Total Delay (ms) & 20 & 10--28 & 1.5 & 1 \\ \hline
	\end{tabular}
\end{table}

\subsubsection{Common Public Radio Interface (CPRI and eCPRI)}
\label{CPRI:sec}
\begin{figure}[t!] 	 \centering
\includegraphics[width=3.5in]{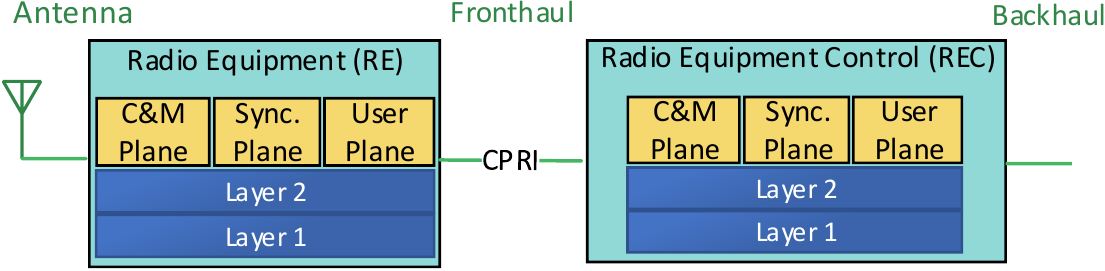}  \vspace{-0.6cm}
\caption{Common Public Radio Interface (CPRI) system
overview~\cite{cpri}: The Radio Equipment Control (REC) connects to the
Radio Equipment (RE) via the CPRI interface. The REC is part of the Base
Band Unit (BBU) and the RE is part of the Remote Radio Head (RRH) in the
Cloud-RAN architecture.}   \label{fig_5g_cpri_sys}
\end{figure}
\paragraph{CPRI}
The Common Public Radio Interface (CPRI)~\cite{de2016overview} is a
digital interface for transporting information between Radio Equipment
(RE) and Radio Equipment Control (REC). The RE resides at the Remote Radio
Head (RRH) and is responsible for the transmission of radio signals
while the baseband signal processing is conducted at the BaseBand Unit
(BBU) which implements the REC.  In particular, CPRI provides the
specifications for packing and transporting baseband time domain
In-phase/Quadrature (I/Q) samples.  Figure~\ref{fig_5g_cpri_sys}
illustrates the connectivity of BBU and REC with the RRH and RE using
the CPRI. CPRI mandates the physical layer (L1) to be optical Ethernet
transmissions over fiber, while the MAC layer can include control and
management, synchronization, and user data. CPRI has been widely
adapted for LTE and 4G deployments due to the protocol simplicity and
readily available dark fiber owned by cellular
operators~\cite{chanclou2013optical}.

5G is expected to support high bandwidth connections up to several
Gbps, resulting in very high effective I/Q CPRI data rates. For
instance, a massive MIMO connectivity with 64 antennas for both
transmission and reception would require more than 100
Gbps~\cite{vardakas2017towards}.  Additionally, the CPRI Service Level
Agreements (SLAs) require delays below 75~$\mu$s.  Therefore, CPRI
poses severe scalability issues as the required data rate increases
drastically with the number of antennas for massive MIMO which are
widely considered for 5G networks~\cite{vardakas2017towards}. Dense
Wavelength Division Multiplexing (DWDM) and Optical Transport Networks
(OTNs) can support the stringent CPRI SLA requirements. However, dense
deployments of 5G radio nodes due to the short mmWave range require
fiber connectivity to large numbers of radio nodes.  Therefore, eCPRI,
an enhanced version of CPRI, has been proposed to address the
scalability issues of CPRI~\cite{monti2017flexible}.  The 5G fronthaul
enabled by eCPRI will not only reduce the required fronthaul
bandwidths, but also relax latency requirements compared to CPRI.

\begin{figure}[t!] 	 \centering
\includegraphics[width=3in]{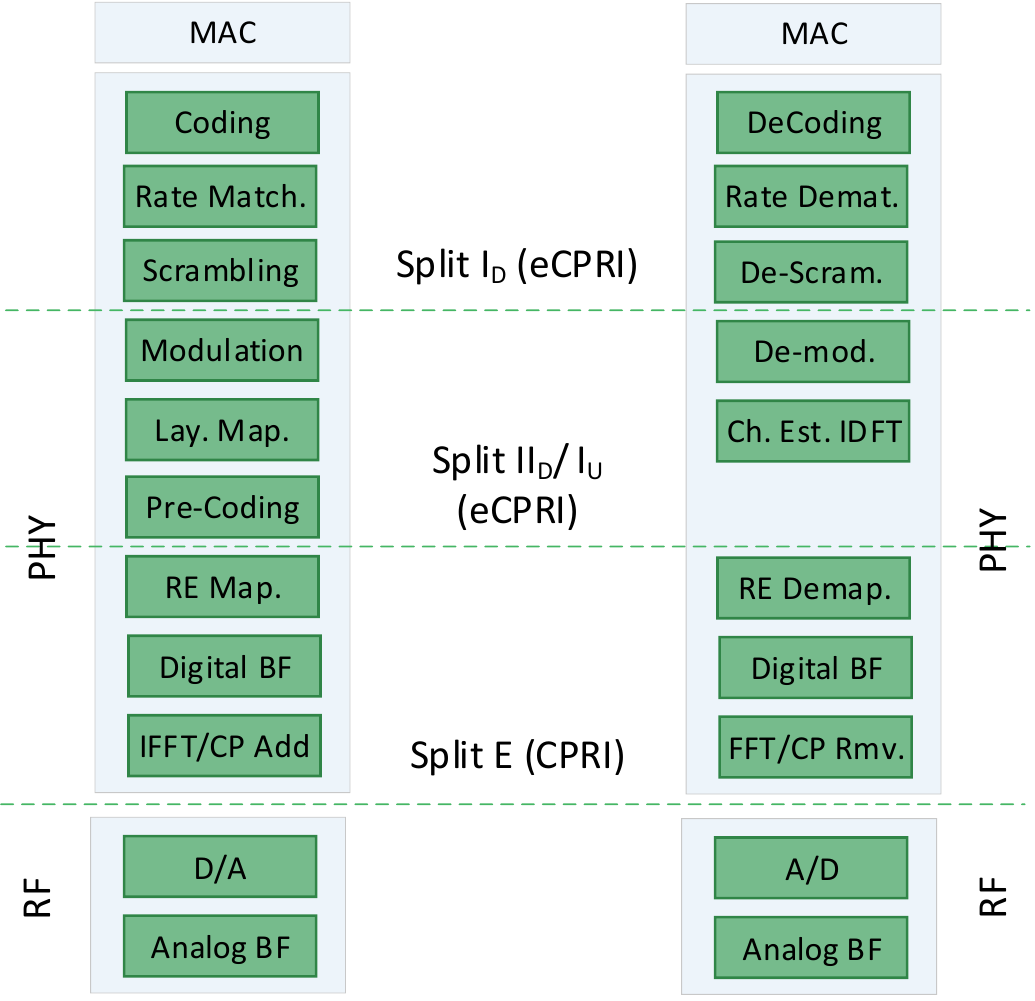}  \vspace{-0.2cm}
\caption{Split options defined by eCPRI the steps above the
    horizontal dashed line are processed at the BBU and the steps
    below the dashed line are processed at the RRH: Split E
  corresponds to the CPRI data, split I$_D$ corresponds to the eCPRI
  downlink data after scrambling, split II$_D$ corresponds to the
  eCPRI downlink data after pre-coding, and split I$_U$ corresponds
  to the eCPRI uplink data after RE-demap~\cite{de2016overview}.}
	\label{fig_5g_cpri_split}
\end{figure}
\paragraph{eCPRI}
eCPRI reduces the effective data rate carried over the L1 connection
between RE and REC. eCPRI also removes the mandatory L1 requirements,
thus allowing operators to implement low-cost Ethernet links.  More
specifically, the data rate reduction is achieved by various
functional split options as shown in
Fig.~\ref{fig_5g_cpri_split}.
The split option defines the allocation of the RF and PHY
  processing steps to the RRH and BBU. The steps above the split
  indicated by a horizontal dashed line in
  Fig.~\ref{fig_5g_cpri_split} are conducted at the BBU, while the
  steps below the split are conducted at the RRH. Accordingly the
  split option governs the type of signal (and its corresponding QoS
  requirements) that has to be transmitted over the fronthaul
  network.  eCPRI primarily defines two split options in the
downlink. The $I_D$ split performs PHY layer bit scrambling at the
BBU, while RF transmissions are modulated at the RRH. Similarly, the
$II_D$ split conducts pre-coding, Resource Element (RE) mapping,
digital Bandpass Filter (BF), and IFFT/FFT and Cyclic Prefix (CP) at
the BBU.  In contrast to the downlink, eCPRI defines only one split
option in the uplink $I_U$, whereby the PHY layer functions, from the
channel estimation to the decoding, are conducted at the BBU, while RE
demapping to RF transmissions are processed at the RRH.
\begin{table}[t] \centering \footnotesize
\caption{CPRI Split E as well as eCPRI splits $I_D$, $II_D$ (downlink),
  and $I_U$ (uplink) one-way packet delay and packet loss
  requirements~\cite{eCPRItransportnetwork}.}
	\label{tab:ecpri:delay}
	\begin{tabular}{llll}
		\hline
		\multicolumn{1}{c}{\textbf{CoS name}} & \multicolumn{1}{c}{\textbf{Example use}} & \multicolumn{1}{c}{\textbf{\begin{tabular}[c]{@{}c@{}}One-way max.\\ packet delay\end{tabular}}} & \multicolumn{1}{c}{\textbf{\begin{tabular}[c]{@{}c@{}}One-way \\ pkt. loss ratio\end{tabular}}} \\ \hline
		\multicolumn{1}{c}{High} & \multicolumn{1}{c}{User Plane} & \multicolumn{1}{c}{$100~\mu$s} & \multicolumn{1}{c}{$10^{-7}$} \\
		\multicolumn{1}{c}{Medium} & \multicolumn{1}{c}{\begin{tabular}[c]{@{}l@{}}User Plane (slow),\\ C\&M Plane (fast)\end{tabular}} & \multicolumn{1}{c}{$1$ ms} & \multicolumn{1}{c}{$10^{-7}$} \\
		\multicolumn{1}{c}{Low} & \multicolumn{1}{c}{C\&M Plane} & \multicolumn{1}{c}{$100$~ms} & \multicolumn{1}{c}{$10^{-6}$} \\ \hline
	\end{tabular}
\end{table}
In contrast to eCPRI, CPRI only carries the output from the IFFT/FFT
and Cyclic Prefix (CP)
at the BBU to the RF Digital to Analog
(D/A) converter at the RRH. The delay requirements for the various
Classes of Service (CoS) for the $I_D$ and $II_D$ splits (eCPRI) and
the E split (CPRI) are summarized in Table~\ref{tab:ecpri:delay}. The
high CoS corresponding to split E (CPRI) requires the one way maximum
packet delay to be on the order of $100$~$\mu$s.  The split E
transports the I/Q data and in-band Control and Management (C\&M)
information. The medium CoS, which supports both the user and C\&M
plane data, requires 1~ms delay. The low CoS for the uplink eCPRI
$I_U$ split requires $100$~ms delay.

The eCPRI services include:
\begin{itemize}
\item[i)] User plane I/Q data transport between
  BBU and RRH, user plane control and management (C\&M), and support
  services, such as remote reset.
\item[ii)] Time synchronization between BBU and RRH.
\item[iii)] Operation and management (OAM), including eCPRI connection
  setup, maintenance, and tear-down.
\end{itemize}

\begin{figure}[t!] 	 \centering
\includegraphics[width=3.5in]{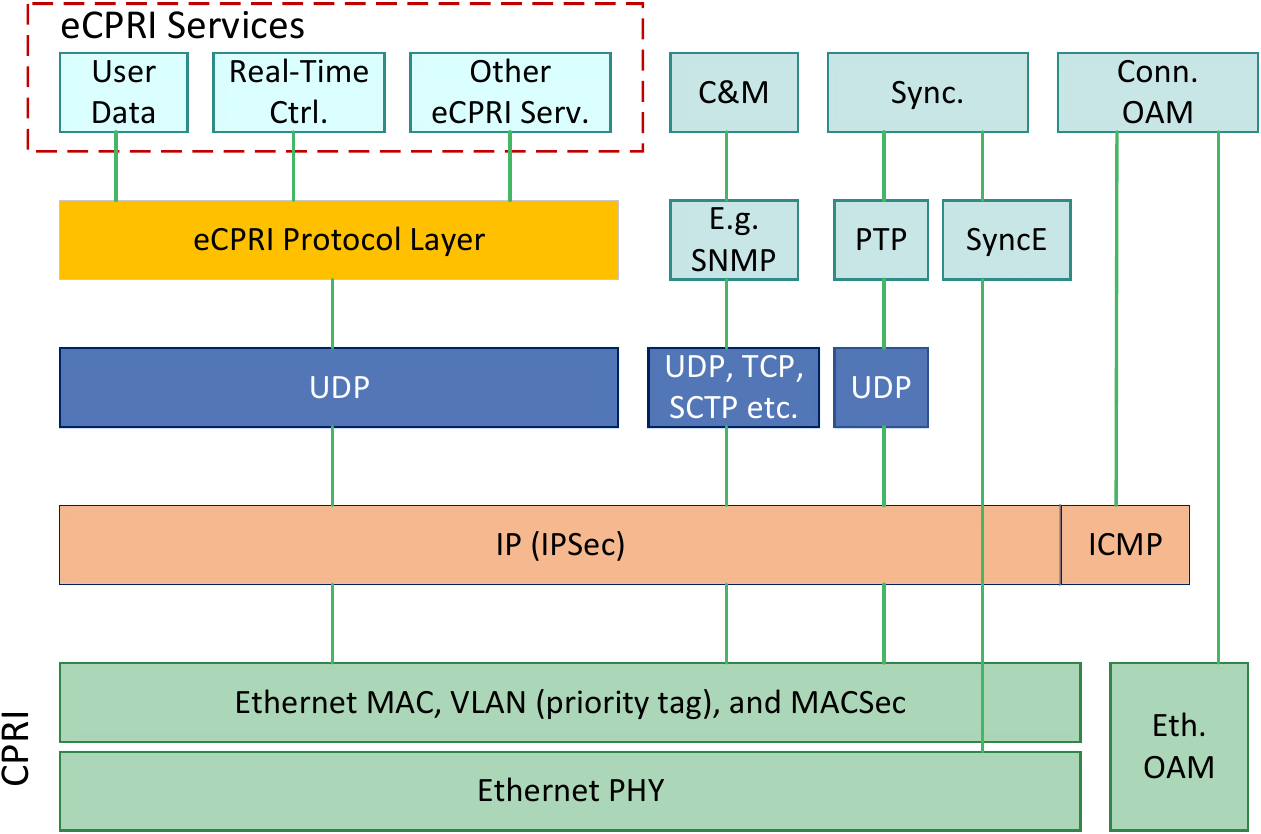} \vspace{-0.2cm}
\caption{The eCPRI protocol stack consists of the eCPRI protocol
  layer, which transports the data from various split options over
  generic UDP and IP protocol layers. The lower layers, PHY and MAC,
  are equivalent to the CPRI protocol. The eCPRI services as well as the
  eCPRI control and management data along with synchronization are
  supported by the eCPRI protocol stack~\cite{eCPRItransportnetwork}. }
	\label{fig_5g_ecpri_protocol}
\end{figure}
eCPRI supports various message formats to transport I/Q data according
to the adopted split option. The protocol stack description of eCPRI
services over IP and Ethernet is illustrated in
Fig.~\ref{fig_5g_ecpri_protocol}.  The eCPRI specific protocol layer
transports the time domain I/Q data for split E, or frequency domain
I/Q data for splits $I_D$ and $I_U$.  eCPRI messages are transmitted
as UDP/IP packets whereby the eCPRI header and data constitute the
UDP packet payload. The UDP packet headers contain both the source and
destination IP addresses of the eCPRI nodes.  Various message types
control the overall operation of eCPRI over Ethernet links, including
one-way delay measurement, remote reset, and event indication.

Unlike CPRI, which requires point-to-point and point-to-multipoint
operation in a master-to-slave configuration, eCPRI is agnostic
to the network topology which may encompass local area
networks, as well as public routers and switches. The logical topologies
that are possible with eCPRI include:
\begin{itemize}
\item Point-to-point, i.e., one BBU to one RRH which is similar to CPRI.
\item Point-to-multi-point, i.e., one BBU to multiple RRHs (supported
  in CPRI as well).
\item Multi-point-to-multi-point, i.e., multiple BBUs to multiple RRHs (mesh
	      configuration), unique to eCPRI.
\end{itemize}

In a generalized Ethernet network carrying multiple traffic types
(including best effort traffic), the user plane I/Q data and the real
time O\&M data require high priority transmissions. TSN mechanisms,
see Sec.~\ref{tsn:std:sec}, can enable Ethernet networks to meet the eCPRI
delay requirements. eCPRI management messages and
user plane data can be regarded as Control Data Traffic (CDT) that
is transmitted with high priority scheduling over the TSN network.
Traffic requirements for user plane data vary for the different split
options, which can be assigned different TSN priority levels.  For
instance, the C\&M data is typically not as delay sensitive as user
plane data; hence, a lower priority can assigned to C\&M traffic.
However for critical C\&M data, such as remote reset while
troubleshooting a Remote Equipment (RE) problem may require higher
priority levels than user data.  Therefore, two priority levels can be
assigned to C\&M traffic, i.e., a priority level higher than user
data and another priority level lower than user data.  These
priority levels can be readily supported by TSN networks, which
accommodate eight independent priority queues.

\paragraph{Summary and Lessons Learned}
5G technology supports diverse applications with a wide range of data
rates and latency requirements, which directly translate to
requirement for a flexible and scalable fronthaul.  CPRI and eCPRI
provide standardized protocols for inter-operating with existing
cellular infrastructures.  CPRI may not be suitable for supporting
massive broadband services due to the very high required I/Q data
rates. Also, the CPRI latency requirements need to be carefully
considered and may require the judicious use of the scheduled traffic
concept~\cite{wan2015per}.  eCPRI overcomes the data rate issue
through functional splits but increases the complexity of remote radio
nodes.  Another shortcoming of eCPRI is that the system considers
asymmetrical OFDM in the downlink and uplink, i.e., single-carrier
OFDM (SC-OFDM) in the uplink.  Symmetrical OFDM systems are being
investigated for increased spectral uplink
efficiency~\cite{vardakas2017towards}. However, there is no specific
split defined for symmetrical OFDM systems in eCPRI.  Remote spectrum
analysis for troubleshooting RF issues is possible in CPRI; whereas,
eCPRI does not provide such remote RF evaluation capabilities,
although splits $I_U$ and $I_D$ allow for remote RF management.
Hence, mechanisms for the transmission of sampled time domain I/Q
samples from the RRH back to the BBU must be developed for advanced
troubleshooting.

\subsubsection{IEEE 802.1CM: Time-Sensitive Networking for Fronthaul}
\begin{figure}[t!] 	 \centering
\includegraphics[width=3.3in]{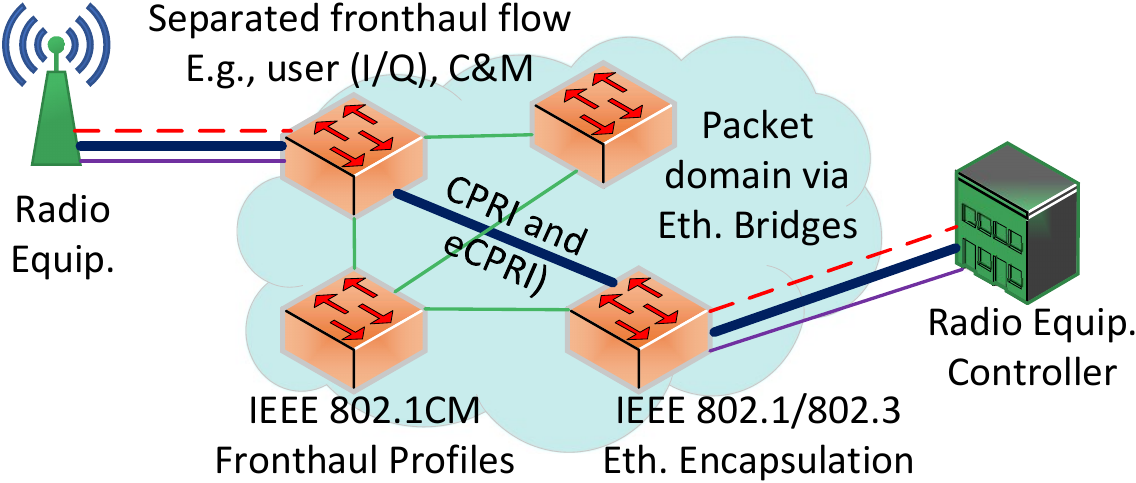}  \vspace{-0.2cm}
\caption{IEEE 802.1 CM defines the support for Ethernet-based
  fronthaul in a bridged network. Flows are separated into different
  classes and a specific fronthaul profile is applied to each class to
  transport the flows over the Ethernet bridges based on the flow
  requirements~\cite{tsn80211cm}.}  \label{fig_tsn_cm_bridged_net}
\end{figure}
The IEEE 802.1CM standard~\cite{tsn80211cm} is a CPRI-IEEE 802.1
collaboration to provide bridged Ethernet connectivity for fronthaul
networks, as illustrated in Fig.~\ref{fig_tsn_cm_bridged_net}.
An 802.1CM bridge must support a data rate of 1 Gbps or
higher on each port.  The IEEE 802.1CM requirements are derived from
CPRI and eCPRI so as to support various splits, such as splits at the
FFT, demapping, and scrambling radio functions.  IEEE 802.1CM defines
mechanisms for end stations, bridges, and LANs to establish Ethernet
networks that can support the time sensitive transmissions of
fronthaul streams.  In current cellular network deployments, the
separation between RRH and BBU requires connectivity with stringent
latency and capacity requirements.  These fronthaul connection
requirements could not be readily provided by today's bridged Ethernet
networks.

IEEE 802.1CM provides specific mechanisms, such as scheduling,
preemption and synchronization mechanisms, to satisfy the fronthaul
requirements. With IEEE 802.1CM, mobile operators can utilize large
segments of existing bridged networks to support 5G fronthaul
networks, reducing capital expenditures.  Moreover, centralized
management mechanisms can be employed for automatic network
reconfigurations, reducing the operational expenditures compared to
manual network configuration.  IEEE 802.1CM distinguishes Class 1
traffic for CPRI and Class 2 traffic for eCPRI. In terms of network
synchronization, the IEEE 802.1 CM standard specifies two mechanisms:
$i)$ packet timing using protocols, such as the Precision Time
Protocol (PTP) for point-to-point synchronization distribution from a
remote common master, and $ii)$ co-located common master for both BBU
and RRH.

\paragraph{Latency Components of a Bridge}
A bridge supporting fronthaul network functionalities needs to tightly
control the latency and synchronize its functions.  The latency for a
single hop in a bridge network is the time duration from the arrival
of the last bit of a given frame at a given bridge port to the
arrival of the last bit of the same frame at a particular port at the next
hop bridge. The main delay component are:
\begin{itemize}
\item[i)] Store and forward delay $t_{\rm SF}$ due to all the elements
  responsible for the internal frame forwarding from ingress to egress
  port.
\item[ii)] Queueing (interference) delay $t_{\rm Queuing}$ due to
  ongoing transmissions of higher priority frames.
\item[iii)] Self queuing delay $t_{\rm Self\_Queuing}$ due to frames
  of the same class that arrive across multiple ports and need to be
  sequentially queued.
\item[iv)] Periodic Constant Bit Rate (CBR) high priority data flow
  delay $t_{\rm MaxGoldFrameSize+Pre+SFD+IPG}$. IQ data flows are referred to as gold flows in IEEE 802.1 CM.
  The CBR data delay
  $t_{\rm MaxGoldFrameSize+Pre+SFD+IPG}$ of a gold frame
  corresponds to an IQ data frame with maximum frame size
  with Preamble (Pre), Start Frame Delimiter
  (SFD), and Inter Packet Gap (IPG).
\end{itemize}
The total worst-case self-queuing delay in a bridge can be evaluated
based on the number $N_p$ of ingress ports that can receive interfering
gold frames which need to be transmitted over egress port $p$, and the
total number of flows $F_{i,p}$ supported between ingress port $i$ and
egress $p$.  Let $G^{i,p}_k$ denote the maximum number of frames
belonging to a gold flow $k$ traversing from ingress port $i$ to
egress port $p$ that can be grouped into a single time window before
the reception of frames at the ingress edge port of the bridge
network.  The resulting worst-case self-queuing delay at port $j$ can
be evaluated as
\begin{eqnarray}
t^{j,p}_{\rm Self\_Queueing} &=& t_{\rm MaxGoldFrameSize+Pre+SFD+IPG} \nonumber \\
& & \ \ \ \times \sum_{i=1, i \neq j}^{N_P} \sum_{k=1}^{F_{i,p}} G^{i,p}_k.
\label{eq:SelfQueueing}
\end{eqnarray}
Without preemption, the maximum queuing delay $t_{\rm Queuing}$
incurred by gold flows depends on the maximum size of the low
priority frame along with preamble (Pre), Start Frame Delimiter (SFD),
and the Inter Packet Gap (IPG), which results in $t_{\rm Queuing} =
t_{\rm MaxLowFrameSize+Pre+SFD+IPG}$.  However with preemption, a high
priority frame is transmitted right after the transmission of the fragment of
the preemptable frame, which includes the Cyclic Redundancy Check (CRC) and
Inter Frame Gap (IFG).
Therefore, the total worst-case delay $t_{\rm MaxBridge}$ for gold flows
in a bridge  can be evaluated as
\begin{eqnarray}
t_{\rm MaxBridge} &=& t_{\rm MaxGoldFrameSize+Pre+SFD+IPG} \nonumber \\
  & & \ \ \ +  t_{\rm SF} + t_{\rm Queuing} + t_{\rm SelfQueuing}.
\label{eq:bridge:delay}
\end{eqnarray}

\paragraph{Fronthaul Profiles}
In general, the fronthaul flows in a bridged network are classified
into High
Priority Fronthaul (HPF), Medium Priority Fronthaul (MPF), and Low
Priority Fronthaul (LPF) flows.  The HPF corresponds to class 1
I/Q data and class 2 user plane data with the requirement of
100~$\mu$s end-to-end one-way latency.  Similarly, the MPF corresponds
to the class 2 user plane (slow) data and class 2 C\&M (fast)
data. The LPF could include the C\&M data of class 1 and 2 traffic.
IEEE 802.1 CM defines two profiles, namely profiles A and B,
to service different
fronthaul technologies supporting both class 1 and 2. The MPF data is typically assigned a priority
level immediately below the HPF; whereas, the LPF data is
assigned a priority immediately below the MPF data. In contrast to
the traffic classes which are designed based on the relative
priorities, the profiles are designed based on the worst-case
end-to-end delay within a given traffic class.

\textbf{Profile A}: The goal of profile A is to simplify the
deployments and support only strict priority, focusing on the
transport of I/Q user data as high priority traffic and C\&M data with
  lower priority.
The maximum fame size for all traffic is 2000 octets.

\textbf{Profile B}: Profile B adopts advanced TSN features, including
frame preemption, as defined in IEEE 802.3br and 802.1Qbu, as well as
strict priorities
to carry I/Q user data as high priority traffic and
  C\&M data as low priority preemptable traffic.
The maximum frame size for user data is 2000 octets, while all other
traffic can have variable maximum frame sizes.

\paragraph{Summary and Lessons Learned}
IEEE 802.1CM primarily supports CPRI and eCPRI connectivity over
bridged networks. IEEE 802.1CM enables cellular operators to use the
existing Ethernet infrastructure reducing the capital and operational
expenditures.  However, the lack of support for generalized fronthaul
networks
 limits the applicability of the
IEEE 802.1CM standard to a wider set of 5G applications, such as
crosshaul~\cite{de2015xhaul}.
The relative performance of the
  low priority C\&M traffic as compared to the high priority I/Q user
  data traffic (i.e., the ULL traffic) still needs to be thoroughly
  investigated to understand the behavior of traffic classes when
  operating at high load levels that approach the link capacities.

Although the delay and synchronization aspects have been specified in
the standards, the security and reliability issues have not yet been
considered in detail.  Hence, security and reliability present a wide
scope for future research and standards development.  These security
and reliability issues should be investigated by the fronthaul task force
which is responsible for the IEEE 802.1 CM standards development.

We note that a cellular operator may choose to change priority levels
as desired. A potential pitfall is that regular (non-fronthaul)
traffic could be assigned higher priority than fronthaul user data or
C\&M traffic.  Such a priority assignment would increase the
self-queuing and queuing delays for the fronthaul traffic.  Thus, the
relative priority levels of the different traffic priority classes
need to be carefully considered in the network resource allocation.

\subsubsection{Next Generation Fronthaul Interface (NGFI)}
\paragraph{Overview}
Although the IEEE 802.1 CM, CPRI, and eCPRI fronthaul protocols provide
implementation directions for fronthaul networks, the lack of
fronthaul architectural standardizations has prompted the IEEE
standards group to commission the IEEE 1914 Working Group
(WG)~\cite{ieee1914} to define the standards for packet-based
Fronthaul Transport Networks (FTN). In particular, the IEEE 1914 WG
has defined two standards: $i)$ IEEE P1914.1 focusing on architectural
concepts related to both data and management fronthaul traffic in an
Ethernet based mobile FTN networks, and $ii)$ IEEE P1914.3 focusing on
the encapsulation of I/Q data for Radio Over Ethernet (RoE).  In comparison
to IEEE 1914.3, the latency impact on the fronthaul deployment is mainly
influenced by IEEE P1914.1. Hence, we primarily
focus on architectural concepts, protocol operations, traffic
management, and requirements as well as the definitions for fronthaul
links as defined by IEEE P1914.1. The goals of IEEE P1914.1 are to
support 5G critical use cases, such as massive broadband services and
to design a simplified fronthaul architecture that can utilize the
existing standard Ethernet deployments of cellular operators. However,
IEEE 1914.1 does not define the functional split
aspects of the fronthaul, while aligning with 3GPP to support
functional splits suitable for 5G.

\paragraph{Two-Level Fronthaul Architecture}
\begin{figure}[t!] 		\centering
\includegraphics[width=3.5in]{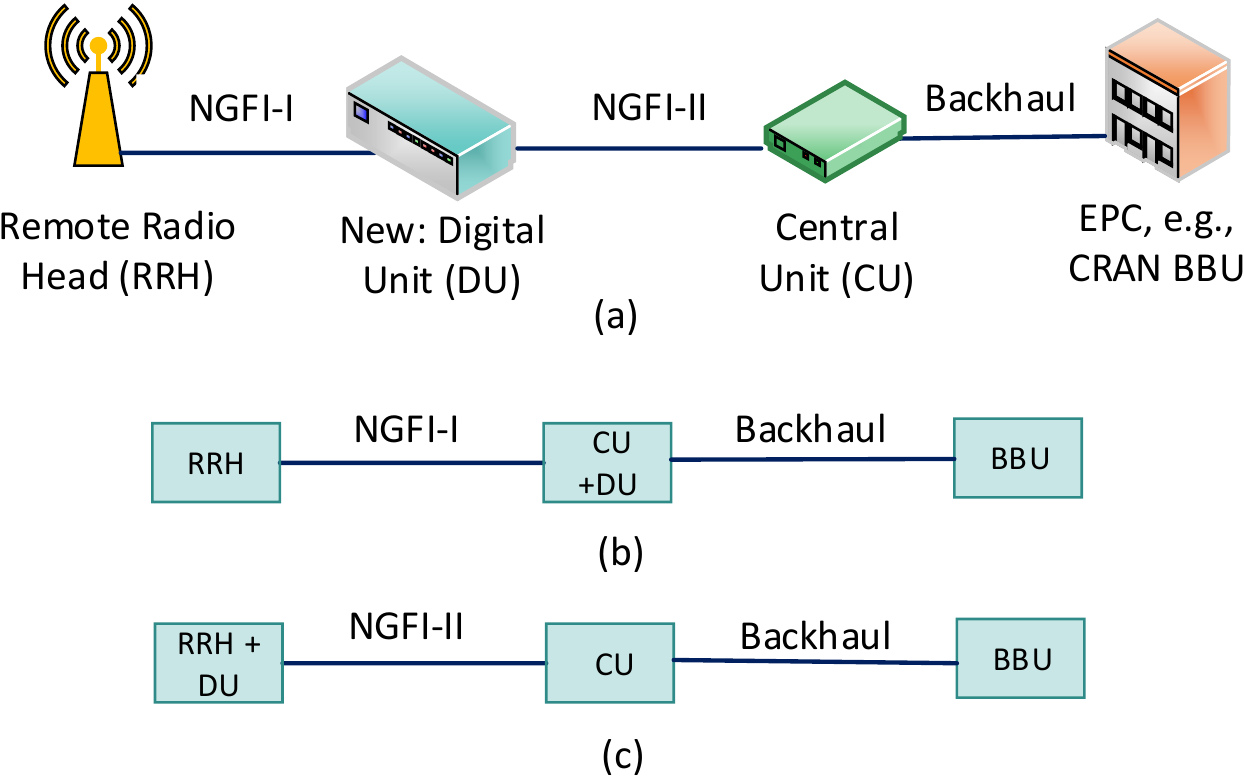}  \vspace{-0.2cm}
\caption{Illustration of two-level architecture options for
next-generation fronthaul transport network: (a) RRH is connected
via NGFI-I fronthaul interface to Digital Unit (DU) and DU is
connected via NGFI-II interface to Central Unit, (b) RRH is
connected via NGFI-I interface to integrated CU and DU, and (c) DU
is integrated with RRH and connected via NGFI-II to CU~\cite{ieee1914}.}
\label{fig_5g_ngfi}
\end{figure}
IEEE P1914.1 defines a two-level fronthaul architecture that separates
the traditional RRU to BBU connectivity in the CRAN architecture into
two levels, namely levels I and II.  Level I connects the RRH via a
Next Generation Fronthaul Interface-I (NGFI-I) to a new network
element, the so-called Digital Unit (DU). Level II connects the DU via
an NGFI-II interface to the newly introduced Central Unit (CU), as
shown in Fig.~\ref{fig_5g_ngfi}(a). Figs.~\ref{fig_5g_ngfi}(b) and (c)
show different deployment options with integrated RRH and DU, and with
integrated CU and BBU, respectively. The purpose of the two-level
architecture is to distribute (split) the radio node (i.e., eNB/base
station) protocol functions between CU and DU such that latencies are
relaxed, giving more deployment flexibilities. In general, NGFI-I is
targeted to interface with the lower layers of the function split
which have stringent delay and data rate requirements. In contrast,
NFGI-II is targeted to interface with the higher layers of the
function split relative to NGFI-I, relaxing the requirements for the
fronthaul link.

The NGFI is designed to mainly address:
\begin{itemize}
\item[i)] Scalability: To enable C-RANs and Virtual-RANs that are
  functional split and traffic independent.
\item[ii)] Resource Utilization: To achieve statistical
  multiplexing by supporting variable MIMO and Coordinated Multipoint
  (CoMP) for 5G.
\item[iii)] Flexibility: To operate in a radio technology agnostic
  manner while supporting SDN controlled dynamic reconfigurations.
\item[iv)] Cost Effective: To utilize existing cellular network
  infrastructure.
\end{itemize}
\begin{table}[t] \centering  \footnotesize
\caption{NGFI Transport Classes of Service; Low split, Med. split, and
  High split are relative to the positioning of the split in
  Fig.~\ref{fig_5g_cpri_split}, whereby the low split is closer to
  the bottom of Fig.~\ref{fig_5g_cpri_split}.}
	\label{tab:ngfi}
	\begin{tabular}{cllll}
		\textbf{Class} & \multicolumn{1}{c}{\textbf{Sub Class}} & \multicolumn{1}{c}{\begin{tabular}[c]{@{}c@{}}\textbf{Max. Lat.} \end{tabular}} & \multicolumn{1}{c}{\begin{tabular}[c]{@{}c@{}}\textbf{Pri.}\end{tabular}} & \multicolumn{1}{c}{\textbf{App.}} \\
		\hline
		\multirow{2}{*}{C\&M} & Sync. & TBD & TBD &  \\
		& \begin{tabular}[c]{@{}l@{}}Low Lat. RAN\\ ctrl. plane\end{tabular} & $100~\mu$s & 2 &  \\
		\hline
		\multirow{5}{*}{Data Plane} & Subclass 0 & $50~\mu$s & 0 & ULL data \\
		& Subclass 1 & $100~\mu$s & 1 & Low split. \\
		& Subclass 2 & $1$ ms & 2 & Med. split \\
		& Subclass 3 & $3$ ms & 3 & High split \\
		& Subclass 4 & $10$ ms & 4 & \begin{tabular}[c]{@{}l@{}}Legacy\\ Backhaul\end{tabular} \\
		\hline
		\begin{tabular}[c]{@{}c@{}}Trans. Net.\\ C\&M\end{tabular} & \begin{tabular}[c]{@{}l@{}}Trans. Net.\\ ctrl. plane\end{tabular} & $1$ ms & 2 & \\
		\hline
	\end{tabular}
\end{table}
Additionally, NGFI supports connectivity to Heterogeneous Networks
(HetNets) by decoupling the transport requirements from the radio
technologies.  Thus, multiple traffic classes, as summarized in
Table~\ref{tab:ngfi}, can be transported by the NGFI network, mainly
to support latencies according to the application demands.  The C\&M
class supports low-latency control plane data for radio node
signalling. Data plane latencies vary according to the different
subclasses 0--4 to support multiple technologies and deployment
versions with multiple split options. Subclass 0 requires the highest
priority with $50~\mu$s of maximum allowed latency, while subclass 4
has the lowest priority and a $10$~ms maximum delay bound.
Subclass 4 can, for instance, be used for the legacy backhaul over the NGFI
interface. The traffic of each subclass is independently transported
between the end points without any mutual interference while achieving
statical multiplexing gain among the subclasses.

\paragraph{Summary and Lessons Learned}
The NGFI primarily addresses the scalability and cost issues with the
current fronthaul solutions, such as CPRI.  With NGFI, connections
between DU and CU can be directly connected by an Ethernet link
supporting IEEE P1914.1 specifications.  The NFGI L2 subclass 0
transport service can readily accommodate the requirements of the
existing CPRI deployments without any changes to the infrastructure
deployments.  Thus, NGFI is expected to play a significant role in the
unification of heterogeneous radio technologies at the transport level
and support converged fronthaul and backhaul networks for converged
and coexisting 4G and 5G technologies.  An important aspect to
investigate in future research is the tradeoff between link
utilization and multiplexing gain for the standard Ethernet networks
while adopting these new fronthaul support architectures and
protocols.

\subsubsection{Backhaul Networks} \label{sec:5g:back}
\paragraph{Overview}
The backhaul networks consisting of core network elements play a
critical role in setting up the end-to-end flows.  Core networks
control the user data scheduling in both uplink and downlink. The
control signalling of the radio technology, e.g., LTE, can contribute
to flow latency when user devices transition among various states,
e.g., idle to active (connected) and vice versa~\cite{che2017mac,meh2016cla,tya2017con}.
For scenarios with
intermittent data activity, devices typically implement a state
transitioning mechanism from active to idle to conserve computing and
wireless resources. For instance, if the inter packet delay is more
than 40~ms, the device can pro-actively change the radio control state
to idle.  Thus, within a single ULL flow session, there can be
multiple user device state transitions between idle and active.  The
core network manages the control plane signalling of the radio
technology whereby advanced methods can be implemented to reduce the
state transition overhead during flow setup, thereby reducing the
latency. For ULL flows, irrespective of whether the traffic is
intermittent or has a constant bit rate, the end-to-end latency should
be minimized for both flow setup and steady state traffic flow.

An efficient backhaul network design can reduce control plane
signalling for both initial ULL flow setup and steady state traffic.
We give brief overviews of the two standardization efforts that
efficiently implement the 5G core network functions for setting up and
supporting ULL flows, namely Control and User Plane Separation (CUPS)
of EPC and Next Generation (NG) Core.

\paragraph{Control and User Plane Separation (CUPS) of EPC}
\begin{figure}[t!] 	 \centering
\includegraphics[width=3.5in]{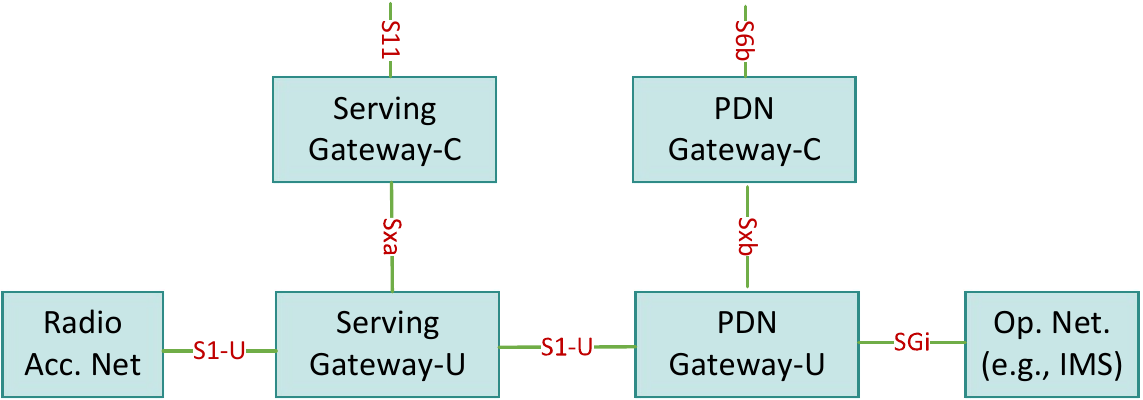}  \vspace{-0.2cm}
\caption{Illustration of Control and User Plane Separation (CUPS) for
  the EPC as proposed by the 3GPP. The Serving-GW (S-GW) functions and
  the PDN-Gateway (P-GW) functions in the EPC are split between S-GW-C
  (i.e., control), S-GW-U, and P-GW-U (i.e., user) to increase the
  flexibility of existing EPC networks~\cite{3GPPcups}.}
	\label{fig_5g_cups}
\end{figure}
The SDN paradigm of separating the control and data plane functions
while centralizing the overall control plane has provided substantial
advantages in traditional networks. The
3GPP has proposed Control and User Plane Separation
(CUPS)~\cite{3GPPcups} for the Evolved Packet Core (EPC) backhaul of
the LTE radio technology, see Fig.~\ref{fig_5g_cups}, to adapt SDN principles in the cellular
backhaul core networks to achieve similar benefits. Current
network deployments are facing increased capital
and operational expenditures when scaling the infrastructures to meet
the capacity demands from the users. This infrastructure scaling
problem is exacerbated by the integrated control and user plane
functions in the existing backhaul networks. CUPS targets $i)$ flexible
deployments in both distributed
and centralized control plane, and $ii)$ independent scaling of
control and user plane functions.

CUPS plays an important role in reducing the overall end-to-end
latency through the cellular operator networks by selecting the user
plane nodes that are close to the RAN node.  In particular, the data
is transported without having to interact with the control plane nodes
for the path setup, which is especially beneficial for user mobility
scenarios.  That is, the flow paths of user plane nodes are
dynamically adapted according to the requirements and mobility,
without having to negotiate with control plane entities, such as SGW-C
and PGW-C. This capability will greatly increase the backhaul
flexibility of the existing LTE radio technology deployments. New
interfaces, namely Sxa, Sxb, and Sxc, see Fig.~\ref{fig_5g_cups}, have
been introduced to communicate between the control and user planes of
the Serving-GW (S-GW). The main advantages of CPUS in comparison to
the existing EPC are:
\begin{itemize}
\item[i)] Removal of GPRS Tunneling Protocol (GTP) and session management
 between control plane entities.
\item[ii)] A cross connection interface between control and user
  plane, such that any control function can interact with any user
  function.
\item[iii)] A UE is served by a single control plane, but the data flow
  path may traverse multiple data plane functions.
\item[iv)] A control plane function is responsible for
  creating, managing, and terminating a flow over the user plane
  functions.  All 3GPP control functions, such as PCC, charging,
  and admission control are supported within control plane function,
  while the user plane is completely agnostic to the 3GPP control
  functions.
\item[v] A legacy EPC consisting of S-GW and PDN-GW can be replaced
  with new user plane and control plane split nodes without any impact
  on existing implementations.
\end{itemize}

\paragraph{Summary and Lessons Learned}
CUPS provides a mechanism to adapt advanced resource management
functions, such as SDN, to existing networks while improving the
flexibility. The reduction of data plane and control plane overhead,
particularly the removal of GTP tunneling, allows user data to be
transported without encapsulation and without GTP sessions. Moreover,
the user device state transitions trigger control plane activities in
the core networks. Therefore, the separation of control and data plane
not only increases the flexibility, but also reduces the radio control
signalling to support ULL flows.  Thus, cellular operators can
incrementally upgrade towards 5G deployments. For distributed
deployments, future research needs to thoroughly examine the placement
and implementation of control and user plane entities without
impacting the overall EPC system behavior.

\subsubsection{Next Generation (NG) Core}
\begin{figure}[t!] 	 \centering
\includegraphics[width=3.4in]{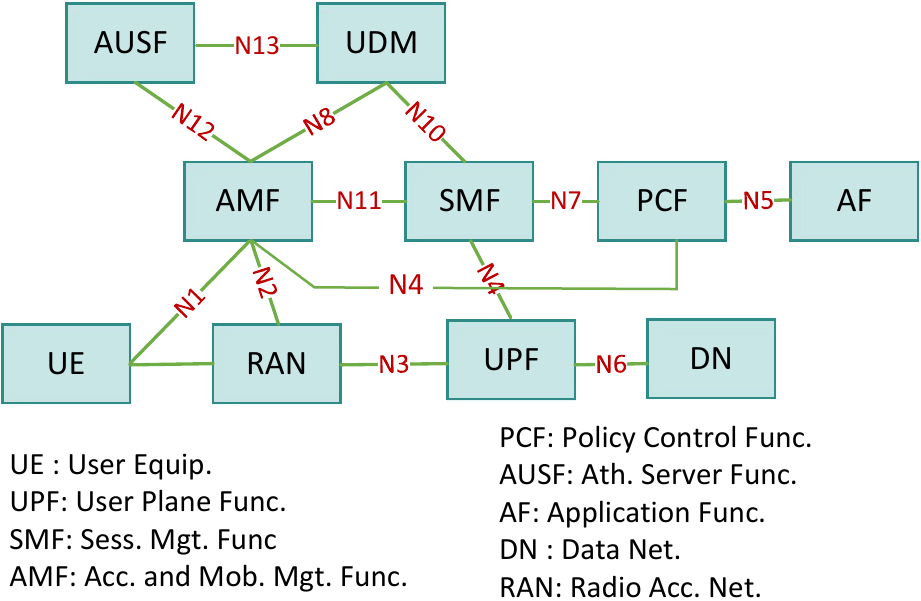}  \vspace{-0.2cm}
\caption{Illustration of 3GPP Next Generation (NG) Core:
Point-to-point reference architecture based on service functions to
support 5G radio nodes~\cite{kim20173gpp}. }  \label{fig_5g_ng_core}
\end{figure}
\paragraph{Overview of NG Core Architecture}
The 3GPP Next Generation (NG) core~\cite{kim20173gpp} is equivalent to
the LTE Evolved Packet Core (EPC).  However, the NG core network has
been redesigned to separate and isolate the network nodes based on
service functions, i.e., functions related to the radio service, such
as user authentication and session management. While the EPC core
provides the network functionality for the LTE backhaul, the NG core
specifically provides the backhaul for the standalone 5G New Radio
(NR) technology~\cite{3GPPNR}. A non-standalone 5G would operate in
coexistence with EPC and LTE support.

The existing EPC core collectively implements the LTE radio service
functions in a combined fashion within the backhaul network gateways,
such as S-GW and P-GW. In contrast, the NG core separates the service
functions at the network nodes level.  The service function concept is
akin to Network Function Virtualization (NFV) in that multiple
virtualized network functions are needed to implement a single service
function.

\paragraph{NG Core Elements}
The point-to-point NG core architecture is based on service functions
supporting the 5G radio nodes, as show in Fig.~\ref{fig_5g_ng_core}.
The fundamental motivation of the NG core is to support advanced
network implementations and network management schemes, such as
network slicing, NFV, network
service function chaining, and SDN to address the scalability and
flexibility of the core network. Each NG core element is connected to
other elements through Nx interfaces. Critical NG core elements
include:
\begin{itemize}
\item[i)] The Access and Mobility Function (AMF) implements the access
  control and mobility aspects of the user context.
\item[ii)] The Session Management Function (SMF) is responsible for the
  data path setup and tracking and terminating based on the policy
  function.
\item[iii)] The User Plane Function (UPF) defines
the data path characteristics based on the users requirements and policy.
\item[iv)] The Policy Control Function (PCF) controls the user policy,
  such as roaming and network resource allocations, for network
  management, including network slicing.
\item[v)] The Unified Data Management (UDM) manages the subscriber
  information which is used for admission control and for defining the
  data path policies.
\item[vi)] The Network Repository Function (NRF) maintains the registry of
  service functions distributed throughout the network.
\end{itemize}

\paragraph{Summary and Lessons Learnt}
The NG core decouples the network service functions from the gateway
nodes, allowing the core network to implement the network nodes based
on service functions, which enhances the deployment flexibility.  As a
result, operators have more freedom in transitioning from an existing
core network to the NG core by separating the core network elements
based on the service functions.  However, future research needs to
thoroughly examine the overhead of the control plane management, e.g.,
virtualization~\cite{hab2018auc}. For instance, the overhead directly
influences power consumption, and network efficiency for the ULL flow
setup in the core network data path, which must be carefully
evaluated. Therefore, performance, resource utilization, and overhead
must be considered while designing the optimal infrastructure
deployment.

\subsubsection{Discussion on 5G ULL Standardization}
In this section we have provided a brief overview of key components in
the 5G standardization efforts that contribute to ULL connectivity.
Several wireless connectivity and signalling optimizations have
reduced the latency overhead in the data and control planes of the
wireless air interface.  Also, the new Radio Resource Control (RRC)
inactive state reduces the signalling for the RRC inactive to active
state transition (compared to the conventional LTE RRC idle to RRC
active transition).  A wide variety of options, e.g., functional
splits of CPRI and NGFI for the fronthaul, exist for meeting the
requirements of 5G components.  Therefore, the design of an end-to-end
5G supported system requires a comprehensive latency analysis across
all segments to select the right candidate set of transport
mechanisms, protocols, and architectural solutions.

Broadly speaking, the improvements that the TSN standards bring
  to bridged networks can feed into novel standard developments for
  Ultra-Reliable and Low Latency Communications (URLLC) in cellular
  networks in two main areas: $i$) backhaul network, and $ii$)
  fronthaul network.  In traditional cellular networks, the various
  backhaul network nodes, such as the Home Subscriber Service (HSS)
  and the Radio Network Controller (RNC), are typically interconnected
  by bridged networks. The adoption of TSN improves the capabilities
  and enhances the performance of the bridged networks that
  interconnect the backhaul nodes.  In contrast, fronthaul nodes, such
  as the Remote Radio Head (RRH) and the Cloud-RAN (C-RAN), were
  typically interconnected by point-to-point optical links (as opposed
  to the bridged networks) as the fronthaul interconnections have very
  strict latency and throughput requirements. The introduction of TSN
  enables bridged networks to provide the strict latency and
  throughput requirements needed for the fronthaul. Thus, TSN can
  enable the end-to-end URLLC support across both the fronthaul and the
  backhaul for cellular networks.

Overall, the adaptability of each solution for 5G deployment could
impact the end-to-end ULL flow latency.  Flexibility could improve the
scalability and network utilization, but the control plane separation
requires careful consideration of control plane overhead and
latency. Similarly, deployments of new architectures, such as NG core,
could result in efficient backhaul management to support ULL mechanism
with minimal overhead, but may require high expenditures for cellular
operators. Nevertheless, as deployment options vary widely based on
the implementation, relative performance evaluation based on distances
between different nodes, interfaces, protocol overhead, transport
mechanisms, and architectural consideration need to be conducted in
future research as ground work towards optimal 5G system design.

\subsection{5G ULL Research Studies}   \label{sec:5g:res}
\begin{figure*}[t!]
	\footnotesize
	\setlength{\unitlength}{0.10in}
	\centering
	\begin{picture}(33,33)
	\put(8,33){\textbf{5G ULL Research Studies, Sec.~\ref{sec:5g:res}}}
	\put(-6,30){\line(1,0){40}}
	\put(15,30){\line(0,1){2}}
	\put(-6,30){\vector(0,-1){2}}
	\put(-10,27){\makebox(0,0)[lt]{\shortstack[l]{			
				\textbf{Fronthaul} Sec.~\ref{5g:res:front} \\ \\
				\ Optical Trans. Tech.~\cite{chanclou2013optical}\\
				\ Freq. Domain Wind.~\cite{liu2015experimental} \\
				\ Pack. and Sch. over Eth.~\cite{chang2016impact,hisano2017gate} \\
				\ TDM-PON DBA~\cite{tashiro2014novel} \\
				\ Traffic Statistics DBA~\cite{kobayashi2016bandwidth}				
	}}}
	\put(15,30){\vector(0,-1){2}}	
	\put(10,27){\makebox(0,0)[lt]{\shortstack[l]{
				\textbf{Backhaul} Sec.~\ref{5g:res:back} \\ \\
				\ Int. Fronthaul and Backhaul~\cite{de2015xhaul} \\
				\ mmWave Backhaul~\cite{gao2015mmwave} \\
				\ In-band mmWave Backhaul~\cite{tao2015poi} \\
				\ TCP over mmWave~\cite{pieska2017tcp}
	}}}
    \put(34,30){\vector(0,-1){2}}
    \put(30,27){\makebox(0,0)[lt]{\shortstack[l]{
				\textbf{Network Management} Sec.~\ref{5g:res:netmgt} \\ \\
				\ Int. Fronthaul and Backhaul Arch.~\cite{jungnickel2014software} \\
				\ Optical-Wireless Net.~\cite{vardakas2017towards} \\
				\ SDN based EPC core~\cite{page2016software} \\
				\ Dynamic Gateway Placement~\cite{lakkakorpi2016minimizing}
    }}}	
	\end{picture}  \vspace{-4.5cm}	
\caption{Classification of 5G Research Studies.}
	\label{5G_class:fig}
\end{figure*}
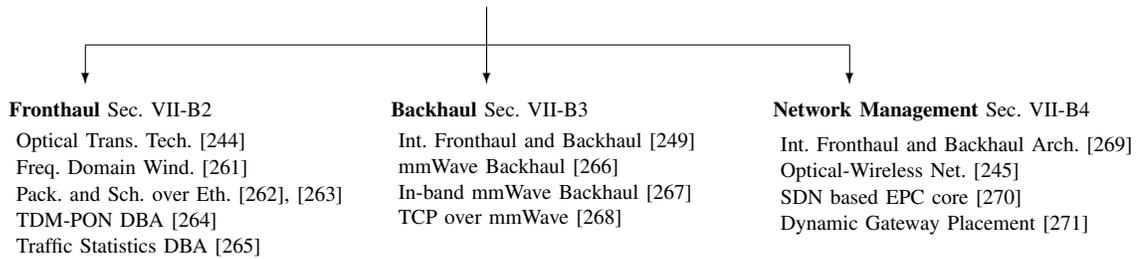

This section surveys the research studies on 5G ULL mechanisms
following the classification in Fig.~\ref{5G_class:fig}.  In particular,
we first give a brief overview of the main ULL research directions in
the 5G wireless access segment and refer to the extensive 5G wireless
access literature for more
details~\cite{agiwal2016next,cai2018mod,dai2018survey,din2017sur,parvez2018survey,sut2018ena}.
Then we survey in detail the research studies addressing ULL in the
fronthaul, backhaul, and network management of fronthaul and backhaul.

\subsubsection{5G Wireless Access ULL Research Studies}
In this section we give a brief overview of the main
research directions on ULL techniques in
the 5G wireless access segment.  Efforts to reduce the latency in the
wireless access segment have been mainly focused on two aspects: $i)$
shortening of the Transmission Time Interval (TTI), and $ii)$ reduced
processing time for each TTI~\cite{Nagata2017}.  The TTI is the
fundamental time unit for the protocol operations, e.g.,
transmissions, in a given wireless technology, e.g., LTE.  A shorter
TII contributes to an overall reduced Round-Trip Time (RTT) due to
shorter cycles. For example, in LTE, the number of OFDM symbols in one
TTI can be reduced from 7 to 2 or 3 OFDM symbol to reduce the
latency~\cite{agyapong2014design}.
In contrast to LTE which uses only Orthogonal Frequency Division
Multiplexing (OFDM) based waveforms, the New Radio (NR) access
technology~\cite{stefan2017NR} for 5G provides a platform to design
and implement more flexible waveforms based on both OFDM and non-OFDM
over a wide range of spectrum resources, including microwave and
mmWave~\cite{bus2018mil}.
\begin{figure}[t!] \centering
\includegraphics[width=3.5in]{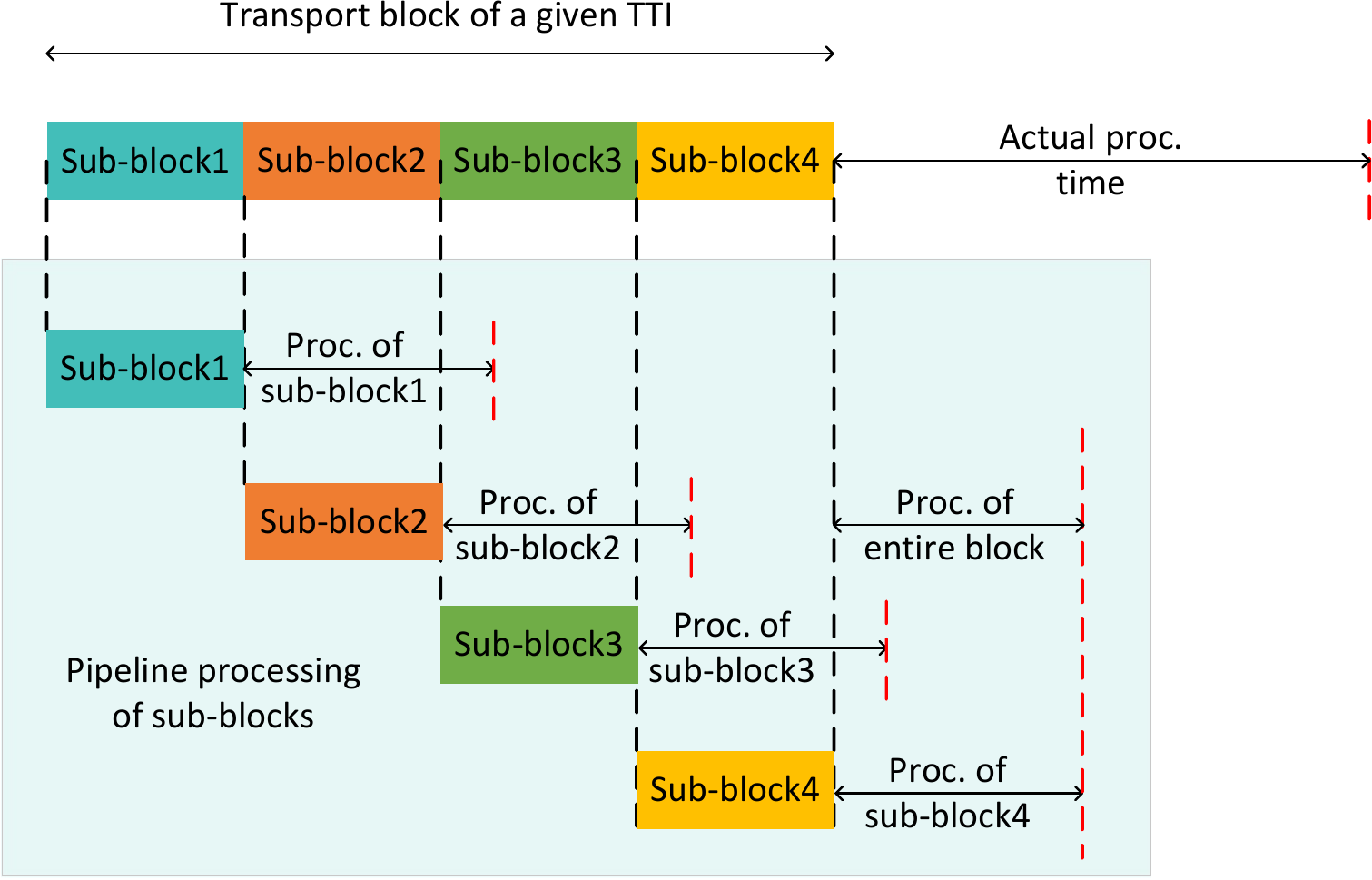} \vspace{-0.2cm}
\caption{A given frame can be divided into multiple sub-blocks.
  Each sub-block is independently processed without having to wait for
  the entire frame to arrive to start the processing, reducing the
  overall latency~\cite{Nagata2017}.}  \label{fig_tti_processing}
\end{figure}

In terms of reducing the TTI processing time, if a given TTI is
divided into multiple sub-blocks, and each block is independently
processed in a pipelined fashion, the overall processing time can be
reduced~\cite{Nagata2017}, as illustrated in
Fig.~\ref{fig_tti_processing}.
The independent processing of sub-blocks
  incurs an overhead in terms of both the physical wireless resources
  (i.e., Resource Element (RE)) mapping and the processing overhead
  for demapping.  The mapping and demapping operations mainly involve
  table lookups and minimal arithmetic computations. Thus, current
  hardware implementation can readily accommodate this mapping and
  demapping processing overhead.
Without pipelined processing, the radio
node has to wait for the entire TTI frame to arrive before starting to
process the symbol, incurring the delay.

Alternatively, the OFDM sub carrier spacing in the frequency domain
can be increased, thus inherently reducing the TTI duration in the
time domain, i.e., reducing the OFDM symbol duration.  However, such
techniques require increased guard bands in both the frequency and
time domains to protect from inter-carrier and inter-symbol
interferences as well as increased hardware complexity in terms of
tight synchronization and sensitive receiver designs.

The next generation Node B radio node in the context of 5G is often
referred to as gNB; this gNB is equivalent to the eNB in 4G LTE.  For
simplicity, we follow the common eNB terminology to refer to the radio
node in both legacy and 5G technology.  The wireless link latency in
5G networks can typically be attributed to two sources: $i)$ user
plane latency when the User Equipment (UE) is in CONNECTED state
(i.e., active radio link is established between UE and radio node
(eNB/gNB)), and ii) control plane latency when device is in idle state
(i.e., no active radio link connectivity exists).  The user plane
latency in the uplink consists of the delays for the scheduling, and
the UE to eNB transport, including the packet processing.  The
wireless control plane latency consists mainly of the delays for the
state change from IDLE to CONNECTED through a signaling process, such
as PAGING and Random Access CHannel (RACH). With increasing numbers of
devices connecting to 5G networks, robust scheduling mechanisms are
essential to preserve the fairness among all the devices in terms of
latency and data rate.  Intermittent data generation, e.g., in IoT,
increases the control plane signaling due to the IDLE to CONNECTED
transitions~\cite{Meng2016}.  Furthermore, in small cell environments,
the device mobility, e.g., for automotive and industrial robot
applications, can result in additional data and control plane
delays. The additional data plane delay in mobility scenarios is
associated with the wireless link discontinuity during the handover
process. Whereas, the control plane delay in the mobility scenarios is
associated with the signaling over the core network due to device
transitions between eNBs.

Robotic systems in industrial networks require ULL for control system
loops.  As compared to unlicensed wireless access (e.g., WiFi), the
licensed LTE and 5G technologies not only provide ultra reliable and
low latency connectivity for a closed ecosystem of industrial
networks, but also support seamless mobility for robotic
systems~\cite{ashraf2016ultra}.  The scheduling of data from the
devices is a MAC layer procedure which incurs significant delays in 5G
wireless networks. To address the scheduling delay, pro-active
granting, similar to Semi-Persistent Scheduling
(SPS)~\cite{seo2012per}, i.e., periodic grants to device for
transmission, can be employed. However, pro-active granting could
reduce the overall link utilization due to over-provisioning of
scheduling resources. In LTE with 1~ms TTI, the Round Trip Time (RTT)
for a Scheduling Request (SR) and GRANT is at least 4~ms, resulting in
data transmission delays of 8~ms or more. Proactive granting can
reduce the packet delay to less than 4~ms by eliminating the SR and
GRANT procedures.

\subsubsection{Fronthaul} \label{5g:res:front}
The fronthaul segment connects the radio nodes, i.e., radio
transmission nodes, to the radio processing nodes, i.e., radio signal
processing~\cite{pizzinat2015things}.  Typically, radio nodes are
referred to as Remote Radio Units (RRUs) and radio processing nodes
are referred to as Base Band Units (BBUs). Cloud-RAN (CRAN)
technology~\cite{checko2015cloud} centralizes and virtualizes the BBU
functions such that a given BBU can connect to and serve several RRHs.
Initial CRAN designs entirely virtualized the BBU functions and
transported only time domain In-Phase/Quadrature (I/Q) samples to
RRHs. However, the time domain I/Q transport technology was limited by
strict delay and bandwidth requirements that hampered the scalability
of deployments.  Recent CRAN designs feature flexible BBU
function separation between CRAN and RRU to meet scalability and
latency demands~\cite{wubben2014benefits, bartelt2015fronthaul}. While
there exist extensive discussions on fronthaul challenges and future
designs~\cite{alimi2017towards, bjo2018han, chang2017flexcran,
  rostami2017orchestration}, we focus on the key aspects of fronthaul
techniques supporting ULL connectivity.

\paragraph{Optical Transport Techniques}
The Common Public Radio Interface (CPRI)~\cite{de2016overview}, see
Section~\ref{CPRI:sec}, imposes an overall fronthaul link delay limit
of 5~ms, excluding the propagation
delay~\cite{de2016overview}. Typically, the distance between BBU and
RRU is 20~km with a delay tolerance of 100~$\mu$s and a frequency
accuracy within 2~ppm (parts per billion). In addition to the CPRI
requirements, the deployment consideration should also consider the
availability of fiber, cost efficiency, CPRI propagation delay, as
well as administration and management, since fiber providers are
typically different from mobile network providers. The main topology
consideration for deployments are the point-to-point, daisy chain,
multi-path ring, and mesh topologies. Point-to-point links provide
dedicated fiber resources for the fronthaul connectivity, but can be
expensive.  The daisy chain topology allows the fiber resources to be
shared among multiple RRUs; however, a link failure can impact all the
connected RRUs. Multipath ring and mesh topologies provide generally a
better balance between fiber availability, cost, and resilience to
link failures. Fronthaul data can be transported through several
optical transport techniques~\cite{chanclou2013optical}:

\textbf{Optical Transport Network (OTN):} The OTN uses a TDM approach
over a single wavelength which can be extended to multiple wavelengths
through dense wavelength division multiplexing (DWDM).  OTN has
relatively high power consumption, as OTN equipment requires power for
the optical transmissions at both receiver and transmitter.

\textbf{Passive Optical Network (PON):} PONs may provide a cost-effective
option for fiber deployments, if PONs are already deployed for fiber
to the home connections. Recent PON developments~\cite{bid2018fir,mikaeil2017performance, chanclou2018mobile, kani2017solutions,mcg2010sho,xu2017fle}
support both high bit rates and low latencies to meet the fronthaul
requirements. PON technology is also power efficient as compared to
the OTN.

\textbf{Point-to-Point with CWDM:} Point-to-point links with a
wavelength multiplexer for Coarse Wavelength Division Multiplexing
(CWDM) are generally cheaper than an OTN with DWDM.  Motivated by
diverse optical transport options, Chanclou et
al.~\cite{chanclou2013optical} have proposed a WDM optical network
solution to meet the data rates and latency requirements of the CRAN
fronthaul.  Automatic wavelength assignment is enabled by passively
monitoring the RRUs through a self-seeded
approach~\cite{saliou2015self} that considers the bit rates,
latencies, jitter and synchronization, as well as fiber availability
of the CPRI links.

\paragraph{Frequency Domain Windowing}
The general 5G end-to-end latency guideline is 1~ms, while the total
fronthaul link (propagation) delay budget is
200~$\mu$s~\cite{liu2016emerging}.  Consider a 20~km fronthaul link,
then the processing delay (for CPRI signal and protocol processing)
would need to be significantly lower than the link (propagation)
delay, i.e., on the order of a few $\mu$s. The general consideration
for the processing delay in the fronthaul is 5~$\mu$s.  In an effort
to further reduce the processing delay of 5~$\mu$s, Liu et
al.~\cite{liu2015experimental} have designed an optical transport
system supporting the CPRI-equivalent rate of 59~Gbps.  48 LTE RF
signals of 20~MHz each were transmitted through a single WDM channel
with an effective RF bandwidth of 1.5~GHz. The processing delay was
reduced through a Frequency Domain Windowing (FDW) technique that
reduces the overall FFT/IFFT size in the process of channel
aggregation and de-aggregation. FDW is applied to each $N$-point IFFT
corresponding to every aggregated channel. The FDW technique
attenuates the high-frequency components such that the inter-channel
crosstalk is reduced. As a result, the effective FFT/IFFT size can be
reduced, thereby reducing the overall processing latency. The
experimental results for the fronthaul distance of 5~km have shown an
overall fronthaul delay reduction from 5~$\mu$s to 2~$\mu$s.

\paragraph{Packetization and Scheduling over Ethernet}
Similar to optical transport of I/Q data from BBU to RRH, I/Q data can
be digitized and packetized for the transmissions over Ethernet. Radio
over Ethernet
(RoE)~\cite{al2017scheduling,al2017mod,ass2016swi,ass2017eth} defines
the process of converting radio signal I/Q data to packets which can
be transported over Ethernet.  The main issues associated with the
packetization process while encapsulating the I/Q data over the
fronthaul link are: $i)$ overhead, $ii)$ packetization latency, and
$iii)$ scheduling delay.  The packetization overhead results from the
frame and packet headers. Therefore, to reduce the overhead, each
frame must be created with the maximum I/Q data possible such that the
overall number of packets and Ethernet frames is minimized. However, a
large frame size adds wait time for the data filling up the maximum
frame size.  Hence, reducing the latency requires the transmission of
short frames.

The scheduling of Ethernet frames can provide multiplexing gain
through resource sharing, however, the scheduling can incur queuing
delays.  Therefore, to achieve low latency the overhead, packetization
latency, and scheduling delay must be carefully considered.  Chang et
al.~\cite{chang2016impact} have evaluated the CRAN performance in
terms of packetization and scheduling on the Ethernet fronthaul.  For
functional splits along layer boundaries, for instance when the
complete PHY layer is implemented in the RRH, or the complete MAC and
PHY layers are implemented in the RRH, an RRH Ethernet gateway has
been introduced to perform the scheduling, aggregate the traffic from
RRH nodes, and discard the packets which are past their deadlines.
For instance, look-ahead depth packetization packs channel estimation
I/Q data such that the channel estimation data precedes regular
payload data in the demodulation.  That is, demodulation does not wait
for all the frame I/Q data to process the I/Q data related to channel
estimation.

In contrast, the prefetch method~\cite{chang2016impact} waits
uniformly over all the I/Q data for the packetization to receive the
Reference Signal (RS) symbols consisting of I/Q for channel
estimation.
More specifically, the packetization process is performed for
  transporting the I/Q data to the base band processing module only when all
  the required I/Q symbols corresponding to the RS within the
  look-ahead depth buffer have been received. Thus, transporting the
  I/Q data needed for the channel estimation has priority as compared to
  regular I/Q data.
Various scheduling policies were applied to study the impact of the
packetization process based on first-come, first-served (FCFS),
shortest processing time (SPT), least remaining bit (LRB), earliest
due date (EDD), and least slack time (LST). The performance analysis
evaluated the maximum number of RRHs supported over the RRH link for a
given Ethernet link capacity, packet size, scheduling policy, and
functional split. The simulation results showed that packetization
techniques (e.g., look-ahead depth and prefetch) while employing the
LRB scheduling policy with packet discarding provided a significant
multiplexing gain and supported the maximum number of RRHs.
In a related research effort, Hisano et al.~\cite{hisano2017gate} have
adapted the gating mechanism (see Section~\ref{802.1Qbv:sec})
to support low-latency 5G fronthaul.

\paragraph{TDM-PON Dynamic Bandwidth Allocation}
\begin{figure}[t!]  \centering
	\includegraphics[width=3.5in]{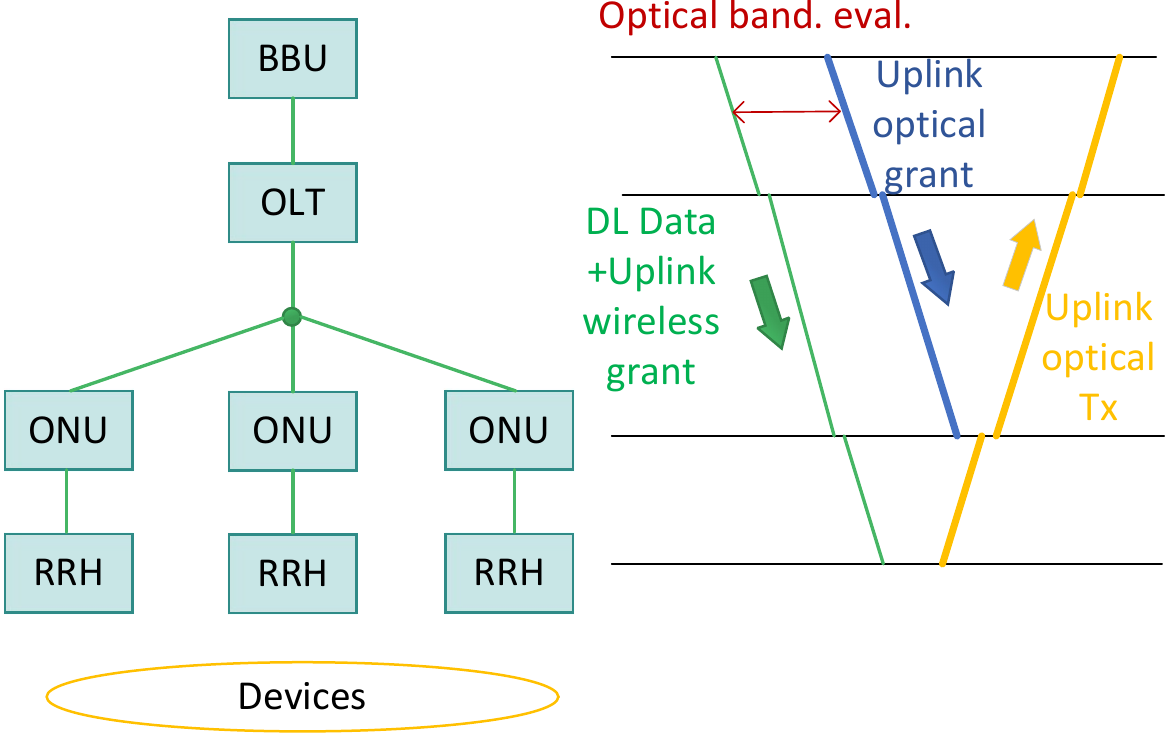}
	\caption{DBA scheme optimizing latency: Grants for the optical
		transmissions are evaluated in advance and sent to ONUs based
		on the wireless uplink information which is known to
		the BBU~\cite{tashiro2014novel}.}
	\label{fig_fronthaul_dba}
\end{figure}
In a PON system, distributed Optical Network Units (ONUs) connects to
a central Optical Line Terminal (OLT) via a shared optical fiber. The
transmissions from the ONUs to the OLT are controlled by a scheduler
implemented at the OLT. In a TDM-PON system, the OLT coordinates the
transmissions from multiple ONUs such that there are no collisions on
the shared fiber.  The Dynamic Bandwidth Allocation (DBA) mechanism
assigns the transmission resources to ONUs based on the QoS deadlines.
For each DBA polling cycle, each ONUs transmits a REPORT message
indicating the queue size to the OLT. The OLT processes the REPORT
messages from all ONUs to determine the transmission schedule. The
transmission schedule is then sent to all the ONUs with GRANT messages
indicating the exact transmission details for each specific ONU. This
polling DBA mechanism consists of reporting the demands and waiting
for the grants from centralized scheduler; therefore, typically, the
total end-to-end PON delay is on the order of
milliseconds~\cite{aurzada2011capacity,mer2013off}, i.e., much higher than the
fronthaul requirements of a few micro seconds.
A PON system in the CRAN framework connects the RRHs to ONUs, and the
BBU to the OLT. Thus, the BBU can schedule transmissions from the
RRHs. Due to the PON delay characteristics, the PON system is not
readily suitable for fronthaul application.

To address the PON delay, Tashiro et al.~\cite{tashiro2014novel} have
presented a novel DBA mechanism specifically for fronthaul
applications.  As the BBU assigns the grants for wireless upstream
transmissions of the devices attached to an RRH (i.e., ONU), the RRH
upstream bandwidth requirements are known at the BBU (i.e., OLT) ahead
in time.  In wireless LTE systems, the request reporting to grant
reception (related to wireless scheduling) is separated by 4~ms in the
protocol operations, similarly the grant reception to RF transmissions
is separated by 4~ms. Hence, the total protocol delay from request to
transmission is 8~ms. As illustrated in Fig.~\ref{fig_fronthaul_dba},
concurrent to the grant evaluation for wireless transmissions, grants
for the optical transmissions of the RRHs (i.e., ONUs) can also be
evaluated and transmitted to the RRHs ahead of time, eliminating the
report and grant cycle between ONUs and OLT. The experimental
evaluation of a TDM PON system with advance scheduling has demonstrated
average end-to-end latencies of less than 40~$\mu$s, and packet
jitters of less than 25~$\mu$s for fronthaul distances up to 20~km.

\paragraph{Traffic Statistics Based Bandwidth Allocation}
Fixed Bandwidth Allocation (FBA) can address the overhead and
scheduling delay incurred by the DBA mechanism, but fixed bandwidth
allocations may waste resources due to over provisioning. For variable
traffic, statistical multiplexing can be employed to increasing the
bandwidth and resource utilization. Based on this principle, Kobayashi
et al.~\cite{kobayashi2016bandwidth} have proposed a TDM-PON bandwidth
allocation scheme based on the traffic statistics of the variable
fronthaul traffic.  The proposed scheme considers the long term
traffic characteristics on the order of several hours. The allocated
bandwidth is then adapted based on the estimated long term mean and
variance,
which can, for instance, be obtained through monitoring
  the packet traffic with software defined
  networking based techniques~\cite{gon2015tow,liu2017cou,yas2015sof},
  the bandwidth allocation requests~\cite{luo2005lim,zhu2008ipact}, or
  monitoring the optical signal levels~\cite{his2017tdm}.
The estimated bandwidth allocation is applied over the
subsequent time period, and a new bandwidth allocation is estimated
for each time period.  The experimental results demonstrated
end-to-end fronthaul latencies of 35~$\mu$s, while the effective link
bandwidth utilization was increased by 58\% compared to FBA.

However, one of the shortcomings of the proposed bandwidth allocation
based on traffic statistics is that it does not consider the specific
fronthaul split option.  For a traditional CRAN, where the RF I/Q
samples are transported from RRH to BBU, a constant bit rate is
required at all times; thus the FBA can efficiently meet the fronthaul
requirements.  Traffic variations according to varying user activity
occur only for higher order functional CRAN splits.  Therefore,
traffic statistics based bandwidth allocation is limited to higher
functional split fronthauls with a split position towards the upper
end of Fig.~\ref{fig_5g_cpri_split}.

\paragraph{Summary and Lessons Learnt}
In a typical CRAN deployment where the RF I/Q is transported
from RRH to BBU, the fronthaul traffic is independent of
the user data which results in a constant bit rate over fronthaul links
at all times to support the normal operations of BBU and RRH.
Hence, there can be significant power consumption overhead
for the CRAN deployment~\cite{carapellese2014energy, tan2017energy}.
Therefore, the new designs of fronthaul solutions should consider the
overall energy consumption in addition to the end-to-end
latency~\cite{wu2018inf}.
Several advanced physical layer techniques, such as,
modulation, detection, and DSP (e.g., I/Q compression) for fiber
transmissions have been proposed as part of energy efficient
designs~\cite{liu2015demonstration, nguyen2017energy, thyagaturu2017r}.
While the higher order functional splits
provide statistical multiplexing gains,
the worst-case delay must be analyzed to ensure that latency is within the
delay budget of the fronthaul link. The fronthaul infrastructure is typically
non-flexible and must support the deployments of future 5G
networks~\cite{gomes2015fronthaul}. Therefore, the fronthaul designs,
such as bandwidth allocation and resource sharing mechanism designs, should be
able to readily accommodate new developments in the 5G technology.
Although several techniques exists to mitigate the delay in fronthaul
networks, there has been no research yet to address the synchronization
of RRH and BBU to a universal timing.
Flexible fronthaul techniques can be developed based on reconfigurable
network functions and physical layer entities, such as
modulators and transparent spectral converters, in the framework of Software
Defined Optical Networks (SDON)~\cite{mao2017dsp}.
For instance, Cvijetic et al.~\cite{cvijetic2014sdn}
have proposed an SDN based topology-reconfigurable optical fronthaul
architecture. The dynamic reconfiguration of fronthaul can support
low latency inter BS communications necessary for
bidirectional Coordinated MultiPoint (CoMP) for inter-cell
interference cancellation and inter-cell D2D.

\subsubsection{Backhaul} \label{5g:res:back}
\paragraph{Integrated Fronthaul and Backhaul}
The backhaul connects Radio Access Networks (RANs) to core networks,
e.g., the LTE backhaul connects the RAN eNB node (base station) to the
Evolved Packet Core (EPC) core network.  Typically, in CRAN
technology, the RRH only implements a split part of the eNB functions,
for instance, the eNB PHY layer is implemented at the RRH, while the
MAC and higher layers are implemented at the BBU. Thus, the RRH, the
BBU, and the fronthaul connecting them, jointly constitute an eNB.
Thus, if the endpoints of a link in a 5G network are the RRH and BBU,
then the link operates as a fronthaul. On the other hand, if the
endpoints are the eNB and EPC, then the link operates as a backhaul.
With the centralization of the computing in the core network, such as
in a CRAN, the BBU and EPC can be implemented at a single physical
location which enables the deployment of a common infrastructure in an
architecture to support both eNBs and RRHs over a common platform.

The crosshaul (Xhaul) architecture~\cite{de2015xhaul} provides a
common platform to support both fronthaul and backhaul using an Xhaul
transport network. In the SDN framework, the Xhaul transport network
provides reconfigurability while operating over heterogeneous switches
and links, such as microwave, mmWave, optical, and high speed
Ethernet.  In an effort to ensure the ULL capability of configurable
integrated fronthaul and backhaul networks, Li et al.~\cite{li2017x}
have proposed an X-Ethernet based on Flexible
Ethernet~\cite{FlexE2017} technology for the Xhaul architecture.  The
experimental demonstration of X-Ethernet has demonstrated an average
latency of 640~ns as compared to 30--50~$\mu$s in a traditional
Ethernet switch, indicating that X-Ethernet can be deployed as a part
of the Xhaul data plane. As the control plane latency of X-Ethernet
for reconfigurations has not been identified, the overall suitability
of X-Ethernet for Xhaul needs further investigations.

\paragraph{MillimeterWave (mmWave) Backhaul}
MillimeterWave (mmWave) radio technology for wireless communications
operates in the spectrum between 30 and
300~GHz~\cite{bus2018mil,mez2018end,wan2018mil}.  mmWaves have
relatively short wavelengths and thus suffer pronounced signal
attenuation with propagation distance and due to obstacles. Also,
mmWaves exhibit high directionality.  Therefore, mmWave technology
exploits beamforming by focusing the signal energy in a narrow spatial
beam to support longer propagation distances.
Nevertheless, the
  typical operational range of mmWave links is in the range of several
  hundred feet. Longer distances require several intermediate
  repeaters which increase the latency.  On the positive side, the high
attenuation property of mmWave signals facilitates geographical
frequency reuse; thus saving the operators spectrum resources by
avoiding co-channel interference.

The availability of high bandwidths in the mmWave spectrum can
  provide high capacity links which are potentially suitable for both
  fronthaul and backhaul.  To date, mmWave research in the context of
  5G networks has mainly focused on the
  backhaul~\cite{deh2014mil,tao2015poi} and we survey the mmWave based
  techniques that specifically target ULL transport. Generally, the
  latency requirements in the backhaul are relaxed compared to the
  very strict latency requirements for the I/Q user data transport in
  the fronthaul. Thus, mmWave transport with its required repeaters
  for covering distances beyond a few hundred feet is generally better
  suited for backhaul. Future research may examine whether it is
  possible to exploit the high capacity mmWave transport for
  fronthaul.  Also, mmWave transport may be suitable for particular 5G
  connectivity scenarios, e.g., for connecting a Customer Premises
  Equipment (CPE) home gateway to an external serving gateway, e.g., a
  5G base station (gNB).

Gao et al.~\cite{gao2015mmwave} have presented a mmWave based backhaul
for 5G using massive-MIMO to support a high number of radio nodes,
i.e., Base Stations (BSs).  The proposed approach exploits Beam
Division Multiplexing (BDM) whereby an independent beam is dedicated
to a BS, thus creating a backhaul link through spatial multiplexing.
Each mmWave beam supports a high capacity link, hence, a Time Division
Multiplexing (TDM) scheduling can be employed to share the resources
within a single beam, supporting multiple BSs over a single
link. However, the scheduling of BDM resources with TDM can incur
significant end-to-end latency as compared to BDM without TDM, and
therefore must be carefully evaluated specific to the backhaul latency
requirements.

\paragraph{In-band mmWave Backhaul}
The in-band mmWave technique shares the spectrum resources with the
wireless access (i.e., BS to device), and backhaul (i.e., BS to BS and
BS to core network).  Since the wireless access and backhaul resources
compete for the same spectrum resources in the in-band communication,
there can be significant overhead in terms of capacity and latency. To
analyze the in-band mmWave communications in terms of capacity, Taori
et al.~\cite{tao2015poi} have conducted a feasibility study and
showed that 25\% of the resources of the mmWave link is sufficient to
support the user data rates over the wireless link up to
0.8~Gbps. Typically, in the in-band backhaul connectivity, the
resources are shared in TDM fashion between wireless and backhaul
applications impacting both wireless and backhaul end-to-end
connectivity during congestion.  Although the suitability of in-band
communication is justified in terms of capacity, the implications of
in-band communication on the latency has not been characterized, and
hence can compromise the performance of the entire end-to-end
connectivity if not carefully considered. Taori et
al.~\cite{tao2015poi} have also proposed a point-to-multipoint
transmission for BS to BS (inter-BS) communication based on in-band
mmWave backhaul connectivity. Inter-BS communication is necessary to
support mobility features, such as handover and redirection, as well
as advanced radio features, such as inter cell interference
cancellation using Coordinated MultiPoint (CoMP) and self organizing
networks.  As the deployments of BS increase to meet the capacity
demands through small cells, the demand for coordination among
neighboring BSs will increase. Hence, inter-BS communication is an
important aspect of 5G that needs be addressed in a flexible, simple
and cost effective manner.  In-band mmWave connectivity provides a
cost effective solution for inter-BS connectivity along with
flexibility due to a wireless connection, as compared to the physical
deployment of optical fiber infrastructure.  Point-to-multipoint
mmWave connectivity results in a simpler and cost effective solution
through a dynamic reconfiguration of mmWave links based on the
requirements.

\paragraph{TCP over 5G mmWave Links}
mmWave links have typically high bandwidths, but are prone to outages
as they require Line-of-Sight (LoS). Thus, there are high chances for
temporary link disruptions, which can result in temporary congestion.
TCP congestion control could negatively impact the overall capacity
and the latency when a link is temporarily interrupted as a result of
buffer bloating. Active Queue Management (AQM) can be applied to
adaptively drop packets from the queue such that the queue size is
contained for a particular flow to keep the end-to-end delay on
average below a threshold.  Control Delay
(CoDel)~\cite{nichols2018controlled} is an AQM technique which ensures
short packet sojourn delays, i.e., short packet delays from ingress to
egress.  Each packet is time-stamped at the ingress and elapsed time
is evaluated for the packet drop decision.
Building on the
  well-known non-linear relationship between drop rate and throughput
  in TCP~\cite{mat1997mac}, the time interval between packet drops is
reduced inversely proportional to the square root of the number of
drops so as to linearly vary the throughput in relation to the drop
count~\cite{nichols2018controlled}.

To investigate the impact of temporary 5G mmWave link disruptions on
end-to-end network connections, Pieska et al.~\cite{pieska2017tcp}
have evaluated the TCP performance tradeoff between
capacity and latency.  The evaluation indicated that the disruption
duration and frequency directly impact the TCP performance in addition
to the aggressiveness of the TCP variant, such as TCP Reno, TCP
Illinois, TCP cubic, and TCP Scalable.  Although CoDel is a promising
technique in curtailing the buffer bloat in regular TCP networks,
Non-LoS (NLOS) occurrences of a mmWave link can result in significant
throughput loss of TCP over mmWave links due to extensive CoDel
packet dropping, especially for a single flow of the TCP Reno variant.
However, the evaluations indicated that CoDel can achieve low latency
and fast recovery for flows with short RTTs and disruption durations.
Nevertheless, to avoid the implications of buffer bloat, new TCP
designs should be able to accommodate short link disruptions,
specifically for 5G mmWave connectivity for access, fronthaul, and
backhaul.

\paragraph{Summary and Lessons Learnt}
Small cells where the devices are close to the radio nodes are widely
adopted to save power and to offload the burden on the macro wireless
cells~\cite{xu2018sur}. However, the small cell traffic needs to be
eventually aggregated at the backhaul, resulting in demanding
requirements for the small cell connectivity with the core networks.
The connectivity can be provided through fiber backhaul links that can
be shared through FiWi techniques among multiple wireless
nodes~\cite{Beyranvand2017}.
mmWave technology is another promising technology for meeting the high
bandwidth and ULL requirements for next generation connectivity, such
as, small cell backhaul supporting 5G, and fronthaul and backhaul
sharing~\cite{kuo2017millimeter}.
mmWave wireless links support
  $i$) high throughputs with short symbol and frame durations, and
  $ii$) high user numbers at a given radio node. Thus, mmWave backhaul
  can increase the overall capacity of cellular networks in terms of
  supported flows with low-latency QoS.
As compared to the power consumption of optical communications, the
power consumption of mmWave links is typically significantly higher
due to the scattering of wireless transmissions as compared to the
guided optical waves in a fiber.  Therefore, mmWave requires new
energy efficient methods in resource management and shared backhaul
and fronthaul for 5G applications.

In contrast, optical wireless
communication~\cite{jungnickel2015optical} utilizes the visible light
with similar characteristics as mmWave. In addition to the
directionality (LoS) and spatial multiplexing properties, optical
wireless communication suffers from interferences due to ambient light
sources.  Similar to mmWave designs, the system design should be
robust to accommodate disruptions due to temporary link
obstructions. Future designs should also ensure synchronization on the
order of 65~ns~\cite{cpri,chitimalla20175g,wan2015per} while
supporting the shared fronthaul and backhaul.

\subsubsection{Network Management} \label{5g:res:netmgt}
ULL mechanisms are closely related to network management for meeting
the flow demands in terms of resource allocation, reliability,
congestion control, and end-to-end QoS. The increasing number of
protocols that support the fronthaul and backhaul connectivity in a
single end-to-end path creates a heterogeneous environment. The
comprehensive (end-to-end) management of this heterogeneous network
environment can be complex without the support of an inter-operative
mechanism. Management mechanisms based on Software Defined Network
(SDN) could provide a single platform for the coordination of
a multitude of protocols~\cite{fer2018lay,jag2015sof,maa2015com,thy2016sdn}.

\paragraph{Integrated Fronthaul and Backhaul Architecture}
Both Distributed-RAN (DRAN) and CRAN offer unique deployment options
for cellular operators to enable cellular connectivity to the users.
DRAN conducts the baseband signal processing at the remote Base
Station (BS). As a result, the BS to core network (backhaul)
connectivity has relaxed QoS requirements and thus can be leased in
the access network domain. On the other hand, CRANs require dedicated
fiber links (typically owned by the cellular operator) for connecting
the radio nodes to the core networks. Therefore, 5G networks are
expected to uniformly support DRAN and CRAN architectures for enabling
cellular connectivity to the users.

Jungnickel et al.~\cite{jungnickel2014software} have proposed an
integrated fronthaul and backhaul based on SDN to commonly support
DRAN and CRAN deployments for cellular operators.  Traditional
Ethernet deployment strategies~\cite{chundury2008mobile}, such as the
E-tree, can be adapted for the CRAN, and the E-LAN for D-RAN based on
their topology support. To utilize the existing fiber, independent
wavelengths can be used to meet the latency and capacity requirements
of the fronthaul and backhaul. For example, the backhaul can use TDM
within a single wavelength that is shared among multiple radio nodes,
and the fronthaul requires a dedicated wavelength between radio node
and CRAN.  However, the sharing of traditional access networks in
E-Tree and E-LAN mode can cause security issues. Nevertheless, SDN
provides both flexibility of statistical multiplexing in both the
optical and electrical domains, and security through the
virtualization of the network infrastructure.  In a similar study,
Ameigeiras et al.~\cite{ameigeiras2015link} have proposed a
hierarchical SDN architecture based on virtualization, as well as
Ethernet and IPv6 technologies focusing on low latency.

\paragraph{Optical Wireless Networking}
\begin{figure}[t!]  \centering
\includegraphics[width=3.5in]{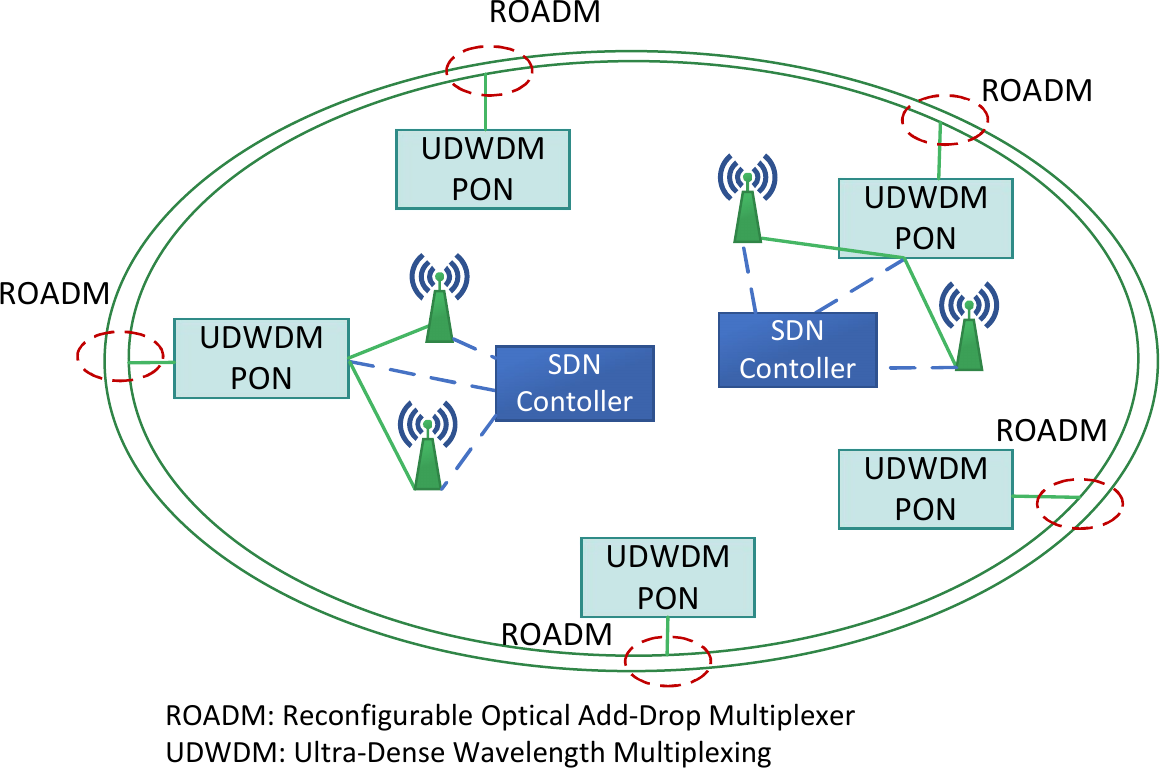}  \vspace{-0.2cm}
\caption{Simplified version of ULL optical wireless architecture where
  WDM ring connects to wireless nodes and SDN controller through PON
  framework~\cite{vardakas2017towards}.}  \label{fig_optical_wireless}
\end{figure}
The inter-working of optical and wireless technologies has been
explored in FiWi networks~\cite{aurzada2014fiwi,che2018fiw,liu2016new}
and in the general context of optical-wireless integration in
access and metro networks~\cite{ahm2012rpr,gha2009sup,mai2003hyb,sch2003wav,yan2004met}.
As next-generation
applications demand ULL and high reliability, there is a great need
to integrate optical and wireless technologies with minimal impact on
the traditional cellular infrastructures, such as 4G LTE. Towards this
end, the 5G STEP-FWD project~\cite{vardakas2017towards} has been
funded by the European Commission to develop novel networking
solutions that closely integrate the optical and wireless technologies
within the 5G framework.

Vardakas et al.~\cite{vardakas2017towards} have proposed a high
capacity and low latency 5G backhaul architecture as illustrated in
Fig.~\ref{fig_optical_wireless}. Network densification is supported by
small cells which are connected to macro BSs through PONs, mainly:
$i)$ Optical Line Terminals (OLTs) connected through fiber links,
$ii)$ point-to-point dedicated links, and $iii)$ local Optical Network
Unit (ONU) connections through a fiber protection ring offered by dark
fiber. The dark fiber utilization provides a cost effective solution
as the infrastructure already exists. The wireless access by the small
cells and backhaul connectivity supported by PONs are controlled by a
unified SDN management framework.  mmWave-UDWDM technology effectively
utilizes the wavelength and space division multiplexing, while PONs
provide effective backhaul connectivity.  The SDN management can
support dynamic reconfigurability to support advanced network
features, such as self-organization and self-healing for
ultra-reliable infrastructure networks.

\paragraph{SDN Based Evolved Packet Core (EPC) Networks}
Pag{\'e} et al.~\cite{page2016software} have presented an SDN
architecture for the LTE Evolved Packet Core (EPC) to support
low-latency towards an evolutionary 5G core network. OpenFlow
technology has been integrated into the switching nodes that connect
the BSs (i.e., eNBs) to the EPC.  The advantages of SDN based
switching include reduced need for protocol based transport services,
such as GTP, elimination of the Serving-Gateway (S-GW) which
conventionally provides flow based services, such as buffering and
connection management. In contrast to the conventional LTE backhaul
connectivity, where the S-GW anchors the connections of the eNBs to
the P-GW, the SDN based EPC is managed by an SDN controller, which
replaces the S-GW control plane functions. The S-GW data plane
functions are replaced by the SDN supported switching nodes. Thus, the
SDN architecture eliminates the data and control bearer based
connectivity~\cite{cox2012introduction} by replacing the large GTP
messages with small SDN control messages.  Additionally, the SDN based
switching nodes can assist in attach and mobility procedures to reduce
the overall load on the EPC core. As a result, the overall end-to-end
latency can be reduced by reducing the data plane and control plane
latency introduced by the intermediate nodes in the EPC core.

\paragraph{Dynamic Gateway Placement}
Lakkakorpi et al.~\cite{lakkakorpi2016minimizing} have proposed a low
latency technique in an SDN based backhaul network architecture that
is fully reconfigurable. The gateway functions and queue management
are configured to achieve low latency by minimizing the flow
reestablishment procedures. The SDN controller dynamically programs
the switching nodes to implement the network functions based on the
flow characteristics. More specifically, an anchor switching node is
dynamically selected to implement the gateway functions and AQM based
on the flow mobility characteristics.  For instance, in case of
frequent handovers, the flow path must often be reconfigured to pass
from one gateway function node to another. Therefore, the gateway
functions can be implemented deeper in the core networks for the
specific flows with frequent handovers, such that only the path
routing is updated during handovers. This implementation of the
gateway functions in the core networks also distributes the gateway
functions across the switching nodes, reducing the overall burden on
the core network.

\paragraph{Summary and Lessons Learnt}
In addition to the optimization of handover latencies in the wireless
access, the backhaul architecture should support lower handover
latencies. Chen et al.~\cite{chen2016efficient} have discussed the
need for efficient backhaul architecture to support ultra-short
handover latencies. However, the discussions are limited to
DBA mechanisms in PONs for optimizing the LTE X2 and S1
interfaces.

In 5G technology, handovers can cause temporary disruptions to large
data flows which can result in buffer-bloat problems across the
network. New congestion control mechanisms must be adapted to
address the short and temporary disruptions due to handovers during
large data transfers.  SDN based strategies can help to address
these challenging handover problems~\cite{yazici2014new}.  However,
existing studies have not considered the control plane latency and
complexity, which may significantly impact the overall end-to-end
latency. Therefore while ensuring the flexibility and reliability in
5G networks, it is also important to consider the end-to-end latency,
through infrastructure based solutions, such as, dense
wavelength-division multiplexed (DWDM) optical ring transport
networks~\cite{wong2017enhancing} using dark fiber, which is both
energy and cost efficient.

\subsubsection{Discussion on ULL 5G Research Studies}
There have been numerous research efforts in the wireless access
segment of 5G networks. However, there is still a need for research to
solve compelling technical challenges~\cite{li20175g} in enabling
ultra-reliable ULL communication. These research challenges include
infrastructure reuse, as well as cost and power efficiency.
Throughout, the implications of wireless access techniques on ULL
  services should be carefully considered.  For instance, the emerging
  5G New Radio (NR) platform proposes new waveform designs. The symbol
  and frame durations as well as the guard band durations (e.g.,
  cyclic prefixes in the OFDM symbol) in these new waveforms would
  directly impact ULL services.  Increasingly complex waveforms would
  require longer symbols and longer frames, not only because of
  limited receiver processing capabilities, but also to maintain the
  synchronous delay between uplink and downlink messages.  Thus,
  increasing the waveform complexity would tend to increase the
  wireless round trip delay.  Moreover, the channel characteristics,
  such as the maximum (mobility) speed of 5G user devices and the cell
  size influence the guard band duration.  For example, a high speed
  train scenario requires a relatively long doppler correction.
  Similarly, rural deployments require large cells.  In both
  situations, a long guard band (cyclic prefix) is preferred such that
  the inter symbol interference can be minimized. A long guard band
  (cyclic prefix) would imply relatively long symbols and frames which
  could negatively affect ULL services. Thus, the new waveform designs
  in the 5G platform should carefully consider the impact on ULL
  services throughout the development process.

With the radio node densification, user mobility between radio nodes
is expected to increase dramatically, which can significantly increase
the control plane complexity in terms of user context updates in the
core networks. Therefore, a light weight (i.e., reduced user context)
user information set must be managed by the core networks, as opposed
to intense policy and security mechanisms that contribute to control
plane complexity.  End-to-end security can reduce the burden of
security measures by the core network. Similarly, user activities can
be tracked by the radio node to enforce the policy and QoS measures
across the network.

SDN plays an important role not only for managing fronthaul, backhaul,
and core networks, but also for reducing the network complexity by
reducing the network function implementation in dedicated entities,
such as policy enforcement and user authentication.  SDN can also
integrate the heterogeneous protocol operations through dynamic packet
header manipulation such that the protocol overheads are
minimized.

Content caching in edge nodes has been widely discussed for reducing
the delivery latency in fog-RAN and edge computing
domains~\cite{Liu2017Cache,sengupta2017fog,kakar2017fundamental,
  radwan2017mobile}. SDN provides a platform for caching content
across the entire network as well as based on user demands, optimizing
both content caching and latency.  Although 5G technology is primarily
focused on power optimization of user devices and wireless radio
nodes~\cite{Miyanabe2015, olsson20135green,sabella2014energy}, the
overall energy consumption of the network responsible for the
end-to-end packet delivery should also be considered in future
designs.

\section{Future Work Directions} \label{fut:res}
In this section we discuss the main open TSN and DetNet research
problems and outline directions for future research efforts in TSN and
DetNet networks.

\subsection{Time Sensitive Networks (TSN)} \label{fut:res:tsn}

\subsubsection{Inter-Scheduler Coordination}
Time aware sharpers implement local scheduling principles specific to
each TSN node.  The end-to-end time sensitive characteristics of a
flow are established under the assumption that each TSN node in the
flow path guarantees the time sensitive characteristics. However, if
an intermediate TSN node fails to enforce the TSN characteristics due
to overload, or due to scheduler or timing inaccuracies, the overall
end-to-end flow characteristics can be compromised.  This situation
may be more likely for TSN nodes that are positioned where multiple
flows can aggregate as opposed to the edge nodes (that are traversed
by only few flows).

To address this shortcoming, future research should develop a robust
inter-scheduler coordination mechanism.  The coordination mechanism
should facilitate interactions between the time aware shapers in the
TSN nodes in a flow path to ensure the overall end-to-end time
sensitive characteristics of the flow. For instance, upon frame
reception at the destination, the overall end-to-end latency can be
estimated and the information can be fed back to the nodes.  The TSN
nodes can then establish a self performance profile.  The interactions
of the time aware shapers would enable inter-scheduler coordination
such that each TSN node can guarantee the time sensitive scheduling
relative to the end-to-end behavior of the flow path similar to
time-triggered scheduling~\cite{meyer2013extending}.

However, time-triggered scheduling depends on time synchronization to
synchronously trigger the scheduling over the entire flow path. In
contrast, the inter-scheduler coordination enables dynamic changes of
the scheduler policies, such as timing adjustments of frame
transmissions (i.e., to delay or advance the transmissions in the
scheduled time slots) correcting the synchronization inaccuracy. Thus,
the time aware scheduler depends not only on the time synchronization,
but also on the end-to-end flow characteristics.  The inter-scheduler
coordination can be enabled through a centralized mechanism.  For
instance, an SDN based control can monitor the end-to-end
characteristics of the flows, and configure the timing advances and
corrections of the time aware schedulers at specific TSN nodes as
required.

\subsubsection{In-band Control Plane Overhead}
Control plane data in TSN network corresponds to the data generated
from the control functions, e.g., for setting up connections,
synchronizing nodes, managing flows, and tearing down connections. The
impact of control plane data in TSN networks has been largely ignored
to date in research and standardization.  Control plane traffic could
be transported with the in-band connectivity of the high priority
Control Data Traffic (CDT) class, which carries time critical
information from data sources, such as sensors.  However, the control
plane traffic would then compete with the CDT traffic.

Resource reservations in TSN networks to enable the deterministic
time-sensitive properties are typically estimated based on CDT traffic
requirements.  Since the control plane traffic rates are generally
significantly lower than the CDT traffic, the in-band control plane
traffic is generally ignored in the system design and resource
reservations.  However, new use TSN cases, such as robotics and
automated drones, may require the establishment of short lived TSN
flows with commensurate frequent triggering of control plane
activities.  Thus, new use cases may significantly increase control
plane data traffic.  Therefore, new resource reservations designs,
especially for the in-band control plane data transport should
consider both the control plane data traffic as well as the CDT
traffic in evaluating the resource reservation requirements.  We
anticipate that it will be particularly challenging to ensure the
requirements of the varying and dynamic control plane data as compared
to the steady CDT traffic.

\subsubsection{Low Priority Deadline Traffic}
TSN nodes preempt an ongoing low priority frame transmission for
transmitting an incoming high priority frame to guarantee the absolute
minimum TSN node transit delay of the high priority frame.  Depending
on the intensity of the high priority traffic, a low priority frame
can be preempted several times. As a result, the end-to-end delay
characteristics of the low priority traffic cannot be guaranteed as
the preemption occurrences depend directly on the high priority
traffic intensity. If the high priority traffic intensity is
significantly higher than the low priority traffic intensity, then
the end-to-end delay of the low priority traffic can be greatly increased.
Generally, low priority traffic carries delay sensitive data, that is
less critical than high priority traffic data, but still should be
delivered within a worst-case deadline. In the current state of the
art, there exists no mechanism in research nor standards to ensure the
worst-case end-to-end delay of low priority traffic under preemption.

Therefore, future research needs to develop new mechanisms to ensure a
bounded worst-case delay for low priority traffic in TSN networks.  A
key challenge in designing a bounded worst-case delay for low priority
traffic is to not degrade the performance of high priority traffic.
Rather, the new mechanisms should opportunistically accommodate low
priority traffic transmissions to meet a worst-case deadline.

\subsubsection{Impact of Synchronization Inaccuracy}
Several techniques for improving the synchronization accuracy while
minimizing the synchronization errors have been developed for TSN
networks.  However, there is a lack of studies that quantify the
implications of synchronization inaccuracies on the TSN network
performance in terms of end-to-end delay and throughout.  For low cost
devices which are typically employed in large scale networks and for
remote applications in IoT scenarios, the synchronization may not be
as accurate as for industrial and robotic applications.  Due to
synchronization errors in TSN nodes, the transmissions scheduled by
the time-aware shaper over a particular time slot, can extend or
advance to adjacent time slots, which can impact the overall
scheduling mechanism in a TSN node. For instance, in a time-triggered
network, where all the TSN nodes schedule a flow based on synchronized
timing information, synchronization errors can offset the
time-triggers which can miss the schedule of a very short frame
depending on the timing offset duration. Therefore, the performance
impact due to synchronization errors for multiple priority traffic
classes, frame sizes, and timing offset durations requires a close
investigation.

\subsubsection{Ingress and Egress Nodes for TSN}
TSN networks are typically implemented in closed environments, such as
automotive and industrial environments. However, most use cases
require external connectivity to inter-operate with other networks.
So far, no mechanism exists for establishing a common platform for the
inter-operation of TSN networks with external non-TSN networks.
We envision the inter-operation of TSN networks with non-TSN networks
in two ways: $i)$ centralized SDN management, and $ii)$ ingress and
egress based management for the TSN network.  In case of the
centralized SDN, a TSN flow outside the TSN network can be
distinguished and apply for resource reservations to ensure the delay
sensitive characteristics. In case of ingress and egress based
management, an outside flow that requires TSN properties while
traversing through a TSN network can be identified and configured over
the entire flow path such that the end-to-end flow integrity is
preserved.

\subsubsection{TSN Performance for 5G Fronthaul Applications}
Fronthaul networks transport the highly delay sensitive
In-phase/Quadrature (I/Q) symbol information between the central base
band processing units and the remote radio heads.  Therefore, typical
deployments prefer optical fiber to establish high capacity and low
latency links. Although traditional Ethernet can meet the capacity
requirements, delay requirements are challenging to achieve with
Ethernet networks. However, due to time sensitive properties, TSN
Ethernet is being considered as a potential candidate L2 protocol for
5G fronthaul applications as an alternative to the Common Public Radio
Interface (CPRI) and eCPRI~\cite{de2016overview} protocols. The
adoption of TSN for existing Ethernet infrastructures could result in
significant capital and operating expenditures for new fiber
deployments.  But, the actual performance of TSN networks for
fronthaul applications has not yet been investigated for the various
fronthaul splits~\cite{chang2017flexcran}. The PHY and sub-PHY splits
require strict deadlines on the order of sub-microseconds.  On the
other hand, function splits in the MAC, Radio Resource Control (RRC),
and higher cellular protocol layers relax the delay requirements to
the order of milliseconds. A comprehensive performance evaluation
considering the full range of aspects of fronthaul applications, such
as relative performance between Ethernet Passive Optical Networks
(EPONs) and TSN Ethernet, packetization, functional split, and
fronthaul distances for a Cloud Radio Access Network (CRAN) system
could provide deep insight towards deployment considerations for
mobile operator networks.
The ULL requirements
of a wide range of 5G wireless network applications and services
have been extensively documented,
see e.g.,~\cite{Abbas2017,Ahmad2017lte,ashraf2016ultra,
	Beyranvand2017,Carvajal2016,chen2015techniques,Choudhury2016,
	Condoluci2017,durisi2015towards,Durisi2016,Dutta2017,Fan2017,
	Lee2017,Lema2017,li20175g,Liu2017Cache,Luvisotto2017,Lu2017Ortho
	,Maier2016,Meng2016,Miyanabe2015,mogensen2014centimeter,Nagata2017,
	patel2015traffic,Pflug2013,pilz2016tactile,ploder2017cross,
	Rimal2017,Salem2011,She2017,Simsek2016,Vu2017,Wei2000,
	Wu2015enable,Zhang2017rfg}.
Thus, there is an extensive need to research latency reductions for 5G
wireless networks.  Investigating the combined impacts of the various
latency reduction techniques developed in future 5G wireless network studies
in conjunction
with TSN based fronthaul is an important direction for future
research.

\subsubsection{TSN Applied to Wide Area Networks}
The time-sensitive protocol mechanisms that are applied to
micro-environments, such as automotive networks, can also be applied
to macro-environments, such as Wide Area Networks (WANs).  In most
situations, the end-to-end network delay is dominated by the wait time
in the queues (buffers) of intermediate forwarding nodes. With the TSN
rules applied to nodes, the overall end-to-end delay of a flow over a
WAN network can be significantly reduced. However, WAN networks
typically handle large numbers of flows and operate at very large
capacities, making the TSN flow management very challenging. Despite
these challenges, WAN networks should, in principle, be capable of
supporting TSN characteristics for specific flows that require strict
end-to-end latency bounds, such as remote surgery in health-care
applications, where a doctor could operate on a patient across a WAN
network.  One possible approach to handle the challenging flow
management could be through SDN based control. The large geographical
WAN area would likely require an SDN control hierarchy consisting of
multiple control plane entities, such as, local and root controllers,
as well an orchestrator.

\subsection{Deterministic Networking (DetNet)}

\subsubsection{Packet Replication and Elimination}
Packet replication inherently increases the flow reliability by
increasing the probability of packet delivery to the end destination.
Additionally, packet replication can reduce the overall end-to-end
latency due to disjoint paths~\cite{Armas2016}.  However, a major
disadvantage of packet replication is the increase in the effective
bandwidth required for a flow. The required bandwidth can be decreased
by reducing the degree of replication, which can effectively reduce
the reliability.  Thus, a balance between bandwidth and degree of
packet replication must be ensured to operate the network within the
required bandwidth (capacity) and latency limitations.

Towards this end, we propose a reverse packet elimination mechanism in
which the destination node triggers an instruction to the nodes in the
reverse path to apply a packet drop action.  For instance, consider
the forward direction of a flow with four disjoint paths, i.e., each
packet is replicated four times.  These replicated packets traverse
independently across the disjoint paths through the network to reach
the common end destination. We can assume that one packet will arrive
earlier than the others, considering that multiple packets will likely
arrive at the destination. In the current implementation, the other
packets are discarded when they eventually arrive at the
destination. Thus, the effective bandwidth is four fold increased in the
forward direction.

In a DetNet/SDN framework, the destination node can be made aware of
the exact nodes traversed by the different paths.
That is, for a given path with node 0 denoting the source node and
node $n$ denoting the destination node, the destination node knows
the  $n-1$ intermediate nodes. If there is sufficient bandwidth in
reverse direction, the destination node can send a short drop-packet
in the reverse direction on paths through which the destination nodes has
not yet received the packets; upon reception drop-packet, the
intermediate nodes drop the forward packet.  This drop-packet would
traverse backwards through the nodes $n-1$, $n-2$, $\ldots$ towards the
source node while applying the rule
to drop until the drop-packet meets the forward packet.  Thus, because
of the reverse back propagation of the drop-packet, some of the
forward direction bandwidth is freed up.
In many networking scenarios, the ratio of uplink traffic to downlink
traffic is low, and therefore the uplink can typically readily
accommodate the reverse back propagation of the small drop-packet
notifications.
Future research would need to conduct a rigorous performance study of
the proposed drop-packet approach for a wide range of network
conditions, such as number of flows, relative delay in diversity
paths, and numbers of intermediate paths.

\subsubsection{Virtualization: L2 Independent Mechanisms}
Although DetNet focuses on the network layer (L3) and higher layers,
DetNet relies on the time sensitive link layer (L2) to establish the
deterministic L3 packet flow properties. Therefore, promoting DetNet
mechanisms which are independent of the time sensitive link layer
could result in the wide adoption of DetNet due to the simple and
cost-effective infrastructure support.  For instance, packet
replication and fragmentation do not require timing information and
can be implemented independently of the link layer. One way to achieve
independence from the link layer is through Network Function
Virtualization (NFV), which can dynamically scale the resource
reservations based on the flow demands. However, such NFV mechanisms
would require hypervisor and control plane
management~\cite{afo2018net,blenk2016survey,ble2016con,era2017app,yi2018com}. NFV
also provides a platform for centralized control plane management
through the SDN framework. Thus, through a unique combination of SDN
and NFV, DetNet can be independently adapted to networks without time
sensitive link layer properties.

\subsubsection{Inter-networking}
The DetNet inter-networking with an external network (i.e., a
non-DetNet network) is still an open issue.  Generally, a DetNet
requires an centralized in-domain controller to establish an
end-to-end placket flow. Therefore, if an external flow needs to
traverse a DetNet network, the flow requirements must be configured
within the DetNet network.  Ingress and egress nodes could be
introduced to manage the configurations for the incoming and outgoing
external flows.  In particular, an ingress node could perform
admission control to make flow accept/reject decisions.  The
ingress node would then also track and manage the packet flows. During
this process, the ingress node could cooperate with the DetNet centralized
control entity, e.g., a Path Computing Element (PCE), to accomplish
the flow setup over the DetNet network.  Thus, the cooperation between
ingress node and PCE would enable the inter-networking of DetNet
and non-DetNet networks. Of particular importance will be
the study of interactions with data center networks.
Latency reduction techniques for data center networks have
received increasing attention in recent years, see for
instance~\cite{alizadeh2012less,alizadeh2014conga,Avci2016,Berisa2013low,cao2017optimal,fujiwara2015swap,Guan2014,Guo2013high,Guo2014,
	he2017low,Khabbaz2017,Kumbhare2015,Liu2012packet,Liu2014inorder,
	Liu2015low,Liu2016delay,liu2017eba,lockwood2015implementing,miao2014low,miao2015sdn,miao2016towards,Perello2013,
	Rezaei2016,saridis2016lightness,shpiner2016race,stephens2014practical,teymoori2016even,Wang2016adap,wang2017priority,Wong2017,Yuang2015}.
Thus, it will be important the extend DetNet into the
data center networking domain.

\subsubsection{Application-adaptive Resource Reservations}
With the increasing number of applications on
end user devices that require network connectivity, the diversity of the traffic types has been
increasing. Traditional data included voice and user-data, such as
files and media, while the present data sources include sensor data
as well as tracking and analytics information. Time sensitive advanced
applications in the automotive and industrial sectors require special
transmission resource reservations to meet their ULL
requirements. Therefore, we believe application-based resource
reservations in L2 and PHY (i.e., proactive grants, periodic grants
and semi-persistent scheduling) across the entire network are a
promising technique to achieve the fundamental limits of ULL
end-to-end latency for the users.

\subsubsection{Integration and Support for 5G Backhaul Networks}
To meet the growing data demands of ubiquitous mobile devices, 5G
networks are expected to increase the infrastructure deployments
through small cells. The small cells are deployed close to the
users/devices, such as in shopping malls, stadiums, and on university
campuses. However, the deployment of large numbers of small cells
increases the backhaul network complexity.  Backhaul for
small cells requires deterministic latency for establishing secure IP
layer connectivity with the core networks.  DetNet can provide
backhaul connectivity for the small cells in 5G networks. However, the
integration of DetNet at the protocol level (e.g., GPRS Tunneling
Protocol (GTP) and IPSec) with the existing cellular networks is yet
to be thoroughly investigated.  Key challenges are to achieve a low
complexity overall control plane management as well as to keep the
impact on the existing 5G control plane minimal.

\subsection{5G Networks}
\subsubsection{Seamless Networks Access}
Although 5G is envisioned to support ULL and high data rates in both
the wireless air interface and the core networks, the seamless network
access across multiple operators and connectivity technologies, such
as cable and DSL networks, is still an open issue in terms of
inter-networking functions.  The inter-networking functions across
multiple networks and technology domains must be able to negotiate the
same set of services while the devices are operating in the 5G domain.

\subsubsection{Network Session Migration}
The current network connectivity technology trends, including the 5G
technology trends, enumerate several network interfaces that concurrently
connect a user device to different networks, such as WiFi, LTE, 3G,
and Ethernet. However, the actual network characteristics of each
interface change over time. For instance, in cellular communications
the transmit power is proportional to the distance from the base
stations.  Hence due to device mobility, the transmit power varies
based on the relative distance between base station and device.  While
there exists a static way of choosing the network interface based on
application requirement~\cite{cap2008ubi}, a dynamic selection based on the
network interface characteristics in real time remains an open
research challenge.  Additionally, once a session is established over
an interface, any changes in the network characteristics that impede
the connection quality would negatively impact the end-to-end latency.
To maintain low latencies, an active session should be handed over to
a different interface without interrupting the session.

\section{Conclusion}
This survey has comprehensively covered networks supporting ultra-low
latency (ULL) applications. Providing ULL support requires specialized
network protocol mechanisms that have been standardized for the link
layer in the IEEE Time Sensitive Networking (TSN) set of standards and
for the network layer in the IETF Deterministic Networking (DetNet)
specifications. In addition, extensive research studies have begun to
investigate in detail the performance characteristics and limitations
of these link and network layer ULL mechanisms.  Aside from this link
and network layer perspective, extensive standardization and research
efforts have approached ULL support from the perspective of the common
wireless device-to-core network communication chain.  In particular,
the emerging fifth generation (5G) wireless systems provide extensive
support mechanisms for ULL applications.

The survey has revealed numerous gaps and limitations of
the existing ULL networking mechanisms that present a wide range of
avenues for future research. Aside from addressing the limitations
of the individual ULL support mechanisms, there is an urgent need
to comprehensively evaluate the cooperation of the various developed
ULL mechanisms. Judicious configuration and cooperation of the various ULL
mechanisms will likely be critical for providing effective ULL services to
the end users.

\bibliographystyle{IEEEtran}


\begin{IEEEbiography}{Ahmed Nasrallah}
	is currently a researcher funded by Kuwait University to pursue
	the Ph.D.~degree in Computer Engineering at Arizona State
	University, Tempe. He received his B.Sc.~degree in Electrical
	and Computer Engineering from the University of Dayton, Dayton
	Ohio, and his M.S.~degree in Computer Engineering from Arizona
	State University. His research is focused on communication
	and multimedia networking.
\end{IEEEbiography}

\begin{IEEEbiography}{Akhilesh Thyagaturu}
	is an Engineer at Intel Corporation, Chandler, AZ, USA,
	and an Adjunct Faculty in the School of Electrical, Computer, and
	Energy Engineering at Arizona State University (ASU), Tempe. He
	received the Ph.D. in electrical engineering from Arizona State
	University, Tempe, in 2017.
	He serves as reviewer for various journals
	including the \textit{IEEE Communications Surveys \& Tutorials},
	\textit{IEEE Transactions of Network and Service Management}, and
	\textit{Optical Fiber Technology}.  He was with Qualcomm Technologies
	Inc., San Diego, CA, USA, as an Engineer from 2013 to 2015.
\end{IEEEbiography}

\begin{IEEEbiography}{Ziyad Alharbi}
is a researcher at King Abdulaziz City for Science and Technology
(KACST), Riyadh, Saudi Arabia. He received his B.Sc. degree in
Electrical Engineering from King Fahd University of Petroleum and
Minerals, Saudi Arabia, and his M.S. in Electrical Engineering from
Arizona State University, Tempe. Currently, he is working towards his
Ph.D. in Electrical Engineering at Arizona State University. He serves
as reviewer for various journals including \textit{IEEE Communications
Surveys \& Tutorials}, \textit{Computer Networks}, and
\textit{Optical Switching and Networking}
\end{IEEEbiography}

\begin{IEEEbiography}{Cuixiang Wang}
is a lecturer in the school of Information Engineering at Yancheng
Institute of Technology, Yancheng, China. Currently, she is also a
visiting scholar in the School of Electrical, Computer, and Energy
Engineering at Arizona State University (ASU), Tempe.  She received
her B.Sc. degree in Computer Science and Technology from Qufu Normal
University, Qufu, China, in 2007 and her M.S. degree in Computer
Software and Theory from Nanjing University of Posts and
Telecommunications, Nanjing, China in 2010. Her research is focused on
wireless network routing and network simulation.
\end{IEEEbiography}

\begin{IEEEbiography}{Xing Shao}
is an associate professor in the school of Information Engineering at
Yancheng Institute of Technology, Yancheng, China. Currently, he is
also a visiting scholar in the School of Electrical, Computer, and
Energy Engineering at Arizona State University (ASU), Tempe.  He
received the Ph.D. in Information Network from Nanjing University of
Posts and Telecommunications, Nanjing, China, in 2013. His research is
focused on wireless multihop networking, routing and network
simulation.
\end{IEEEbiography}

\begin{IEEEbiography}{Martin Reisslein} (S'96-M'98-SM'03-F'14)
is a Professor in the School of Electrical, Computer, and Energy
Engineering at Arizona State University (ASU), Tempe. He received the
Ph.D. in systems engineering from the University of Pennsylvania in
1998. He currently serves as Associate Editor for the \textit{IEEE
  Transactions on Mobile Computing}, the \textit{IEEE Transactions on
  Education}, and \textit{IEEE Access} as well as \textit{Computer
  Networks} and \textit{Optical Switching and Networking}. He is
Associate Editor-in-Chief for the \textit{IEEE Communications Surveys
  \& Tutorials} and chairs the steering committee of the \textit{IEEE
  Transactions on Multimedia}.
\end{IEEEbiography}

\begin{IEEEbiography}{Hesham ElBakoury}
is a thirty five year veteran in the telecommunications and data
networking industry with an extensive background and expertise in the
architecture, design, and development of Distributed Systems and
Broadband Access, Enterprise and Telco Communications Systems. He is a
Principal Architect in Futurewei focusing on advanced technology
research and standards in the network research Lab.
Prior to joining Futurewei, Mr. ElBakoury was Chief Systems Architect
in Hitachi-CTA EPON Access Systems Division, and Chief Systems
Architect in Nortel and Bell-Northern Research where he led the
architecture, design, and development of several very successful
Switching/Routing, Security and Carrier Ethernet products. In
Nortel/BNR, Mr. ElBakoury initiated and led the Autonomic Network
research project in the Enterprise division, and the Software Design
and Code Reuse Project in the Data Networking division.
Mr. ElBakoury has been active in different standard groups, including
IEEE 802, IEEE 1904, IETF, ONF, OIF, MEF, the SCTE Energy 2020
program, and CableLabs where he has been heavily involved in IEEE
802.3/802.1, DPoE/DPoG, DOCSIS 3.1, Full-Duplex DOCSIS 3.1, SDN/NFV,
Distributed CCAP Architectures, and Business Services projects. He
holds an M.S.C. degree from Waterloo University, ONT, Canada.
\end{IEEEbiography}

\end{document}